\documentclass[letter,apj,twocolappendix,numberedappendix,appendixfloats]{emulateapj}

\usepackage{natbib}
\usepackage{amsmath}
\usepackage{color}
\usepackage{graphicx}
\usepackage{epstopdf}
\usepackage{threeparttablex} 
\usepackage{booktabs}        

\usepackage{graphicx}

\newcommand{\nsn}     {213}

\usepackage{url}
\usepackage{hyperref}
\hypersetup{
            pdftitle={Light Curves of \nsn\ Type Ia Supernovae from the ESSENCE Survey},
            pdfauthor={Narayan et al.}
            bookmarks=true,
            bookmarksnumbered=true,
            pdfpagelabels=true,
            colorlinks=true,
            anchorcolor=blue,
            citecolor=blue,
            linkcolor=blue
           }

\newcommand{\lt}{<}
\newcommand{\gt}{>}

\newcommand{\snia}    {SN~Ia}

\newcommand{\sn}      {SN}

\newcommand{\chisq}{\ensuremath{\chi^2}}
\newcommand{\chisqnu}{\ensuremath{\chi^2/{\rm dof}}}

\def\lsim{\hbox{\rlap{\raise 0.425ex\hbox{$<$}}\lower 0.65ex\hbox{$\sim$}}}
\def\gsim{\hbox{\rlap{\raise 0.425ex\hbox{$>$}}\lower 0.65ex\hbox{$\sim$}}}
\newcommand{\about}{$\sim\!\!$~}
\newcommand{\RNum}[1]{\textsc{\expandafter{\romannumeral #1\relax}}}

\def\arcmin{\hbox{$^\prime$}}
\def\arcsec{\hbox{$^{\prime\prime}$}}
\def\arcdeg{\mbox{$^\circ$}}

\newcommand{\gri}     {\protect\hbox{$gri$} }
\newcommand{\ri}      {\protect\hbox{$RI$} }
\newcommand{\griz}    {\protect\hbox{$griz$} }

\shorttitle{ESSENCE 6-Year Data Release}
\shortauthors{Narayan et~al.}

\begin{document}
\title{Light Curves of \nsn\ Type Ia Supernovae from the ESSENCE Survey}
\author{G. Narayan\altaffilmark{1,2,3}, A. Rest\altaffilmark{4}, B. E. Tucker\altaffilmark{5}, R. J. Foley\altaffilmark{6,7}, W. M. Wood-Vasey\altaffilmark{8}, P. Challis\altaffilmark{2}, C. Stubbs\altaffilmark{2,3}, R. P. Kirshner\altaffilmark{2}, C. Aguilera\altaffilmark{9},  A. C. Becker\altaffilmark{10}, S. Blondin\altaffilmark{11}, A. Clocchiatti\altaffilmark{12}, R. Covarrubias\altaffilmark{10}, G. Damke\altaffilmark{13}, T. M. Davis\altaffilmark{14}, A. V. Filippenko\altaffilmark{15},  M. Ganeshalingam\altaffilmark{15}, A. Garg\altaffilmark{2}, P. M. Garnavich\altaffilmark{16},  M. Hicken\altaffilmark{3}, S. W. Jha\altaffilmark{17}, K. Krisciunas\altaffilmark{18}, B. Leibundgut\altaffilmark{19}, W. Li\altaffilmark{15,20}, T. Matheson\altaffilmark{1}, G. Miknaitis\altaffilmark{21}, G. Pignata\altaffilmark{22}, J. L. Prieto\altaffilmark{23}, A. G. Riess\altaffilmark{4,24}, B. P. Schmidt\altaffilmark{5},  J. M. Silverman\altaffilmark{25}, R. C. Smith\altaffilmark{8}, J. Sollerman\altaffilmark{26}, J. Spyromilio\altaffilmark{19}, N. B. Suntzeff\altaffilmark{18}, J. L. Tonry\altaffilmark{27}, and A. Zenteno\altaffilmark{9}}
\altaffiltext{1}{National Optical Astronomy Observatory, 950 North Cherry Avenue, Tucson, AZ 85719, USA} 
\altaffiltext{2}{Harvard-Smithsonian Center for Astrophysics, 60 Garden Street, Cambridge, MA 02138, USA} 
\altaffiltext{3}{Department of Physics, Harvard University, 17 Oxford Street, Cambridge, MA 02138, USA} 
\altaffiltext{4}{Space Telescope Science Institute, 3700 San Martin Drive, Baltimore, MD 21218, USA}
\altaffiltext{5}{The Research School of Astronomy and Astrophysics, Australian National University, Mount Stromlo Observatory, via Cotter Road, Weston Creek, ACT 2611, Australia}
\altaffiltext{6}{Department of Astronomy, University of Illinois at Urbana-Champaign, 1002 W. Green Street, Urbana, IL 61801, USA} 
\altaffiltext{7}{Department of Astronomy, University of Illinois at Urbana-Champaign, 1010 W. Green Street, Urbana, IL 61801, USA} 
\altaffiltext{8}{Department of Physics and Astronomy, University of Pittsburgh, Pittsburgh, PA 15260, USA}
\altaffiltext{9}{Cerro Tololo Inter-American Observatory, National Optical Astronomy Observatory, Casilla 603, La Serena, Chile}
\altaffiltext{10}{Department of Astronomy, University of Washington, Box 351580, Seattle, WA 98195-1580, USA}
\altaffiltext{11}{Aix Marseille Universit\'e, CNRS, LAM (Laboratoire d'Astrophysique de Marseille) UMR 7326, 13388, Marseille, France}
\altaffiltext{12}{Pontificia Universidad Catolica de Chile, Instituto de Astrofisica, Casilla 306, Santiago 22, Chile and Millenium Institute of Astrophysics, Chile}
\altaffiltext{13}{Department of Astronomy, University of Virginia, Charlottesville, VA 22904-4325, USA}
\altaffiltext{14}{School of Mathematics and  Physics, University of Queensland, Brisbane, QLD 4072, Australia}
\altaffiltext{15}{Department of Astronomy, 501 Campbell Hall, University of California, Berkeley, CA 94720-3411}
\altaffiltext{16}{Department of Physics, University of Notre Dame, 225 Nieuwland Science Hall, Notre Dame, IN 46556-5670, USA}
\altaffiltext{17}{Department of Physics and Astronomy, Rutgers, The State University of New Jersey, Piscataway, NJ 08854, USA}
\altaffiltext{18}{Department of Physics and Astronomy, Texas A\&M University, College Station, TX 77843-4242, USA}
\altaffiltext{19}{European Southern Observatory, Karl-Schwarzschild-Strasse 2, D-85748 Garching, Germany}
\altaffiltext{20}{Deceased 2011 December 12.}
\altaffiltext{21}{Fermilab, P.O. Box 500, Batavia, IL 60510-0500, USA}
\altaffiltext{22}{Departamento de Ciencias Fisicas, Universidad Andres Bello, Avda. Republica 252, Santiago, Santiago RM, Chile}
\altaffiltext{23}{Astronomy Nucleus, Faculty of Engineering, Universidad Diego Portales, Ej\'ercito 441, Santiago, Santiago RM, Chile }
\altaffiltext{24}{Johns Hopkins University, 3400 North Charles Street, Baltimore, MD 21218, USA}
\altaffiltext{25}{Department of Astronomy, University of Texas, Austin, TX 78712-0259, USA}
\altaffiltext{26}{Oskar Klein Centre, Department of Astronomy, AlbaNova, Stockholm University, 10691, Stockholm, Sweden}
\altaffiltext{27}{Institute for Astronomy, University of Hawaii, 2680 Woodlawn Drive, Honolulu, HI 96822, USA}
\email{gnarayan@noao.edu}

\begin{abstract}
The ESSENCE survey discovered \nsn\ Type Ia supernovae at redshifts $0.1 < z <
0.81$ between 2002 and 2008. We present their $R$ and $I$-band photometry,
measured from images obtained using the MOSAIC~II camera at the CTIO 4~m Blanco telescope, along
with rapid-response spectroscopy for each object. We use our spectroscopic
follow-up observations to determine an accurate, quantitative classification
and a precise redshift.  Through an extensive calibration program we have
improved the precision of the CTIO Blanco natural photometric system. We use
several empirical metrics to measure our internal photometric consistency and
our absolute calibration of the survey. We assess the effect of various
potential sources of systematic bias on our measured fluxes, and we estimate that the
dominant term in the systematic error budget from the photometric calibration
on our absolute fluxes is \about1\%.  
\end{abstract}

\keywords{cosmology: observations --- methods: data analysis --- stars: supernovae: general, surveys}


\section{Introduction}\label{sec:intro}

We present the calibrated photometry of \nsn\ Type Ia supernovae (\snia)
measured by the ESSENCE (Equation of State: Supernovae trace Cosmic Expansion)
survey between 2002 and 2008. Our report more than doubles the sample presented
by \citet{miknaitis07} and \citet{wv07}. We have made a significant effort to
improve the photometric calibration of the survey.  As ESSENCE observed in only
two passbands, our measurements of luminosity distance are strongly correlated
with extinction in the host galaxy of the \snia\ and very sensitive to the
systematic error budget from photometry. In particular, the light curves in
this work are computed using data taken only with the Blanco 4~m telescope at
the Cerro-Tololo Inter-American Observatory, eliminating cross-telescope
systematics present in the calibration by \citet{miknaitis07}. A companion work
\citep[][submitted]{Tucker16} reports on properties of the host galaxies of our
\snia\ sample. In future work, we will use this sample along with low-redshift
\snia\ from the literature to perform a full cosmological analysis and improve
constraints on the nature of dark energy.

Since the discovery of the luminosity-width-color relation \citep{phillips93}, \snia\
have been our most precise standardizable candles at cosmological distances.
The initial Cal\'an-Tololo sample of 29 \sn\ in 4 colors \citep{hamuy96}
enabled the development of various algorithms capable of correcting the
dispersion in the intrinsic brightness of \snia\ and inferring the luminosity
distances to \about10\% per object \citep{riess96,phillips99,goldhaber01}.
These light-curve fitters have been refined as the size of the nearby sample
has increased and its photometric precision has improved; current algorithms
can determine the luminosity distance to well-observed \snia\ to \about5\%
\citep{JRK07,guy07,conley08,Mandel11}.
 
The distance moduli derived for these \snia\ indicated that the Universe's expansion is
\emph{accelerating} \citep{riess98,perlmutter99}.  \snia\ observations have
remained our most sensitive cosmological probe of the expansion history. The
accelerating expansion has been modeled by introducing a fluid with negative
pressure, called ``dark energy,'' into the Friedmann equation: 

\begin{equation}
\begin{split}
h^{2}(a) = h_{0}^{2}\left(\frac{\Omega_{M}}{a^{3}} + \frac{\Omega_{\nu}}{a^{4}} + \frac{\Omega_{k}}{a^{2}} + \Omega_{\text{DE}}\exp[3(1+w)] \right),
\label{eqn:friedmann}
\end{split}
\end{equation}

\noindent where $h$ is the Hubble parameter, $h_{0} = {\rm H}_{0}/100$~km s$^{-1}$
Mpc$^{-1}$, $a$ is the scale factor, and $\Omega$ is the total energy density
of matter ($M$), photons ($\nu$), curvature ($k$), and dark energy (DE),
respectively. Several groups have focused on measuring the ratio of pressure to
density --- the equation-of-state parameter of this fluid, $w = P/(\rho c^{2})$ --- to distinguish between different models of the dark energy. 

High-redshift \snia\ surveys \citep{wv07,riess07,guy10,betoule14,sako14} have
independently reported measurements of $w$ consistent with $-1$, in good
agreement with a classical cosmological constant. However, despite the rapidly
growing number of \snia, the precision of the measurement of $w$ has stubbornly
remained at the 10\% level, dominated by various sources of systematic
uncertainty. Several groups have attempted to reduce the effect of systematic
errors in \snia\ measurements on the dark energy figure of merit
\citep[FoM;][]{albrecht06}, by either incorporating new sources of data or
improving the calibration of existing data. 

Early work by \citet{krisciunas00} demonstrated uniformity in the evolution of
near-infrared (NIR) colors of \snia, and the potential of NIR measurements for cosmology
\citep{krisciunas04}. Using increasingly larger and better-calibrated samples of
nearby \snia\ with $JHK_{s}$ measurements, \citet{woodvasey08},
\citet{mandel09}, and \citet{baronenugent12} have shown that the NIR light
curves of \snia\ span a smaller range in luminosity than in the optical.
Because distance moduli derived from NIR measurements are less susceptible to
host-galaxy dust absorption, the residual scatter in a Hubble diagram generated
from NIR light curves alone is comparable to the scatter derived from
\emph{light-curve-shape corrected} optical data. Consequently, high-redshift
surveys have increasingly attempted to probe further into the rest-frame NIR.
\citet{freedman09} presented the first NIR Hubble diagram to redshift $z \approx 0.7$, but
were limited by a relatively small sample size, systematic uncertainties in
their photometric calibration, and the difficulty of obtaining rest-frame NIR 
data at high $z$, where the light is redshifted to even longer wavelengths. 
Future high-$z$ surveys, such as RAISIN (PI R. P. Kirshner, {\it HST} GO-13046),
will provide valuable high-$z$ \snia\ measurements that probe the rest-frame 
NIR.

\citet{kelly10} illustrated that in addition to demographic differences between
\snia\ in passive and star-forming hosts, the Hubble-diagram residuals are
correlated with derived host-galaxy size and stellar mass. This correlation
indicates that the empirical luminosity-shape relations employed by \snia\
light-curve fitters do not fully account for the spread in peak
luminosity. In an effort to reduce this dispersion, \citet{lampeitl10b}
employed a simple linear correction based on host-galaxy stellar mass and found
an improvement in statistical fit to the \snia\ measurements.
\citet{sullivan10} used different \snia\ absolute magnitudes for high-mass 
and low-mass hosts in their cosmological fits, finding a significant improvement in
\chisq\ over using a relation expressed as a function of host-galaxy stellar
mass. 

However, although metallicity, extinction properties, and specific 
star-formation rate correlate with host-galaxy mass, the fundamental relation
underlying this correlation with \snia\ luminosity is not well understood.
These relations may be an artifact of the treatment of \snia\ color by 
light-curve fitters; \citet{Scolnic14b} found that the strength of correlation of the
host-galaxy properties with Hubble residual was reduced by \about20\% when
\snia\ are treated as having an intrinsic color scatter for a fixed luminosity
distance, rather than an achromatic scatter in peak luminosity.  In addition,
there are challenges in deriving host-galaxy properties from broadband optical
photometry at high redshift in a manner that does not introduce additional
systematic uncertainty into \snia\ measurements. ESSENCE has undertaken a
significant effort to determine host-galaxy morphology and properties for our
sample, to appear in \citet[][submitted.]{Tucker16}. 

Many authors \citep{Wang09,blondin11,nordin11,foley11a,walker11,silverman12}
have found that measurements from spectra of \snia\ correlate with the residual
intrinsic color dispersion after light-curve shape correction. They further
find that these measurements, typically derived from pseudoequivalent widths
of Ca or Si, can be used to improve the precision of distance moduli, although
\citet{blondin11} find that the improvement is not statistically significant
($\lt 2\sigma$). While promising, this approach is limited by the need for
high signal-to-noise ratio (S/N) spectra of \snia.  
Additionally, the dependence on measuring the Si
\RNum{2} $\lambda$6355 feature limits its use at high $z$, where the redshifted Si
features are often not covered by the high-throughput low-dispersion
spectrographs used by \snia\ surveys. 

Surveys such as ESSENCE, the Supernova Legacy Survey (SNLS), the Sloan
Digital Sky Survey (SDSS), and the Panoramic Survey Telescope and 
Rapid Response System (Pan-STARRS) have now produced well over
1000 well-sampled \snia\ light curves that span the redshift range over
which the transition from cosmic deceleration to acceleration occurred.  The
crucial measurement for characterizing the nature of dark energy is mapping out
luminosity distance versus redshift, to constrain the parameters of
Equation~\ref{eqn:friedmann}. The precision of photometric calibration is now
the dominant term of the \snia\ survey systematic error budget. \citet{wv07}
found that systematic uncertainties from the photometry alone could lead to a
\about4\% change in $w$. Exploiting the improved statistics from these large
samples requires a corresponding improvement in the photometric calibration
across diverse instruments, detectors, and filters.  

The most important aspect of this calibration challenge is to establish a
well-grounded understanding of flux measurements made in different broad
optical passbands. This in turn requires adopting a spectrophotometric
standard that serves as the metrology basis for relating fluxes across the
bands being used. In essence, we need to be able to distinguish cosmological
evolution in the luminosity distance vs. redshift relation from 
cross-band calibration issues.

There are two methods in use or in development for flux calibration at CCD wavelengths.
\begin{enumerate}
\item Adopt an astrophysical source, particularly Vega, as a celestial transfer
standard, with ground-based blackbody emitters as the fundamental calibration
sources. This is a long-standing method, serving as the basis for Vega-based
magnitudes \citep{oke70,Hayes75}, and it underpins the \citet{landolt92} standard-star network. 
\item Use well-calibrated laboratory standards (such as silicon photodiodes
from NIST) as the foundational metrology layer, and measure the system
throughput in comparison to these devices. This was the approach
advocated by \citet{stubbs06}, and it is now in various stages of implementation by
Pan-STARRS \citep{tonry12,Rest13}, SNLS \citep{regnault09}, the joint efforts of
SDSS and SNLS \citep{betoule13,betoule14}, the Dark Energy Survey (DES), and
the Large Synoptic Survey Telescope (LSST).  
\end{enumerate}

While the first method is well established, \snia\ surveys require a higher
level of precision than is possible with existing standard-star networks. The
second method is still nascent, and systems to measure the atmospheric
component of the throughput are under active development \citep{altair14}. No
purely laboratory-standard-based magnitude system yet exists. Several surveys,
including ESSENCE, have elected to use a combination of both methods; the first
to determine the absolute-flux calibration, and the second to determine precise
relative system throughputs.

\citet{kessler09a} demonstrated that measurements of $w$ are extremely
sensitive to the calibration of the $U$ band at low redshift: inclusion of 
rest-frame $U$-band data at all redshifts causes a 0.12 mag shift in distance
moduli, corresponding to an enormous 0.3 change in $w$. 
The $U$-band anomaly might arise from differences between the
spectral energy distribution (SED) of \snia\ that correlate with host-galaxy
properties or between objects at low and high redshift
\citep{foley12,maguire12}. Additionally, $U$-band measurements of the same
nearby \snia\ from different telescopes often exhibit differences that are
inconsistent with the stated photometric uncertainties and system-throughput
measurements. \citet{Krisciunas13} have demonstrated that careful modeling of
the $U$-band transmission with appropriate $S$-corrections can resolve the
differences between \snia\ measurements. The size of the systematics associated
with the $U$ band, however, has led most high-$z$ surveys to downweight or
discard rest-frame ultraviolet (UV) observations. 

Larger, more precisely calibrated nearby samples
\citep{Ganeshalingam10,Stritzinger11,Hicken12}, along with better calibration
of high-$z$ \snia\ surveys, offer the most direct path to reducing the
systematic uncertainty on $w$,
Wide-field deep surveys such as Pan-STARRS and DES will
obtain \snia\ measurements over $0 < z < 1.2$ \citep{Rest13}, further reducing
systematic uncertainties from photometry by avoiding any errors associated with
cross-telescope calibration and weakening the sensitivity of $w$ to the overall
photometric calibration of the survey \citep{Scolnic13}.  Recognizing the need
for precision calibration to reduce systematics \citep{stubbs06, tucker2007},
and following the example set by SDSS \citep{ivezic07,Padmanabhan08}, 
current surveys have undertaken
ambitious calibration programs. These efforts combine high-precision
measurements of system throughput calibrated to laboratory standards, with
atmospheric data and repeated observations of stellar standards to obtain
$\lt1$\% photometry over much of the sky \citep{stubbs12, schlafly12, tonry12}.
The work presented here details the calibration of the
ESSENCE survey, with a focus on
minimizing the systematic error budget from photometry. 

We provide a brief overview of the ESSENCE survey in \S\ref{sec:survey},
followed by our photometric data reduction and calibration in
\S\ref{sec:datared}. Our spectroscopic follow-up observations and 
classification are described in \S\ref{sec:spec}. We illustrate our \snia\ light curves, compare and
contrast our methodologies for light-curve fitting, and detail the properties
of the full ESSENCE 6-year sample in \S\ref{sec:lccomp}. Our photometric error
budget from various sources with systematic uncertainty is discussed in
\S\ref{sec:syserr}.  We conclude in \S\ref{sec:conclusions}. The Appendices
contain further information on the computation of illumination corrections, the
properties of the CTIO Blanco natural-magnitude system employed in this work,
tables containing the photometry of ESSENCE \snia\ and likely \snia\ without
spectroscopic confirmation (hereafter,``Ia?'') during the year of
discovery, and light-curve-fit parameters using the two most common
methodologies.
\section{The ESSENCE Survey}\label{sec:survey}

Previous ESSENCE publications have described the survey strategy, fields, data
processing \citep[][hereafter M07]{miknaitis07}, spectroscopic selection
criteria and follow-up observations \citep{matheson05, foley09}, performed a preliminary
cosmological analysis \citep[][hereafter WV07]{wv07}, and scrutinized exotic
cosmological models \citep{davis07}. The \snia\ search was carried out on the
CTIO 4~m Blanco telescope (hereafter, Blanco) over 197 half-nights in dark and gray
time between September and January from 2002 to 2008. Science images were
obtained using the 64 Mega pixel MOSAIC~II camera with an Atmospheric
Dispersion Corrector (ADC) through two primary filters (denoted $R$ and $I$)
similar to Cousins $R_{C}$ and $I_{C}$. The field of view of the system is 0.36
deg$^{2}$ on the sky at the $f/2.87$ prime focus.  

The imager consists of eight 2k $\times$ 4k CCDs arranged in two rows of four.  Each CCD
is bisected along its length, and each section is read out in parallel,
resulting in 16 amplifier images for every science exposure. Readout times are
$\sim 100$~s.  Each pixel subtends 0.27\arcsec\ at the center of
the field. Optical distortions cause a radial variance of \about8\% in the
pixel scale. 

\begin{deluxetable}{cccc} 
\tabletypesize{\scriptsize}
\tablewidth{0pt}
\tablecolumns{4}
\tablecaption{Primary ESSENCE Fields\label{tab:fields}}
\tablehead{
    \colhead{Field} &
    \multicolumn{2}{c}{$\alpha$ (J2000)  $\delta$(J2000)} &
    \colhead{$N$ Images}                                  \\
    \colhead{}                                          &        
    \colhead{{h}\phn{m}\phn{s}}                         &
    \colhead{\phn{\arcdeg}~\phn{\arcmin}~\phn{\arcsec}} &
    \colhead{}                                         
}
\startdata
waa & 23:27:27 & $-$09:51:00 & 172\\
wbb & 01:12:00 & $-$00:20:17 & 275\\
wcc & 02:07:41 & $-$04:55:00 & 289\\
wdd & 02:28:36 & $-$08:24:17 & 293
\enddata
\end{deluxetable}

The survey covered a set of 4 primary fields (listed in Table~\ref{tab:fields},
together with the number of times each field was observed), each consisting of 8 subfields
clustered spatially. Fields were selected to be equatorial but outside the
Galactic and ecliptic planes, in regions with low Milky Way extinction and
minimal IR cirrus, and with coverage from existing surveys (including SDSS, the
NOAO Deep Wide-Field Survey, and the Deep Lens Survey) where possible. The
fields were spaced to ensure that science images could be taken at low airmass.
Fields were divided into two sets, and each set was imaged in both filters every
other observing night resulting in a typical cadence of 4 days. Science frames
are exposed for 200~s in $R$ and 400~s in $I$. The original $I$
filter (NOAO code c6005) sustained significant damage on 2002 Nov 10, severely
degrading the image quality of $I$-band data in CCDs 1 and 2 (amplifiers 1--4).
The filter was replaced on 2003 May 25. CCD~3 failed shortly before the start
of the 2003 observing season, resulting in a 12.5\% loss in efficiency until it
was replaced in 2004.

Survey images were reduced at CTIO using the ``\texttt{photpipe}'' pipeline developed
for use on the CTIO Blanco by the SuperMACHO survey
\citep{Rest2005,miknaitis07,garg07} that operated contemporaneously with the
ESSENCE survey.  Each science image was calibrated and aligned with a fixed
astrometric grid. We subtracted a reference template for each field,
constructed using deep images from previous observations. Point-spread function (PSF) photometry from
the resulting difference image was combined to identify sources that had varied
over multiple epochs, while eliminating sources of contamination such as
difference-image artifacts and diffraction spikes from saturated stars. With
limited time for spectroscopic follow-up observations, we were forced to employ various cuts
and selection criteria in order to determine the most promising candidates.

The spectroscopic follow-up observations of ESSENCE candidates are described in
\S~\ref{sec:spec}. All candidates were visually inspected to select 
spectroscopic targets to classify them and obtain redshifts. We produced a
preliminary reduction of all spectra in real time, using standard
\texttt{IRAF}\footnote{\texttt{IRAF} is distributed by the National Optical Astronomy
Observatory, which is operated by AURA, Inc. under cooperative agreement with the
NSF.} routines, and some custom \texttt{IDL} routines to facilitate data processing for
the various instruments.  Estimates of the redshift and classification were
obtained onsite using \texttt{SNID} \citep{blondin07}. When preliminary classifications
were unclear, we relied on the experience of the observers to determine if
additional spectroscopic observations were warranted. 

Fields containing candidates
with a clear classification as \snia\ were monitored for the remainder of the
observing season. Following survey operations, all data were transferred,
initially to the Hydra Computing Cluster maintained by the Smithsonian
Institution, and later to the Odyssey Compute Cluster, hosted by the Research
Computing Group at Harvard University for the analysis presented in this paper.
All data are also available through the NOAO
archive\footnote{\url{http://archive.noao.edu/nsa/}}.
\section{Data Reduction}\label{sec:datared}

\subsection{Image Detrending}\label{sec:detrend}

The eight CCDs of MOSAIC~II were read out in pairs, through two amplifiers per
chip, by four Arcon controllers.  The crosstalk between the amplifiers is
subtracted using the $xtalk$ task from the $mscred$ package in IRAF. All CCD
images are debiased and trimmed, and masking was applied to bad pixels and
columns. The mask is propagated through all subsequent reduction stages.

All science images are flat-field corrected using dome flats. These flats
accurately corrected for pixel-to-pixel variations, but large-scale variations
were introduced as a result of uneven illumination of the dome screen and stray
light paths in the optical system.  While the precision obtained from dome-flat
images alone is suitable for many projects, we required higher precision for
\snia\ cosmology and strived to minimize potential systematic errors in our
photometry. We therefore accounted for large-scale illumination variation by
constructing an illumination correction from the science images, as described
below.

We applied the nightly dome-flat image to all science images to construct a
temporary preliminary flattened image. The resulting images were masked to
remove contamination from all astrophysical sources, normalized to have the
same sky value, and then averaged. The derived calibration image was inverted,
smoothed with a large kernel, and scaled to have a mean of unity. This
illumination correction was applied to the dome-flat images to take out
residual large-scale gradients. The science images were reprocessed with this
final flat-field image. 

\begin{figure*}[htb]
\centering
\includegraphics[width=\textwidth]{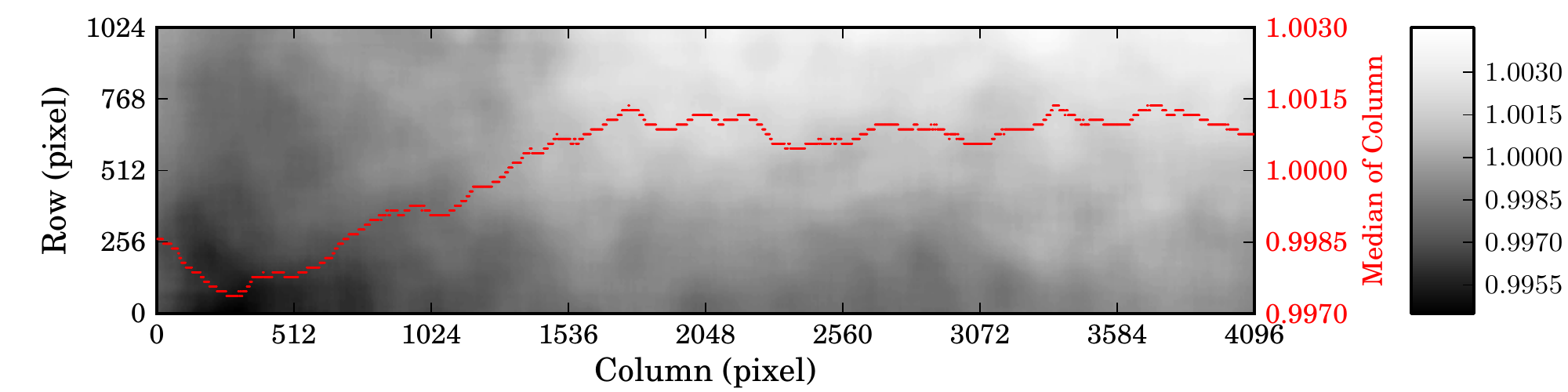}
\caption[Representative Illumination Correction Frame]{A representative $R$-band
illumination correction for Amplifier 6 of the MOSAIC~II. The primary structure
in the illumination correction is a \about0.5\% gradient from left to right and
top to bottom. The median value of each column is indicated in red. The bar at
right indicates the grey-scale values.}\label{fig:illum}
\end{figure*}

To estimate the night-to-night stability of the illumination correction, we
took the ratio of the correction image between different nights of a single run
--- a period of time during which MOSAIC~II was continuously mounted on the
telescope, typically one lunation. We found that the gradient pattern (a
representative example is shown in Fig.~\ref{fig:illum}) was very stable within
a lunar cycle. The standard deviation of the ratio without sigma-clipping was
typically less that 0.1\%, and the absolute value of the maximum difference
between the ratio and the average of the ratio image was $<0.003$. Therefore,
on nights with few science images of sparse fields or with excess stray light
--- either from insufficient baffling or around the time of full moon --- we exploited the
stability of the gradient pattern to estimate the illumination correction from
nearby nights. This estimation and temporal stability of the illumination
corrections is examined in further detail in Appendix~\ref{sec:illum}.

Surveys that use master flats constructed for each run are susceptible to
systematic trends, such as long-period variations in amplifier gain. By
contrast, our procedure avoids such effects: science frames were normalized
with nightly flat frames and primarily used illumination corrections determined
from the same, or at the least extrapolated only from nearby nights.

\subsection{Astrometric Calibration}

In order to construct difference images to search for and measure the flux of
variable and transient objects, we first imposed a consistent astrometric
solution and warped all the science images to a consistent pixel coordinate
system. The transformation between the local image pixel coordinate system and
the FK5 World Coordinate System is dominated by optical distortions that are
well described by a low-order polynomial in radius from the field center. We
determined the polynomial terms of the distortion function from images of dense
Large Magellanic Cloud fields using the \texttt{IRAF} task \texttt{msctpeak}. The distortion terms were used in
combination with the \texttt{IRAF} task \texttt{msccmatch} to derive a world coordinate system (WCS) solution for each
field. The distortion terms were recomputed monthly as they vary over 
timescales of 6~months. If left uncorrected, this variation would introduce
systematic offsets at the \about0.01\arcsec\ level.

With the distortion modeled, the astrometric solution for any image with the
equatorially mounted Blanco reduces to determining the linear rotation matrix with
respect to the center.  We used the \texttt{IRAF} task \texttt{mscmatch} from the \texttt{mscred}
package to match pixel coordinates for objects in the image to an existing
catalog of the field with precise astrometry. We generated an initial
astrometric solution for the survey using reference catalogs derived from the
SDSS DR7 \citep{DR7} wherever possible, defaulting to
astrometry from the USNO CCD Astrograph catalog 2
\citep[UCAC;][]{Zacharrias04} where SDSS coverage was unavailable. As the SDSS
is itself tied to the UCAC, and as we only require precise \emph{relative}
astrometric calibration to precisely position the PSF and measure flux, errors
caused by the differences of the astrometric solution between the two different
reference catalogs are negligible. We used this initial solution to generate
secondary astrometric catalogs using our multiple observations of each field. 

Finally, we used the astrometric solution and the \texttt{SWarp} \citep{Bertin02}
package to resample each image to a common pixel coordinate system using a
flux-conserving Lanczos-windowed sinc kernel. We generated weight maps for each
image to account for the change in the noise properties produced by
resampling. Some covariance between pixels is introduced as a result of the
resampling process and we accounted for it during difference imaging.

\subsection{Flux Measurement}

We used the \texttt{DoPHOT} photometry package \citep{schechter93} to
identify and measure sources within the warped images. \texttt{DoPHOT} is
appropriate for point-source photometry. \citet[][submitted]{Tucker16} will
report on photometry of extended sources.

\subsection{Photometric Calibration}

High-redshift \snia\ surveys typically report observations in their natural
photometric system, relating magnitudes to measured flux via

\begin{equation}
\begin{split}
m_{T,i} &= -2.5\log_{10}(\phi_{{\rm ADU},T,i}) + \text{ZP}_{T,i},
\end{split}
\end{equation}

\noindent where $m$ is the natural magnitude, $\phi$ is the measured flux, and
ZP$_{T,i}$ is the instrumental zero point of image $i$ observed through
passband $T$. 

Natural magnitudes have several advantages: they allow surveys to schedule
observations in different passbands independently, as the \snia\ colors at
every epoch are not needed, and they avoid the additional photometric errors
that arise from converting the observed SN flux to a standard system.
These transformations are nontrivial, as the simple linear transformations
derived for stars are not directly applicable to \snia\ with their more complex
SEDs.  However, as these measurements are reported in a nonstandard magnitude
system, surveys must establish a network of stellar calibrators in the natural
system of the telescope to derive accurate and precise zero points. In
addition, an accurate model of the survey throughput in each passband is
required so that measurements in the natural system can be compared to synthetic
fluxes generated from models derived from \snia\ measurements at low redshifts
in the standard system. We have developed various metrics to quantify our
internal photometric consistency, and we verified our zero-point consistency using
the SDSS. We detail the improvements to the photometric calibration for the
survey in the next subsections.  

\subsubsection{Aperture Corrections}

The extended aureole of astrophysical objects has a surface-brightness profile
that roughly follows $r^{-2}$, and a large fraction of the flux is
outside the seeing disk. Thus, an aperture larger than the seeing disk is
necessary for the enclosed flux to be a reliable estimator of the true source
flux. However, the larger the aperture, the higher the uncertainty from sky
subtraction, and the higher the probability of enclosing contaminating sources.
We follow the standard technique of addressing this tradeoff by measuring the
flux in a fixed aperture, and determining an aperture correction to correct for
its finite size.

\begin{figure*}
\centering
\includegraphics[width=\textwidth]{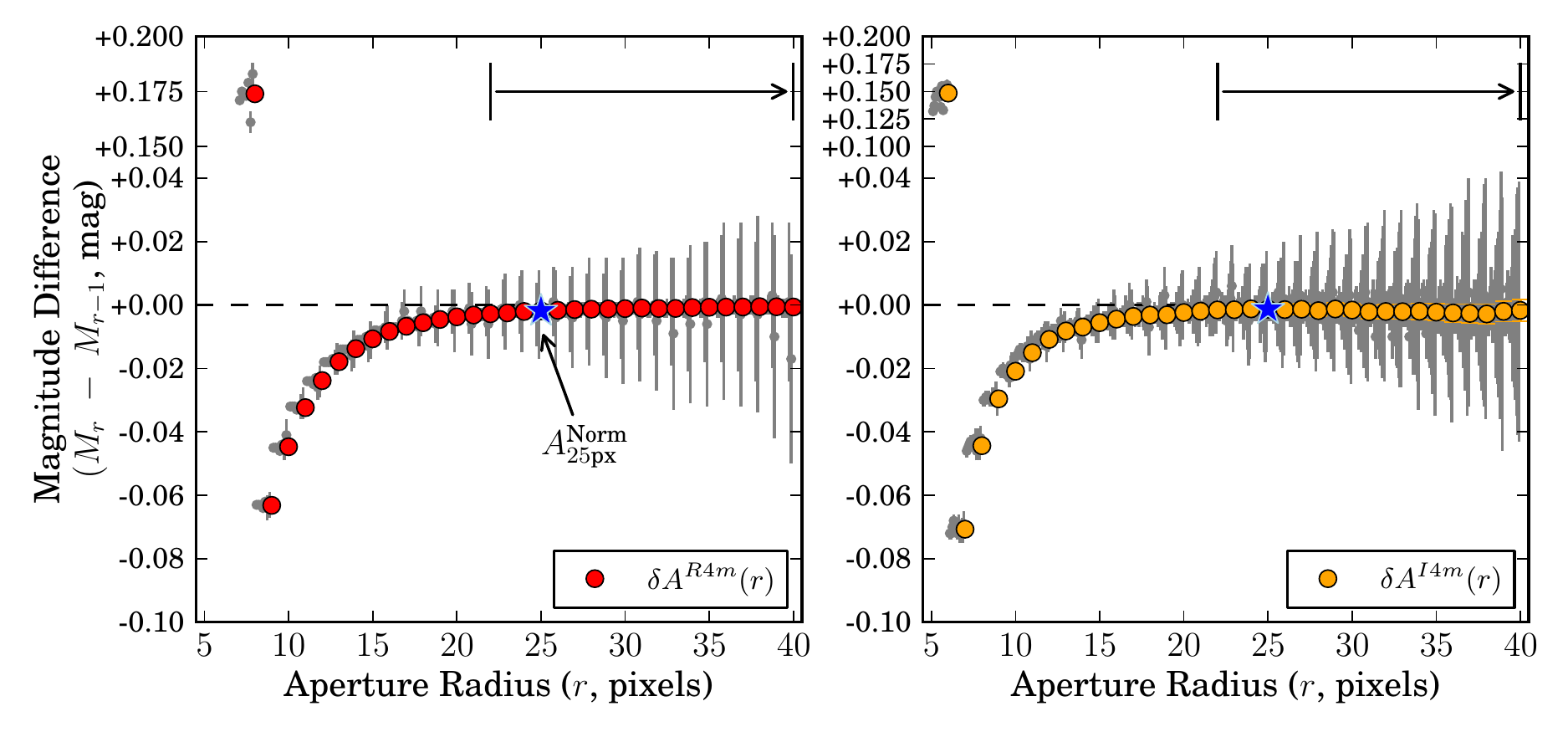}
\caption[Representative Curve of Growth]{Typical differential curves of growth
for $R$ (left, red) and $I$ (right, orange) on 20071103 (YYYYMMDD), for Amplifier 4 (both
randomly selected).  The point at the smallest physical aperture is the
difference between the \texttt{DoPHOT} magnitude of the object and magnitude with an
aperture radius of 5 pixels. We have used a piecewise ordinate-axis scale to show
the full range of the data without compressing local variations. We
plot the individual isolated stars in grey, offseting each individual star
slightly from the aperture through which the flux is measured along the
negative abscissa
direction for clarity. We checked that the growth curve is consistent with a
constant for apertures larger than 22~pixels, indicated by vertical lines with
an arrow in between. Uncertainties in the average measurement at each aperture,
$\delta A$, are typically smaller than the plot symbols. We propagated the
covariance matrix between apertures to determine the final aperture correction
at a radius of 25~pixels (indicated with a blue star, and labeled in the left
panel).}\label{fig:apercorr} \end{figure*}

\begin{figure}
\epsscale{1.2}
\plotone{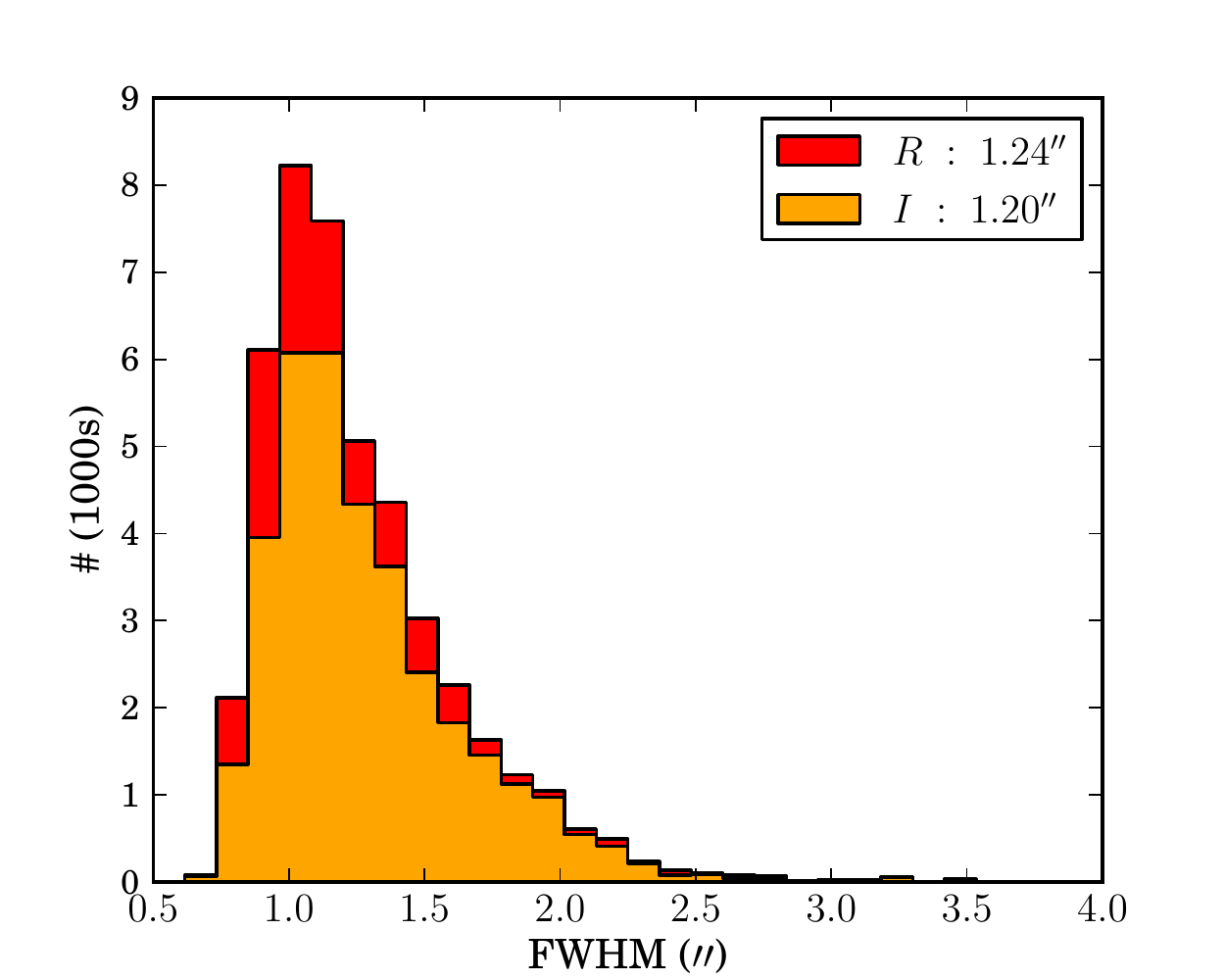}
\caption[FWHM Histogram]{FWHM distribution of $R$ and $I$ science
images from the survey. The mean FWHM is 1.24\arcsec\ for the $R$ band and
1.2\arcsec\ for $I$.}\label{fig:fwhm}
\end{figure}

Accurate aperture corrections are critical to establishing a consistent
photometric system across the survey. We have significantly refined the
algorithm used to generate aperture corrections for images. For each subfield,
we identified several isolated objects (typically 10--25 per amplifier) with
S/N $>$ 20 that are consistent with a point-source PSF in multiple images. We
took care to eliminate instances where we found flux measurements from
isolated, but nonstellar objects in the growth curves computed for M07. The 
flux of each star was measured using aperture radii from 5 to 40 pixels,
accounting for the weight map and any flux lost to masked pixels. We
constructed differential growth curves for each image (a representative example
is provided in Fig.~\ref{fig:apercorr}). The growth curves of individual stars
that indicate contamination by a secondary source (cosmic rays, stray
reflections, streaks) were removed. If more than 25\% of the stars were
clipped, the aperture correction for the image was flagged ``bad''.  We checked
that the growth curves asymptotically approached a constant value for all
apertures larger than 22 pixels, and flagged those that did not. We measured
the total aperture correction to an aperture radius of 25~pixels, or
\about13.5\arcsec\ in diameter, chosen to effectively enclose most of the flux
of each object for all ESSENCE images, which have a typical PSF full width at 
half-maximum intensity (FWHM) of \about1.2\arcsec\ in both passbands (see
Fig.~\ref{fig:fwhm}). 

\subsubsection{Choice of Standard-Star Network and a Fundamental Spectrophotometric Standard}

\begin{figure}
\epsscale{1.2}
\plotone{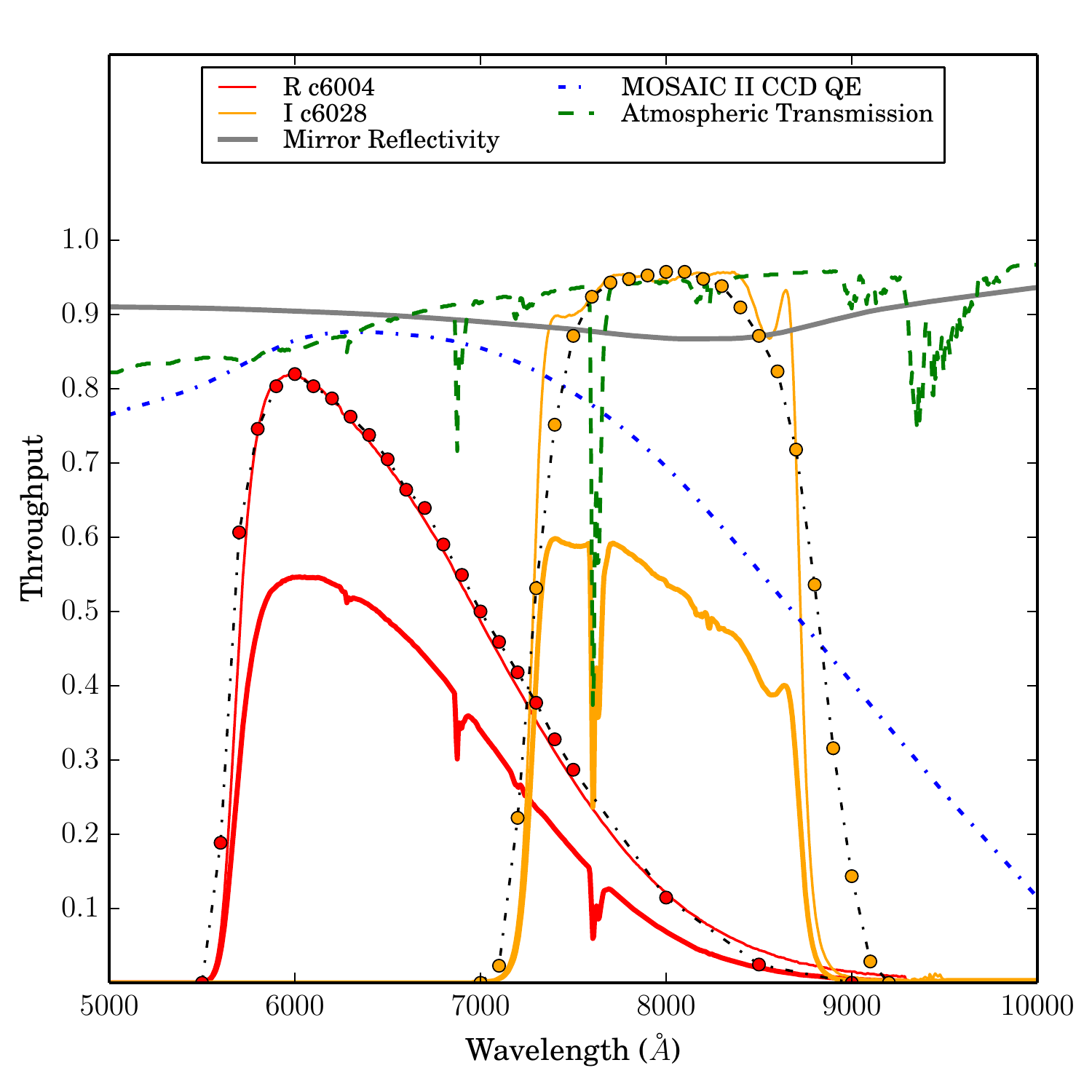}
\caption[System Throughput Curves]{Throughput curves for CTIO Blanco $R$ and $I$
passbands (thick red and orange, respectively), representing full system
throughput including wavelength dependence of CCD quantum efficiency
(dot-dashed blue), aluminum reflectance of the mirrors (solid grey) in the 
Blanco telescope, the optical filters (thin red and orange), and a model of the
atmosphere (dashed green) generated using the \texttt{MODTRAN4} code at an airmass of 1
with 2~mm of water vapor at an altitude of 2~km and a contribution from
aerosols appropriate for the CTIO Blanco site. The measurement of the various
components of the system throughput is discussed in \S\ref{sec:systrans}, and
the response curves are listed in Table~\ref{tab:systemthroughput}. The Bessell
$R$ and $I$ filter curves (red and orange circles, joined by dot-dashed black
lines, and normalized to have the same peak transmission) are shown for
comparison.}\label{fig:passbands}
\end{figure}

While several standard stellar catalogs report broadband magnitudes
in different photometric systems through a range of passbands
\citep{landolt83,stetson2000,ivezic07,landolt07}, the standard-star network of
\citet{landolt92} extended by \citet{stetson2005} remains the most obvious
choice to tie to the Johnson-Morgan-Cousins photometric system. The $R_{C}$ and
$I_{C}$ Cousins filters are broadly similar to those used on the Blanco (see
Fig.~\ref{fig:passbands}), and the magnitudes reported by low-redshift \snia\
surveys are converted into the Johnson system using observations of the Landolt
network stars. This allows us to minimize systematic uncertainties when
comparing our data to the nearby sample.

The choice of standard-star network and the transformation equations
derived between the natural and standard systems also play a critical role in
determining the absolute throughput of each passband. This calibration enables
SED models of \snia\ generated from low-redshift observations to be converted
into the Blanco natural-magnitude system via

\begin{equation}\label{eqn:synphot}
\begin{split}
m_{T} &= -2.5\log_{10} \left( \int \! F(\lambda)T(\lambda) \frac{\lambda}{hc} \, \mathrm{d}\lambda \right) + \text{ZP}_{T}.
\end{split}
\end{equation}

This equation is inverted to determine the zero point, ZP$_{T}$, for the full
optical system (detector, optics, filter, and atmosphere) with dimensionless
total photon efficiency, $T(\lambda)$, using a star with a well-measured SED,
$F(\lambda)$\footnote{The formalism employed throughout this work represents
SEDs as power per unit wavelength as a function of wavelength, while the system
throughput is represented as a dimensionless photon efficiency. The former is
typically provided in erg s$^{-1}$ cm$^{-2}$ \AA$^{-1}$. If the system
throughput is provided in erg \AA$^{-1}$, then the extra factor of the inverse
energy, $\lambda/hc$, must be dropped to account for the
Jacobian of the transformation.}, whose magnitudes ($m_{T}$) are known in the
natural system --- a ``fundamental spectrophotometric standard.''

Unfortunately, most well-measured spectrophotometric standards are too bright
to be imaged directly by the Blanco. We must therefore infer the Blanco natural
magnitudes of the fundamental standard using the star's standard magnitudes.
The most direct way of achieving this is to define the transformation equations
such that the Landolt and natural-system magnitudes agree at some color.

Historically, the choice for the fundamental standard for \snia\ surveys has
been $\alpha$~Lyrae (Vega), either implicitly when the rest-frame \snia\ model is
constructed from low-$z$ data, or explicitly when defining the passband zero
points for high-$z$ surveys
\citep{astier06,miknaitis07,Hicken09a,Contreras10,Stritzinger11,Hicken12}. Vega was one of
six A0~V stars used to establish the color zero point on the
photometric system of \citet{johnsonmorgan1953} by defining the mean $U-B$ and
$B-V$ colors of the six stars to be zero, and this definition was further extended to
Cousins $R_{C}-I_{C}$.  Vega's SED was tied to tungsten-ribbon filament lamps
and laboratory blackbody sources employed as fundamental standards
\citep{oke70,Hayes75}.  With the widespread adoption of the Landolt standard-star
network to tie instrumental photometry to the Johnson system, the use of Vega
as the fundamental spectrophotometric standard became ubiquitous. 

However, as discussed by \citet{regnault09}, Vega is far from an ideal choice
for the fundamental standard. \citet{taylor86} found that in order for several
sources of synthetic and observed Cousins $R_{C}-I_{C}$ measurements to agree,
the $I_{C}$ transmission curve had to be shifted to the red by 50--100~\AA.
With this shift, the synthetic color of Vega was found to be 0.006~mag.
\citet{fukugita1996} report a similar value. Furthermore, the Landolt
$(R-I)_{L}$ color of Vega is significantly more blue than the average for the
Landolt standard-star network (with $(R-I)_{L} \approx 0.47$~mag) and
consequently, any systematic error in the color term or the Landolt $(R-I)_{L}$
color of Vega has a much larger systematic effect on the $RI$ natural
magnitudes than would a standard with a color closer to the average Landolt
standard.  Vega may exhibit some photometric variability \citep{Fernie81}. In
addition it's SED is punctuated with several unusually shaped absorption lines.
Vega has an excess of NIR emission longwards of 1--2 $\mu$m, likely a result of
its dust ring \citep{Bohlin14b} and possibly its  rapid rotation
\citep{Peterson2006}  It also has an excess of UV emission relative to a 9400K
model (a result of its rapid rotation \citet{Bohlin14}). These may introduce
systematic errors when models are used to extend the observed SED of Vega into
the UV and IR.

Following several groups including SDSS \citep{ivezic07} and SNLS
\citep{regnault09}, we instead select the sdF8~D star, BD+17\arcdeg4708, as our
fundamental spectrophotometric standard. At $(R-I)_{L}=0.32$~mag, the color of
BD+17\arcdeg4708 is considerably closer to the average Landolt network star
than Vega. The {\it Hubble Space Telescope (HST)} CALSPEC program 
has measured the SED of BD+17\arcdeg4708
covering 0.17--1~$\mu$m with an uncertainty of $<0.5$\% in the flux
calibration derived from the three primary {\it HST} white-dwarf 
standards and \about2\% in
the relative flux calibration over the entire wavelength range.

\subsubsection{Transformation Between Landolt Network and the CTIO Blanco Natural System}

In order to calibrate the natural system of the Blanco, we obtained several images
of three Landolt standard fields (L92, L95, Ru149) directly with the Blanco/MOSAIC~II over 63 nights in 2006 and 2007. 
The images covered a wide range of airmass
and exposure time, and the calibration fields were dithered across the entire
field of view. With this large dataset, we robustly determined extinction and
color terms between the Blanco ($4m$) and the Landolt network using the relations

\begin{eqnarray}\label{eqn:phot}
\begin{aligned}
\displaystyle R^{\text{Ins}}_{4m} + A_{i} =& R_{L} + k_{R_{4m}}(X_{i}-1)\\
\displaystyle & + c^{R_{4m}}_{(R-I)_{L}}((R-I)_{L}-0.32) - Z_{i}, \\
\displaystyle I^{\text{Ins}}_{4m} + A_{i} =& I_{L} + k_{I_{4m}}(X_{i}-1)\\
\displaystyle & + c^{I_{4m}}_{(R-I)_{L}}((R-I)_{L}-0.32) - Z_{i},
\end{aligned}
\end{eqnarray}

\noindent where $R$ and $I$ denote the $R$- and $I$-band magnitudes in the Landolt
($L$) and Blanco instrumental ($4m$) systems, and $A$, $X$, and $Z$ denote the
aperture correction, airmass, and zero point of image $i$, respectively.
These relations are defined such that at the color of BD+17\arcdeg4708, the
calibrated magnitudes of the Blanco system match those of Landolt.

We expect differences in the aperture corrections between science and
calibration-field frames. Images of the calibration fields were generally short
exposures ($<60$~s) and often unguided, while science images are 200~s in $R$
and 400~s in $I$. We found typical systematic differences of 1--3\% between the
aperture corrections measured in the calibration fields and the mean aperture
correction of all science fields observed on the same nights. The aperture 
correction differences are correlated with the PSF size and ellipticity
measured in the calibration fields. We accounted for these aperture correction
differences while extrapolating zero points between images to construct the
tertiary photometric catalogs in \S\ref{sec:tercatzpt}.

The average offset between Landolt magnitudes for catalog stars and measured
instrumental magnitudes was calculated for each field, fitting for a single
linear term in Landolt $0.3 < (R-I)_{L} < 0.8$ color. As there were
insufficient stars covering the full color range in any single image, the
weighted mean color term for all calibration field images with at least 20
stars in $I$ and 50 stars in $R$ was computed. Computing the color term image
by image allowed us to look for trends in the color term with time and airmass.
While this procedure leads to slightly higher statistical uncertainties than if
a single color term was determined simultaneously for all images, it produces a
robust estimate of the color term, and as shown in \S\ref{sec:syserr}, the
systematic uncertainties in the photometric calibration are dominated by the
uncertainty in determining the absolute zero points.

We found color terms of $c^{R}_{R-I} = -0.030 \pm 0.001$ and $c^{I}_{R-I} =
0.022 \pm 0.001$. These values are in good agreement with measurements by
observatory staff\footnote{\url{http://www.ctio.noao.edu/mosaic/ZeroPoints.html}} for the
Blanco. The dispersion about the fitted value is \about2.5\% in $R$ and \about1.5\%
in $I$. While this dispersion is significantly larger than the photon noise,
this is not unexpected. We seek a single linear color term that is applicable
over a range of color, and in a variety of observing conditions that reflect
the conditions under which science images were acquired. As we compute these
color terms image-by-image, the dispersion about the mean value reflects
unmodeled variation in site conditions, as well as any variation in the sample
of stars used to compute the color term for any given image. This procedure is
preferable to one in which a subset of images is designated as having been
acquired in ``perfectly photometric'' conditions and is used for
calibration, as any difference between conditions on photometric nights and
the mean condition of science nights will lead to systematic errors in the
photometric calibration.

The value in $R$ is the same as that used by M07, while we
find $c^{I}_{R-I}$ to be lower by $0.008 \pm 0.003$ than that work. We
attribute this difference to the different methodology used and the redder
color range of stars selected for photometric calibration in the M07 analysis.

Several imagers show a strong radial dependence on the color term. Any relative
error in the photometry between the center and periphery of the detector can
affect the color term. Such effects can arise because of errors in the
illumination correction or chromatic effects. Other wide-field imagers often
include devices from different suppliers, and the quantum efficiency (and
therefore the color term) is a function of position on the detector. Neither
factor is a major consideration for MOSAIC~II, and there is no evidence of this
effect being a significant concern in other studies with this instrument.
Nevertheless, we elected to look for any systematic CCD-to-CCD variation in the
color terms. We found that the color term had a standard deviation of $0.005$
in $R$ and $0.003$ in $I$ about the mean. However, as standard fields were not
observed over the full duration of the survey, and we might expect any
low-level CCD-to-CCD variation to change as the instrument was mounted,
unmounted, and cleaned, we cannot determine if this variance is systematic.
Consequently, we elected to use a single color term for the entire imager, as in
M07, and absorb this into our systematic error budget in \S\ref{sec:syserr}.

We looked for systematic trends in the residuals between the image-by-image
color terms and the mean color term over time, but found that these were not
statistically significant.  The increasing accumulation of dust on the optical
surfaces leads to a changing zero point but does not significantly affect the
color terms.

\begin{figure*}[htb]
\includegraphics[width=\textwidth]{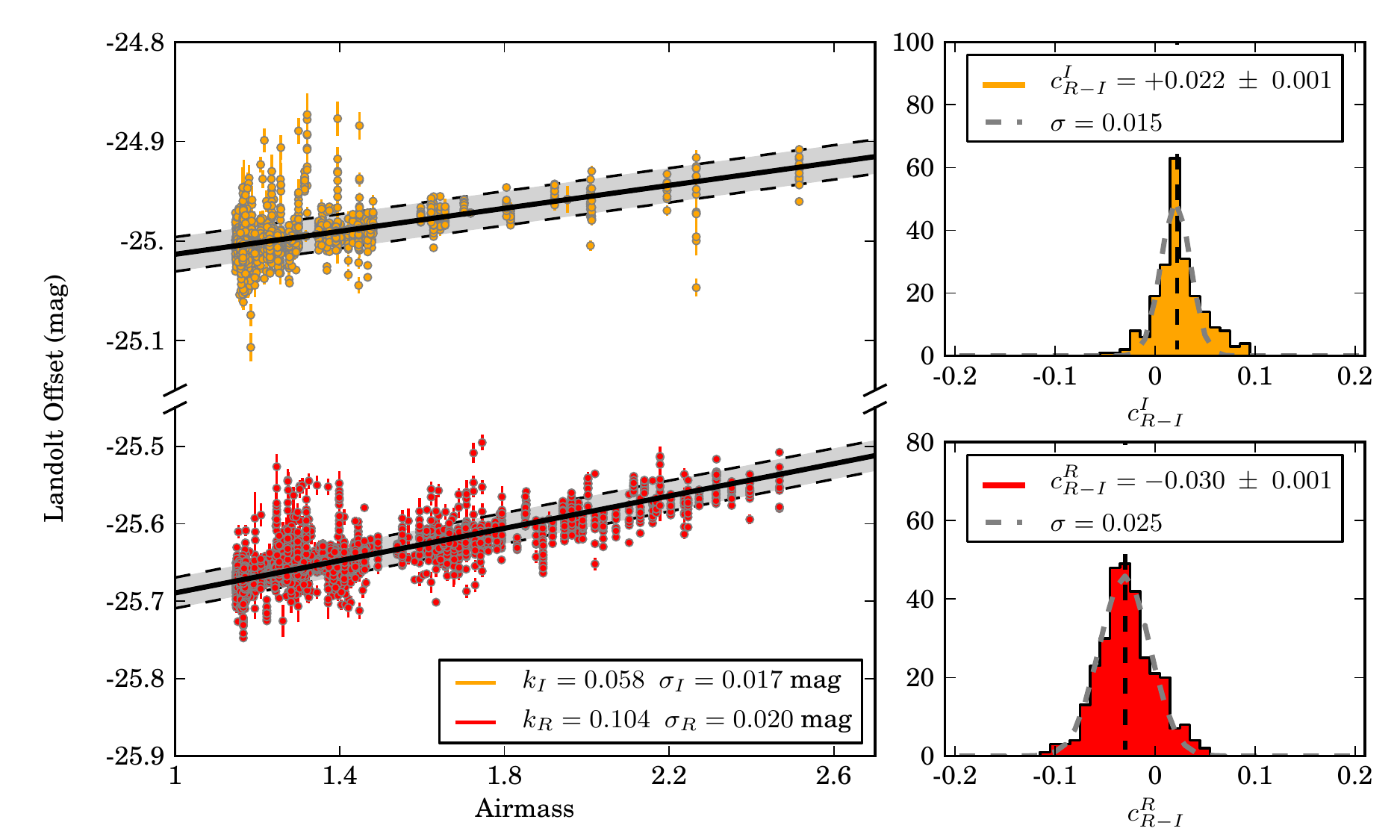}
\caption[Extinction and Color Relations]{(Left) Extinction relation for CTIO
Blanco system in $R$ and $I$ using calibration data for three Landolt fields (L92,
L95, and Ru149) imaged during the 2006 and 2007 observing seasons. The vertical
axis is the difference between instrumental aperture magnitudes and Landolt
catalog magnitudes, corrected for exposure time and variation with Landolt
$R-I$ color. We exclude any data taken in nonphotometric conditions. We find
extinction-law slopes of $0.104$~mag/airmass and $0.058$~mag/airmass in $R$ and
$I$, respectively. (Right) Distribution of color terms, determined per image, to
Landolt $R-I$ for the CTIO Blanco system in $I$ (above) and $R$ (below), using
calibration data from 2006--2007. Only images with at least 50 stars in $R$ and
at least 20 stars in $I$ were used in the analysis.  As there are typically
insufficient stars spanning the full color range in any single image, the
weighted mean color term for all the images is computed (indicated by dashed
vertical lines) and used for all further analysis. We find color terms of
$c^{R_{4m}}_{(R-I)_{L}} = -0.030 \pm 0.001$ and $c^{I_{4m}}_{(R-I)_{L}} = 0.022
\pm 0.001$.}\label{fig:airmass} \end{figure*}

The offset was refit with the color term fixed to this value and the aperture
correction was added.  Thus, the offset represents the average difference
between the Landolt-catalog magnitudes and our instrumental magnitudes through
a consistent 25~pixel aperture. These aperture-corrected zero points were then
regressed against the airmass to determine the slope of the extinction law and
the intercept.

We found no improvement in allowing the extinction term to vary between survey
years. Rather, we found that we could sufficiently account for year-to-year
changes in the overall transparency at the CTIO site by decomposing the survey
zero point into a dominant constant term with a small night-to-night variation.
We measured extinction-law slopes of 0.104~mag/airmass and 0.058~mag/airmass in
$R$ and $I$ (respectively), with dispersions of \about0.02~mag about the fitted
linear relation. The airmass relation and color terms determined are shown in
Fig.~\ref{fig:airmass}. Additionally, we used the \texttt{RANSAC} algorithm
\citep{Fischler81} to determine both the extinction and color terms, to ensure
that our fits were not sensitive to outliers. We found differences at the $10^{-5}$
level for the extinction coefficient, and typically at the $10^{-4}$ level for
the image-by-image color terms, consistent with the uncertainties on these
quantities.


\subsection{Tertiary Catalogs and Zero Points}\label{sec:tercatzpt}

Having calibrated the amplifiers within the footprint of the Landolt standard
field, we derived an extended standard catalog covering the entire field of
view of MOSAIC~II. As this catalog was generated by extrapolating the zero
point to other amplifiers of the same image, we accounted for the differences
in the aperture correction between amplifiers. This procedure prevented any
systematic errors arising from PSF variation, a misestimation of the
extinction coefficient, or short-timescale variations in transparency from
affecting the extended standard catalog. 

\begin{figure*}[htb]
\includegraphics[width=\textwidth]{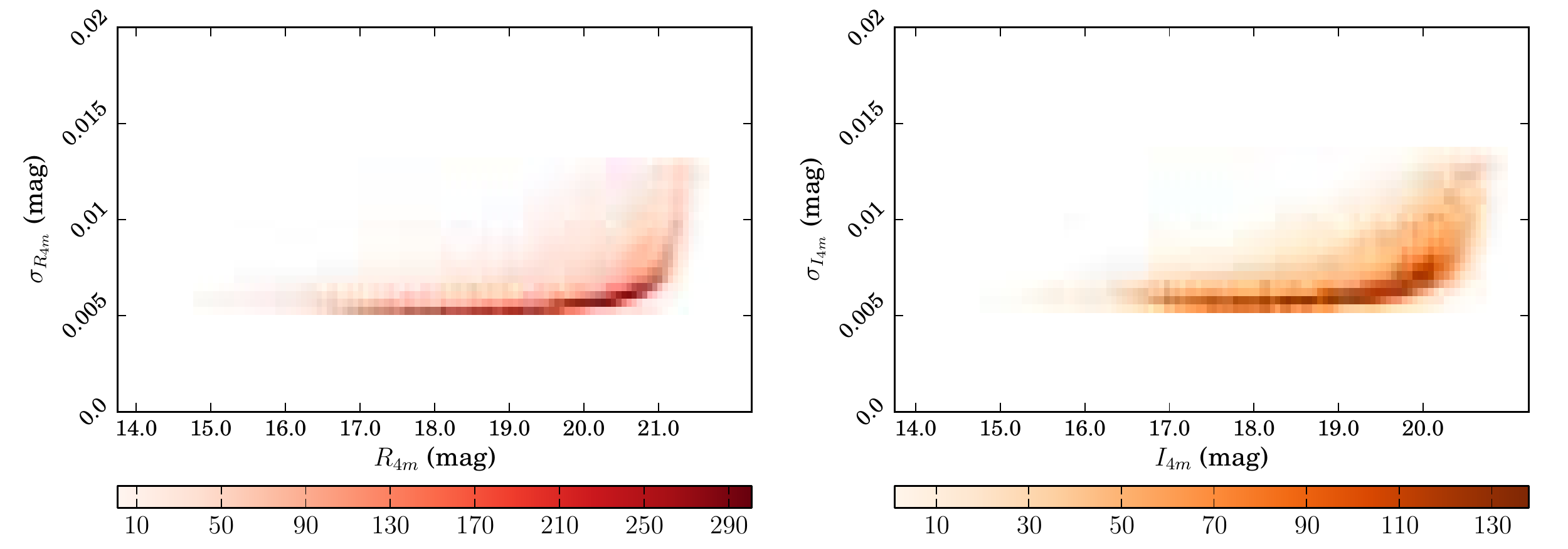}
\caption[Tertiary Catalog Magnitude-Error Distributions]{Uncertainty in the Blanco
photometry of ESSENCE reference catalog stars as a function of magnitude for
$R$ and $I$. The color of each bin indicates the number of stars in that bin.
Individual stars require at least 3 measurements in each filter. The systematic
error arising from an error in the airmass or color term, determined in
\S\ref{sec:syserr}, has been added in quadrature with the statistical
uncertainties. }\label{fig:calstar} \end{figure*}

The zero points were then redetermined using the extended catalog, without any
additional color correction applied. We extrapolated these zero points to
science images on the same nights as the calibration images, adjusting for
differences in exposure time, aperture correction, and airmass. For each star
in the science fields, we determined the 3$\sigma$-clipped error-weighted mean
magnitudes to generate our final photometric reference catalog for each field.
Stars with a high root-mean square (RMS) scatter relative to their mean magnitude errors were
rejected as variable.  The resulting catalogs typically have \about30 stars per
amplifier, with at least 3 observations in both filters, and a median of 8
observations each in $R$ and 5 in $I$. A 0.4\% uncertainty was added in
quadrature to all stars, in order to make the average reduced \chisq\ unity.
The error-magnitude distribution of the reference catalog stars is shown in
Fig.~\ref{fig:calstar}. 

\begin{figure*}
\centering
\includegraphics[angle=90,origin=c,width=0.85\textwidth]{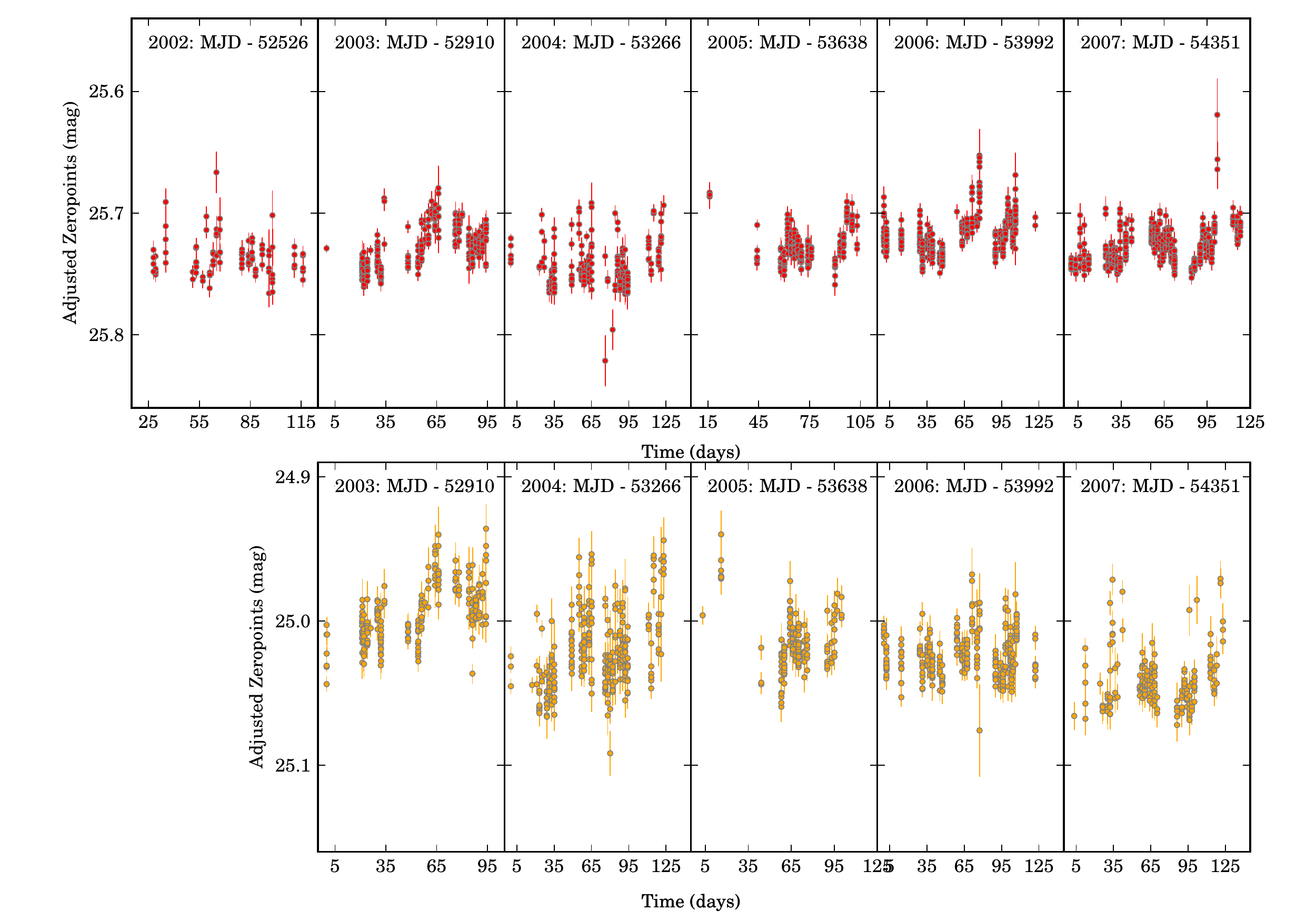} 
\caption[Survey Zero Point Evolution]{Average zero points for images, adjusted
for differences in exposure time, aperture correction, and airmass over the
full duration of the ESSENCE survey in $R$ (left) and $I$ (right). In 2002, the $I$
filter (NOAO code c6005) was damaged and replaced. The zero-point evolution is
correlated in both $R$ and $I$, and the short-timescale variations correspond
to changes in weather conditions at CTIO, whereas the gradual drift in zero
points is likely caused by the increasing accumulation of dust in the optical
system.  }\label{fig:zpt_time} 
\end{figure*}

\begin{figure*}
\centering
\includegraphics[width=\textwidth]{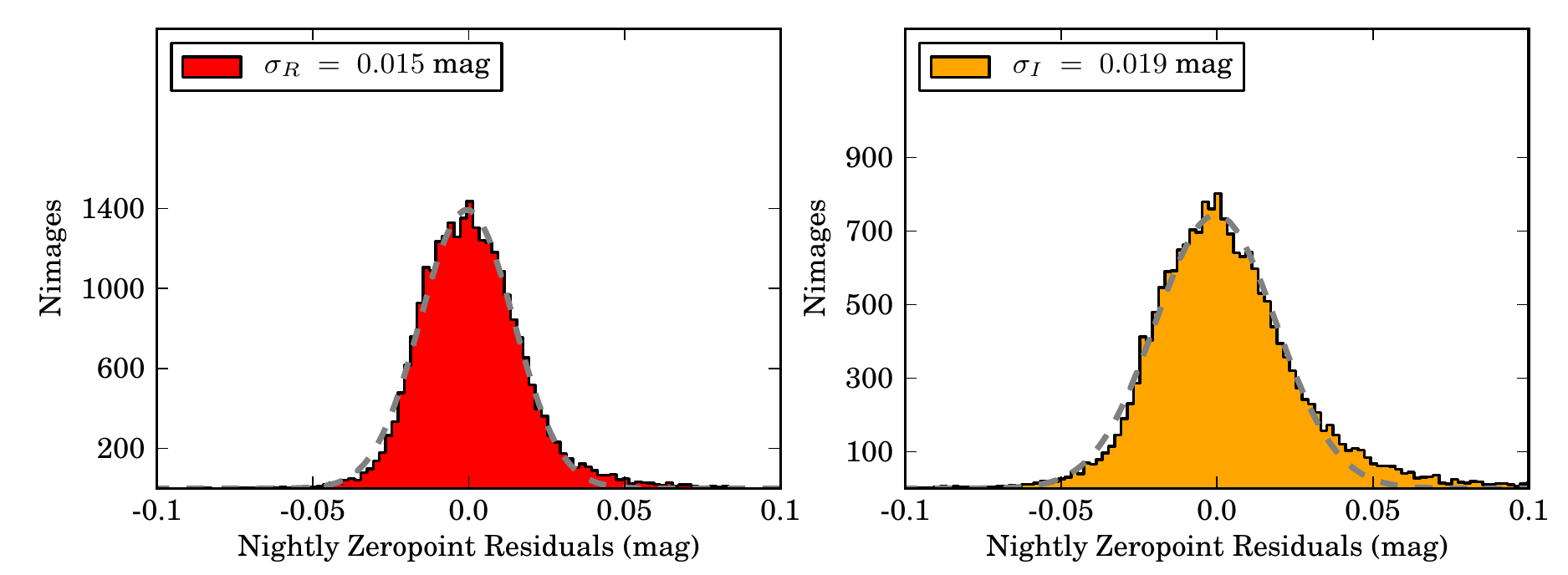}
\caption[Nightly Zero Point Residuals]{Histograms of the residuals of $R$
(left) and $I$ (right) zero points to the average nightly zero point, adjusted
for differences in exposure time, airmass, and the aperture correction. The
measured scatter in the nightly zero-point residuals is $<2$\% in both passbands,
consistent with the standard deviations derived from a fit to a Gaussian
distribution (dashed grey lines), and very comparable to the values found by
M07, illustrating that zero points are very consistent from field to
field.}\label{fig:zpt_stab} \end{figure*}

These reference catalogs were used to determine zero points for all science
images. To examine the temporal stability of the zero points, we adjusted them
for differences in aperture correction, airmass, and exposure time, but not
nightly variations in transparency or variation between different amplifiers.
The adjusted zero points of all available amplifiers were averaged together to
construct the average adjusted zero point for a given image. In
Fig.~\ref{fig:zpt_time}, we plot this quantity as a function of the time since
the start of the each year's observing season; the conditions at the Blanco
remained very stable over the entire duration of the survey. We also
constructed the nightly average zero point, and the histogram of residuals to
the nightly average zero point is plotted in Fig.~\ref{fig:zpt_stab}. The
residual scatter in the nightly zero-point residuals is $<2$\% in both $R$ and
$I$.

\begin{figure*}
\centering
\includegraphics[width=0.7\textwidth]{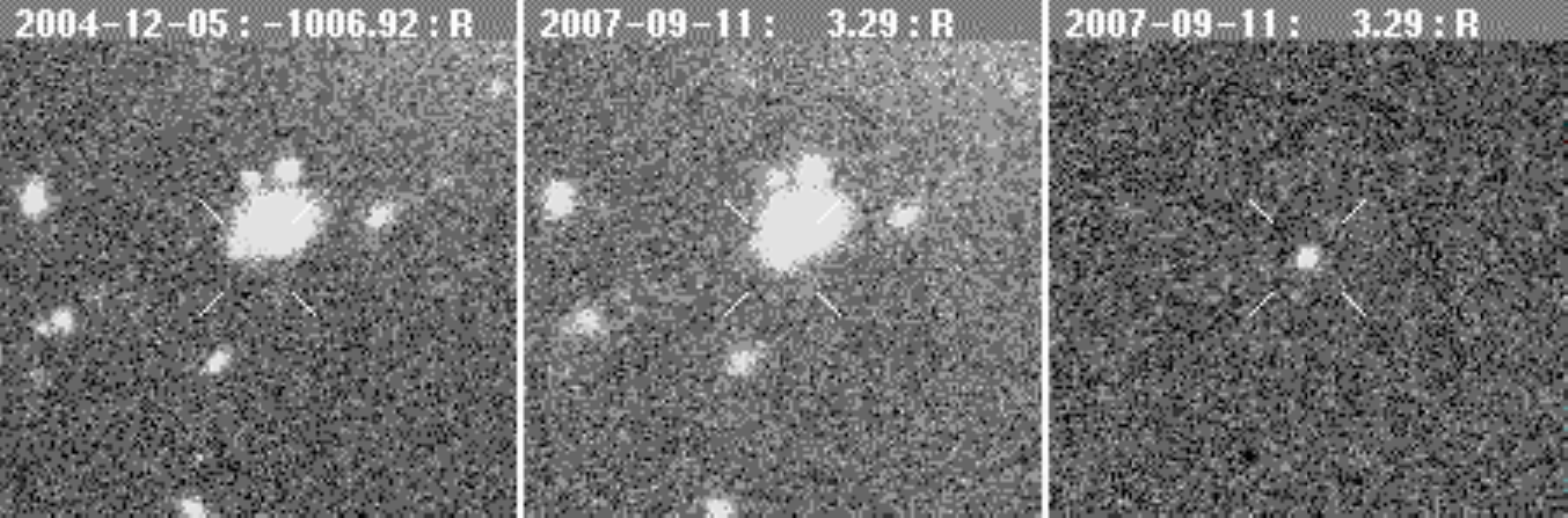}\\
\includegraphics[width=0.7\textwidth]{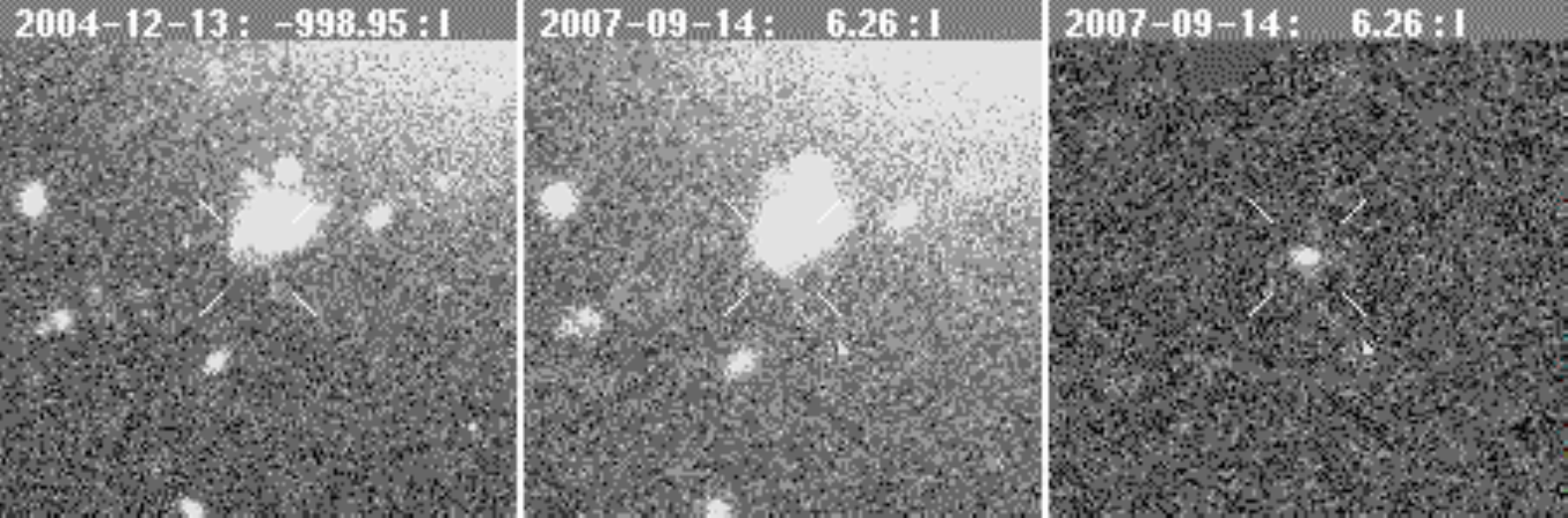}
\caption[Difference Images]{Representative difference imaging ``postage
stamps'' in $R$ (top) and $I$ (bottom) for \emph{x025}, a \snia\ at $z=0.35$,
near the median redshift of the survey. In this instance, \texttt{HOTPANTS} has
convolved the PSF of the reference (left) to match the science image (middle).
The reference is subtracted to produce the difference image (right). Despite
the complex gradient in the background, and clear differences in PSF and depth
between the reference and image, the difference-image background is extremely
uniform and free of artifacts.}\label{fig:diffimps}
\end{figure*}

\subsection{Image Subtraction} 

Having established zero points for each science image, we used image subtraction
to remove the background light of the host galaxies. Prior to subtraction, the
PSF of each image was first determined from field stars. We used the ``High Order
Transform Of PSF And Template Subtraction''
(\texttt{HOTPANTS})\footnote{\url{https://github.com/acbecker/hotpants}}
package to determine the convolution kernel between each image and template
pair. For each pair, the image with the narrower PSF was convolved to match the
image with the broader PSF. All $N(N-1)/2$ possible pairs of image and
reference templates from at least three observing seasons were used to create
difference images for each object (the ``NN2 process''), 
following the algorithm of \citet{barris05}.
We used a version of \texttt{DoPHOT} \citep{schechter93}, modified to use the PSF and
flux calibration of the image with the broader PSF, to measure flux in the
difference image. The flux calibration of the difference image was adjusted by
the normalization of the convolution kernel. The position of the SN was
measured by taking the weighted mean of all detections with S/N $> 5$.
The derived positions are accurate to 0.02\arcsec. The flux in each difference
image was measured with the PSF centroid fixed to the position of the SN.
A representative example of our image subtraction is provided in
Figure~\ref{fig:diffimps}.

As described by M07, the uncertainties in flux in our difference image are
underestimated owing to pixel-pixel covariance introduced during the resampling
process. Rather than scale the noise in each image up by a constant factor of
1.2 as in M07, we determined a correction for each individual difference image
using flux measurements across the frame. We convolved the PSF on a regular
grid across the difference image, measured the standard deviation of the
distribution of flux/$\sigma_{\text{flux}}$, and scaled each noise image by
this factor. This process effectively accounts for the small residual
pixel-pixel covariance introduced by deprojecting each image onto a common
astrometric grid and by the PSF convolution.

Additionally, we constructed a light curve for each object using a single deep
reference image, observed in photometric conditions with excellent seeing, to
identify any potential problems introduced in processing the thousands of
difference images produced by the NN2 process. We found excellent agreement
between the fluxes measured in the single template and in the NN2 process, with
the uncertainty in the flux being lower in the latter, as is expected by the
use of multiple images to measure the galaxy template and sky background at
each epoch. 

\section{Spectroscopy}\label{sec:spec}

Our full sample consists of all  \snia\ for which we were able to obtain a
positive spectroscopic identification. If possible, slits were aligned to
obtain spectra of the host galaxies of the SN candidates in order to obtain a
more accurate redshift.  The first two years of spectroscopic data from ESSENCE
were presented by \citet{matheson05}, while M07 detailed our selection criteria
and classification algorithms. The spectroscopic observations for the objects
included in M07 were presented by \citet{foley09}. The 6-year spectroscopic
sample from the ESSENCE survey is presented in this work, together with a
summary of the spectroscopic observations, data reduction, and the process of
candidate classification and redshift determination.

\subsection{Selection Criteria for Candidates}

As discussed in \S\ref{sec:survey}, during the 6~years of survey operation,
ESSENCE detected thousands of objects exhibiting variability over multiple
epochs, at a significance of S/N $> 5$. Given the limited spectroscopic
resources for follow-up observations, 
it was impossible to obtain spectra of all candidates.
We employed various selection criteria to narrow the list of candidates from
the imaging search to the subset with the most promise of being \snia. The
first set of these selection criteria was implemented as software cuts in our
search pipeline. We required the following.

\begin{enumerate}
\item Candidates detected in difference images have the same PSF as stellar
objects in the source image that was convolved by \texttt{HOTPANTS}.
\item Candidates exhibit no significant negative flux ($<30$\% of the total
number of pixels within an aperture of radius 1.5 $\times$ FWHM around the
detection) to select against difference-image artifacts, such as dipoles
resulting from slight image misalignment.
\item Candidates did not exhibit significant variability in ESSENCE data from
previous years, to reject variable stars and active galactic nuclei (AGNs).
\item Candidates in the difference image are not within 1 pixel (0.27\arcsec)
of objects in the template image, as these are frequently AGNs and
spectra of such candidates suffer from excessive host-galaxy contamination,
making classification very uncertain.
\item Candidates exhibit at least two coincident detections with S/N $> 5$,
in at least two passbands or within a 5-night window in a
single passband, to reject moving objects within the Solar System.
\item Detector and image-reduction artifacts were excluded by visual inspection.
\end{enumerate}

To select \snia\ from the resulting list of candidates, we fit
preliminary light curves using a $BV$ template of a normal \snia\ ($\Delta
m_{15} = 1.1$~mag) constructed from well-sampled low-$z$ \snia. This template
is a good match to \snia\ observed in $RI$ at $z \approx 0.4$, typical for the
ESSENCE survey. Using \chisq\ minimization, we determined the time of $B$
maximum, the $RI$ magnitudes at maximum, and the light-curve stretch ($s$).
These factors allowed us to determine an approximate photometric redshift for
the object, which, along with the $R-I$ color, and the rise-time information 
where available, was used to select likely \snia. 

An additional level of selection cuts was imposed by the observers onsite.
Observers tended to favor candidates thought to be in elliptical or low surface
brightness hosts, as the former are reliably \snia, while the latter aid in
extraction of a clean spectrum. As the various facilities and instruments have
different capabilities, and reach different depths, our faintest objects were
preferentially observed at larger-aperture facilities.

We obtained spectra using a range of facilities including the
Blue Channel spectrograph on the MMT \citep{schmidt89}; IMACS on Baade
\citep{dressler04}, and LDSS2 \citep{ldss294} and
LDSS3\footnote{\url{http://www.lco.cl/telescopes-information/magellan/instruments/ldss-3/}}
on Clay at the Las Campanas Observatory; GMOS on Gemini North and South
\citep{hook03}; FORS1 on the 8~m VLT \citep{appenzeller98}; and LRIS
\citep{oke95}, ESI \citep{sheinis02}, and DEIMOS \citep{faber03} at the W. M.
Keck Observatory.

Spectra were processed and extracted using standard \texttt{IRAF} routines.  Except for
the VLT data, all spectra were extracted using the optimal algorithm of
\citet{horne86}. VLT spectra were extracted using a novel two-channel
Richardson-Lucy restoration algorithm developed by \citet{blondin05} to
minimize galaxy contamination in the target spectra. Spectra were wavelength
calibrated using calibration-lamp spectra (usually He-Ne-Ar) fit with low-order
polynomials, and were flux calibrated using a suite of \texttt{IRAF} and \texttt{IDL} procedures,
including the removal of telluric lines using the well-exposed continua of
spectrophotometric standards.

To avoid relying on subjective assessments of noisy data, we employed the
SuperNova Identification (\texttt{SNID}) algorithm \citep{blondin07} to determine SN classifications 
objectively and reproducibly.  \texttt{SNID} is based on the cross-correlation
techniques of \citet{tonry79}. The input spectrum is compared to a large
library of template spectra at zero redshift, including nearby \sn\ of all
types (SN~Ia, Ib, Ic, II) and subtypes (e.g., SN~Ia-pec, SN 1991T, and SN 1991bg; see \citealt{filippenko97} for a review of SN spectral classification), as well as
other astrophysical sources such as luminous blue variables (LBVs) and other
variable stars, galaxies, and AGNs.  Where the redshift of the host galaxy is
available, we forced \texttt{SNID} to look for correlations at that redshift ($\pm0.02$)
to determine the SN classification.  In general, the spectra of SN with $z > 0.5$ have
low S/N, and thus ambiguities between classifications occurred mainly in that redshift
range. 

The \texttt{SNID} algorithm has been presented by \citet{matheson05} and
\citet{foley09}, and we refer the reader to these publications for further details.

\begin{figure}[htb]
\plotone{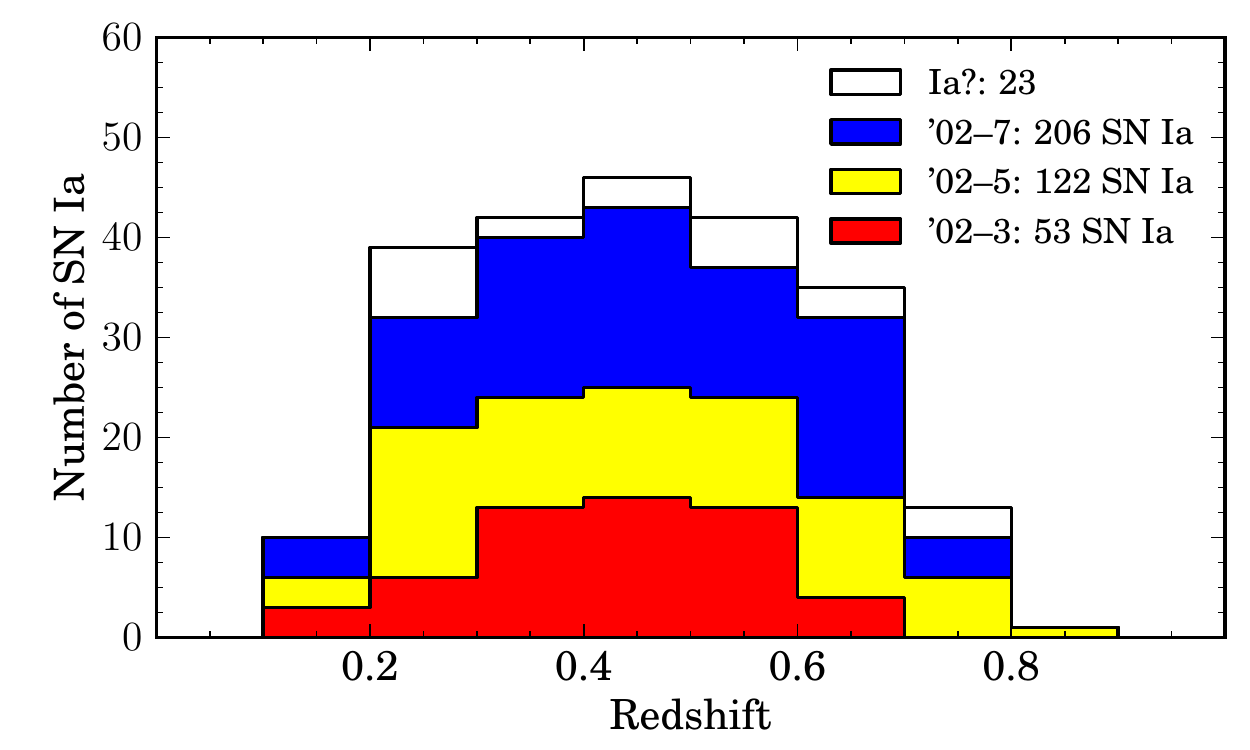} 
\caption[The ESSENCE Redshift Distribution]{The redshift distribution of
spectroscopically identified \snia\ from the ESSENCE survey. Candidates which
have a high confidence of being of Type Ia (all objects whose \texttt{SNID} correlations
with \snia\ templates exceed 50\%) are plotted in the shaded region. The
histogram is shown for observing seasons spanning 2002--2003 (red), 2002--2005
(yellow), and 2002--2007 (blue), along with {\it cumulative} totals, to
illustrate the evolution of the redshift distribution over the course of the
survey. Candidates for which we have less confidence have been classified as
``Ia?'' objects. Several of these have well-measured redshifts from their host
galaxies and are shown in the open region.} \label{fig:redshifthist}
\end{figure}

A list of all objects selected for spectroscopic observations is provided in
Table~\ref{tab:object}. An analysis of the spectroscopic efficiency of the
ESSENCE survey was presented by \citet{foley09}. The redshift distribution of
all ESSENCE \snia\ is shown in Figure~\ref{fig:redshifthist}. 
\section{\snia\ Light Curves from the ESSENCE Six-Year Sample}\label{sec:lccomp}

Of the 422 objects listed in Table~\ref{tab:object}, 233 were considered \snia\
candidates based on their preliminary light curves. Spectra were obtained for
229 of those 233 objects. 206 objects have been definitively classified as
\snia\ using the observed spectra. 

Eight objects were observed in nonstandard fields, without overlap with the
calibration fields used in this paper. Additionally, a few objects were
discovered near bright stars, or near the edge of the detector, and suffer from
repeated difference-imaging failures. We have excluded these objects from
further analysis. Despite being classified as a \snia\
(IAUC~8251\footnote{\url{http://www.cbat.eps.harvard.edu/iauc/08200/08251.html}}), an
analysis of the spectra of object e315 with \texttt{SNID} indicates that it
does not meet the criteria used to classify an object as a \snia\ employed by this
work.

The final \ri\ photometry of \nsn\ of the original 233 candidate \snia\ 
presented in this paper is listed in Table~\ref{tab:lightcurve}. Full light
curves (including non-\snia\ objects) and measurements of the baseline flux
will be made available as machine-readable tables\footnote{Available through
FAS Research Computing at Harvard ---
\url{http://telescopes.rc.fas.harvard.edu/index\_w.html}} along with this work.
Photometry is presented in linear flux units, $\phi$, in the Blanco natural system
for each passband, $T$.  Fluxes can be converted to calibrated magnitudes via

\begin{equation}
\begin{split}
m_{T} &= -2.5\log_{10}(\phi_{T}) + 25.
\end{split}
\end{equation}

\begin{figure}[htb]
\centering
\epsscale{1.21}
\plotone{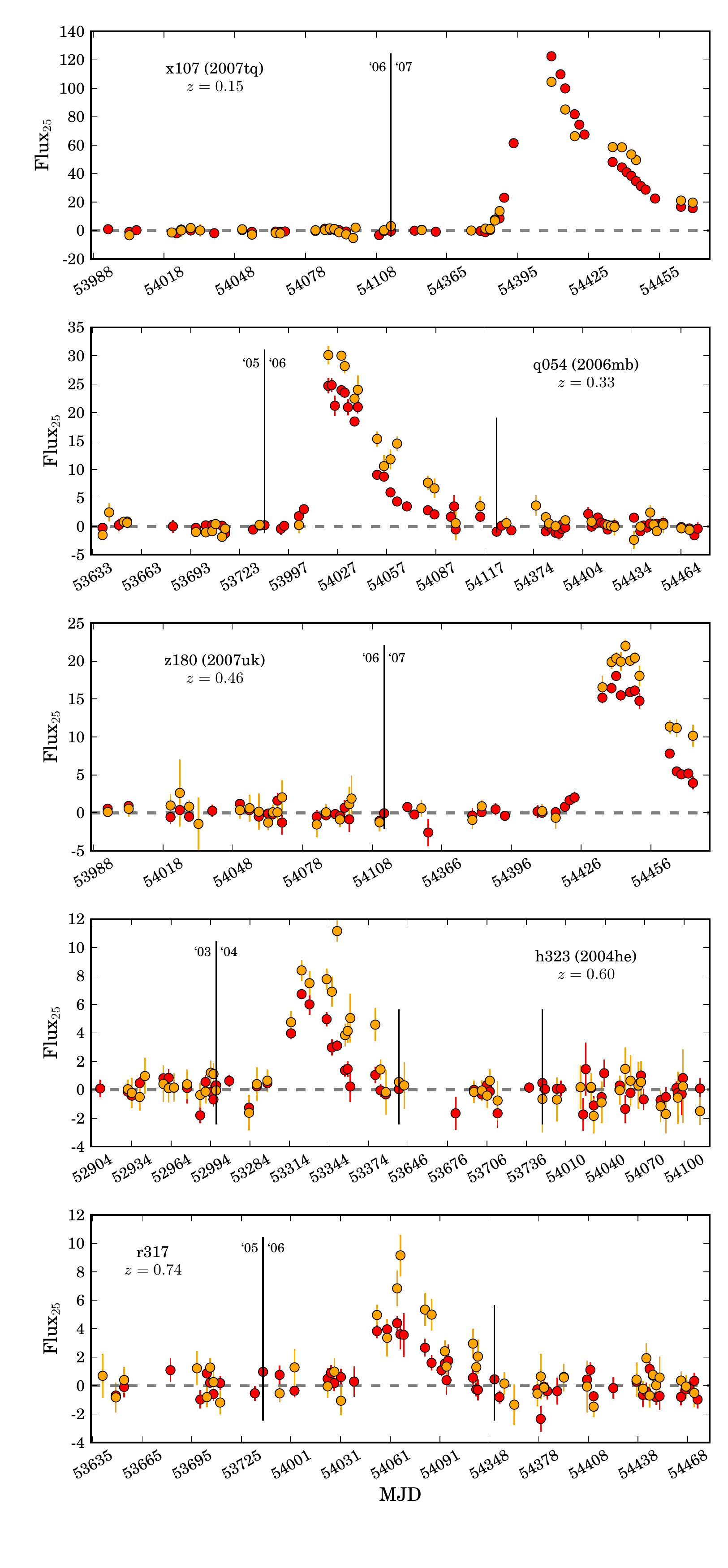} 
\caption{Example ESSENCE $R$ (red) and $I$ (orange) light curves, in units of
linear flux, scaled such that a flux of unity corresponds to magnitude 25. Gaps
between observing seasons have been removed, and the reported MJD is
discontinuous at the locations of the vertical black lines. The first of these
lines is elongated and the year of the observing season is indicated to the
left and right of it.}\label{fig:lcplot}
\end{figure}

\noindent The system-throughput curves and zero points required to derive
magnitudes in our passbands from SED models using Equation~\ref{eqn:synphot}
are provided in Appendix~\ref{sec:natsys}. The ESSENCE \snia\ and ``Ia?'' light
curves are illustrated in Figure~\ref{fig:lcplot}. 

\subsection{Light-Curve Shape and Color Distributions}

Several different algorithms to fit \snia\ optical photometry exist, including
MLCS2k2 \citep{JRK07}, BayeSN \citep{Mandel11}, SALT2
\citep{guy07, guy10}, SiFTO \citep{conley08}, and Dm15
\citep{prieto06}. Each of these corrects for the shape and color relations, but
they diverge when making two choices: how  to train their spectral
models and how to account for intrinsic and extrinsic color
variations. This divergence results in a dichotomy between a physical model,
where color variation is decomposed into an intrinsic variance and a reddening
attributed to extinction from dust (MLCS2k2 and BayeSN), vs. an
empirical model, where all color variation is directly correlated with
luminosity (SALT2 and SiFTO). In the following subsection, we
compare the color and shape parameter distributions derived using
MLCS2k2 and SALT2 for the ESSENCE light-curve sample presented in
this work.

\subsubsection{Light-Curve Quality Cuts}

While all the light curves are fit with both techniques, not all the fits are
reliable, as several objects lack high-significance flux measurements pre- or
post-maximum light, and these typically exhibit a high \chisq\ per degree of 
freedom (dof). Furthermore, objects in
Table~\ref{tab:object} without determined redshifts are not fit. 

Some selection cuts are common to all \snia\ surveys, and are required to
ensure that the light-curve fit is well constrained. These cuts are typically
expressed in terms of the rest-frame phase in rest-frame days $\Phi =
(T_{\text{Obs}}-T_{\text{Max}})/(1+z)$.  K09 required at least one measurement
with $\Phi < 0.0$. \citet[][hereafter
G10]{guy10}\footnote{\url{http://supernovae.in2p3.fr/~guy/salt/}} employed a more
flexible cut, only requiring a single measurement in the range of $-8 \lt \Phi
\lt +5$~ days, and found that this provided a comparable constraint to the K09
cut. Similarly, WV07 required at least one observation with $\Phi \le +5$~days
for both MLCS2k2 and SALT, but also demanded that the observation
had S/N $> 5$, while requiring that the uncertainty on the fit time of maximum 
brightness ($\sigma_{T_{\rm max}}$) be $\lt2$~days. The WV07 cut is effective 
at ensuring that the time and the flux of the peak are well constrained, and we adopt
it here for ESSENCE \snia. The compilation of 441 \snia\ presented by
\citet[][hereafter, C11]{conley11} uses the weaker G10 cut on observations
near maximum brightness.  In addition, G10 do not impose any cut on S/N.  
However, these
objects have observations in more passbands than ESSENCE, and the more
conservative cut is appropriate.

When the cut on pre-maximum measurements is not applied, both the
MLCS2k2 and SALT2 light-curve shape parameters ($\Delta$ and $x_1$, 
respectively) exhibit a significantly increased scatter as a result of 
light-curve fits being ill constrained with only the post-maximum decline.
\citet{Scolnic13} also report that $x_1$ shows a trend toward bigger values
for $z>0.4$ if the pre-maximum data are excluded. G10 did not find such a trend
with high-S/N \snia\ at  $z \lt 0.4$, illustrating how the effect of 
light-curve quality cuts varies with median redshift and therefore 
with survey. 

WV07 and K09 also required that the fit statistic, \chisqnu, be $\lt3$ for both
light-curve fitters. C11 did not impose any quality-of-fit cut, as they felt
that the reported uncertainties for low-$z$ photometry are frequently
inaccurate, rendering such a cut misleading. They also suggested that several
light curves contain the occasional outlying photometric observation that
drives \chisqnu\ to artificially high values, despite having little to no
effect on the derived light-curve shape and color parameters. C11 also argue
that any \chisq-based cut has an asymmetric effect with a \snia\ sample, and
therefore can potentially introduce a bias with redshift.  This in turn could
lead to a systematic bias on $w$. While there is merit in this argument, upon
visual inspection of our light-curve fits we concluded that the \chisqnu\
statistic did accurately represent the quality of the fit, and that this cut
was well motivated. In future work, we will use Monte Carlo simulations to
assess any bias in cosmological inference that results from this cut.

Another common cut is on the minimum number of degrees of freedom. Both WV07
and K09 require $N^{\text{min}}_{\rm dof} \ge 5$. C11 do not explicitly state
such a requirement, but the compilation they presented nevertheless satisfies
that requirement. We adopt $N^{\text{min}}_{\rm dof} \ge 5$ for MLCS2k2;
however, we found that this cut had the consequence of biasing us towards
intrinsically brighter objects. WV07 also required one observation with $\Phi
\ge +9$~days for MLCS2k2.  This cut was intended to ensure that the
post-maximum decline is well sampled.  As MLCS2k2 also imposes its own cut by
requiring observations with S/N $\gt 5$, this cut is considerably more
stringent than was intended. This requirement causes a total of 44 \snia\ and
``Ia?'' objects to fail the selection cuts --- by far the single largest cut on
our MLCS2k2 fits. In addition to eliminating observations of faint sources, or
sources at high $z$ with extremely well-sampled declines, the S/N cut imposed
by MLCS2k2 causes several light-curve fits to fail the selection cuts as a
result of insufficient observations, given the requirement of
$N^{\text{min}_{\rm dof}} \ge 5$ in the MLCS2k2 fit.  

By contrast, WV07 only required one observation after $B$-band maximum for
SALT, and only three objects in our sample do not meet this cut. We
believe that this demonstrates that MLCS2k2 is being needlessly
conservative by requiring that all observations have S/N $\gt 5$. However, the
intent of the cut on the number of post-maximum observations is to ensure that
the light-curve extinction or color is well constrained, and that the location of
the peak is bounded. We are wary of the relatively weak effect of the
post-maximum cut on our SALT2 light-curve fits, and require a stricter
$N^{\text{min}}_{\rm dof} \ge 8$ for that fitter. With the ESSENCE 4~day cadence,
this effectively ensures that there are at least four measurements of the
observer-frame $R-I$ color. As a result, the number of objects that fail the
$N^{\text{min}}_{\rm dof}$ cut for MLCS2k2 and SALT2 are similar, and
some of the most egregious outliers in $x_{1}$ and $c$ are eliminated.

Based on the results of G10, C11 imposed a restriction on the SALT2
color parameter, and  required $-0.25 \lt c \lt 0.25$~mag. WV07 did not
explicitly impose an equivalent cut on $A_{V}$ for MLCS2k2. Several
groups have used multicolor photometry of highly extinguished low-$z$ \snia\
to demonstrate that the extinction law in the host galaxies of these objects
appears to follow the \citet{odonnell94} extinction law with a significantly lower $R_{V}$ than the
Milky Way \citep{Hicken09b,Folatelli10,Mandel11}. 

Additionally, \citet{Scolnic13} employs a requirement that $-3 \lt x_{1} \lt 3$
for the Pan-STARRS1 \snia\ sample. Both of these cuts are well motivated, as there
are few \snia\ in the SALT2 training sample outside these ranges, and
the fits are likely to be ill-conditioned there. WV07 adopted a requirement of
$-0.4 \le \Delta \le 1.7$. All objects in our ESSENCE \snia\ sample that fail
this requirement also fail other selection cuts. 

A summary of the number of light curves that fail each cut for both
MLCS2k2 and SALT2 is provided in Table~\ref{tab:lccuts}.

\begin{center}
\scriptsize
\begin{ThreePartTable}
\begin{TableNotes}
\footnotesize
\vfill
\item [a] The number of \snia\ and ``Ia?'' objects in the ESSENCE sample that are removed by each selection criterion. Each cut is imposed independently. Many objects fail multiple cuts.
\item [b] We require $N^{\text{min}}_{\rm dof} \ge 8$ for SALT2, rather
than the weaker cut of 5 for MLCS2k2, as the last phase cut is very
ineffective with SALT2 when our well-sampled NN2 light curves are fit in
flux space.
\item [c] While at first glance it appears that more objects fail the
cut on pre-maximum imaging with SALT2 than with MLCS2k2, this is not the 
case upon closer inspection. MLCS2k2 merely fails catastrophically
for objects without pre-maximum imaging, and consequently does not report
$T_{\rm max}$ at all.
\item [d] WV07 did not employ an extinction cut. While we have not used
one in this work to faciliate comparison to WV07, it is likely that we will
employ a reasonable cut on this value to remove any highly reddened objects at
low $z$ from the sample for a cosmological analysis, as there is considerable
uncertainty about the nature of the dust in the host galaxies of highly
extinguished \snia. 
\end{TableNotes}

\begin{longtable}{lrr} 
\caption{Effect of Light-Curve Quality Cuts\tnote{a}\label{tab:lccuts}}\\

\hline \hline
Cut & 
MLCS2k2 &
SALT2 \\
\hline
\endfirsthead

\hline 
\hline
\insertTableNotes
\endlastfoot

Fit failed                                       & 7  & --- \\
\chisq\ $\gt 3$                                  & 10 & 20  \\
$N^{\text{min}}_{\rm dof}$\tnote{b}       & 18 & 20  \\
$\Phi_{\text{First S/N} \gt 5}$\tnote{c}  &  9 & 17  \\
$\sigma_{T_{\text{max}}} $                       &  5 &  3  \\                       
$\Phi_{\text{last}} $                            & 44 &  3  \\
$x_{1} | \Delta$ cut                             & 14 & 18  \\
$c$ cut\tnote{d}                          & NA & 33  \\
\end{longtable}
\end{ThreePartTable}
\end{center}

\subsubsection{MLCS2k2 Light-Curve Analysis}

We employ ``v007'' of MLCS2k2 with the ``tweaked-slowz'' vectors.  These
vectors, and the corresponding matrix of model uncertainties (denoted $S$), are
trained using the low-$z$ Hubble-flow sample in \citet{JRK07} (hence ``slowz'')
and ``tweaked'' with small magnitude offsets (typically $\lt 0.005$~mag) to
match the color-extinction distribution zero point, and extended to $-20$~days
prior to $B$-band maximum. We follow WV07 in using the ``glosz'' prior on
extinction, and we assume $R_{V} = 3.1$. 

\begin{figure*}[htpb]
\includegraphics[width=\textwidth,keepaspectratio=true]{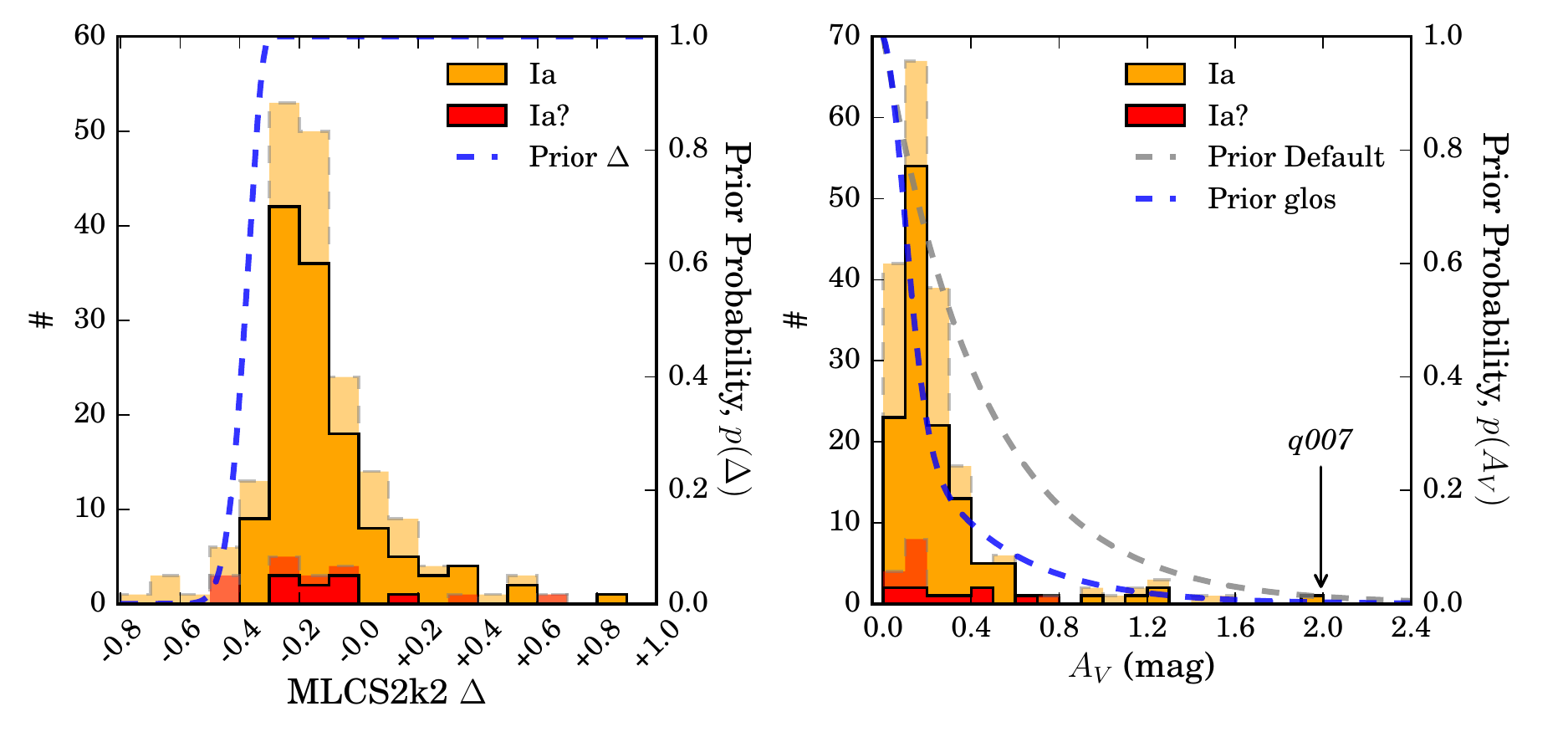} 
\caption[MLCS2k2 $\Delta$ and $A_{V}$ distributions for ESSENCE
\snia]{The light-curve shape ($\Delta$, left panel) and extinction ($A_{V}$,
right panel) distributions estimated by MLCS2k2 for ESSENCE \snia\
(orange) and ``Ia?'' (red). Objects that pass the selection cuts
(except the cuts on the parameter itself being plotted) imposed by WV07 are
indicated in the solid regions, while objects that fail are shown in the light
regions bounded by dashed lines. The MLCS2k2 priors employed in the
light-curve fitting are shown as dashed blue lines. The ``default'' prior is
the extinction distribution derived from low-$z$ \snia\ during the training
procedure.  }\label{fig:MLCSfitp} 
\end{figure*}

The MLCS2k2 light-curve shape ($\Delta$) and extinction ($A_{V}$)
distributions for ESSENCE \snia\ and ``Ia?'' objects are shown in
Figure~\ref{fig:MLCSfitp}. The MLCS2k2 light-curve fit parameters for
the ESSENCE sample are provided in Table~\ref{tab:MLCSfitp}. Additionally, we
have indicated if the objects pass the light-curve quality cuts used by WV07
for the 4-year sample.  From our spectroscopically confirmed \snia\ sample, 126
objects pass these cuts and are useful for cosmological inference. This doubles
the 4-year ESSENCE sample of 60 \snia.

Objects with $\Delta \gt 1.0$ are underluminous relative to normal \snia, and
they are more rare. Consequently, we are extremely unlikely to find any at the
redshifts probed by the ESSENCE survey. Objects that appear to be extremely
overluminous (very negative values of $\Delta$) relative to the training sample
of \citet{JRK07} typically have little or no high-significance flux
measurements pre-maximum, but have well-measured declines post-maximum. Without
a good constraint on the peak and time of maximum, light-curve fitters
typically explore unphysical regions of parameter space. The \chisqnu\ of these
light-curve fits is often relatively high ($\gt 3$) and all fail the quality
cuts of WV07, either owing to a high \chisqnu\ or because of insufficient
observations pre-maximum brightness. 

The $A_{V}$ distribution for ESSENCE \snia\ is consistent with the ``glos''
model employed by WV07. The distribution is significantly narrower
than the MLCS2k2 ``default'' distribution, derived from nearby \snia, as
we are unlikely to find highly extinguished and therefore faint objects at
high $z$. As MLCS2k2 is a magnitude-based fitter, it rejects
measurements with S/N $\lt 5$.  Most of the objects that fail the selection cuts
in the right panel of Fig.~\ref{fig:MLCSfitp} are extremely faint or at
high $z$. 

Note that Figure~\ref{fig:MLCSfitp} shows the distribution for all recovered
fits, and several of these objects do not have light-curve fits that meet the
quality cuts of WV07. Quality cuts are imposed to select spectroscopically
confirmed \snia, with several high-S/N measurements over rest-frame phase $-5
\le \Phi \le 20$~days, to ensure that the derived distance moduli are unbiased,
whereas derived light-curve shape and extinction are generally less susceptible
to poor phase coverage.

\begin{deluxetable*}{lccccrrrcrcc}
\tabletypesize{\scriptsize}
\tablecolumns{12}
\tablewidth{0pt}
\tablecaption{MLCS2k2 Light-Curve Fit Parameters for ESSENCE \snia\ and ``Ia?'' Objects\label{tab:MLCSfitp}}
\tablecomments{This table is published in its entirety in the electronic edition of the journal. A portion is shown here for guidance regarding its form and content.}
\tablehead{
   \colhead{ID} &
   \colhead{$\mu$\tablenotemark{a}} &
   \colhead{$\sigma_{\mu}$} &
   \colhead{$T^{B}_{\text{max}}$} &
   \colhead{$\sigma_{T^{B}_{\text{max}}}$} &
   \colhead{$\Phi_{\text{First}}$\tablenotemark{b}} &
   \colhead{$\Phi_{\text{Last}}$} &
   \colhead{$\Delta $} &
   \colhead{$\sigma_{\Delta}$} &
   \colhead{$A_{V}$\tablenotemark{c}} &
   \colhead{$\sigma_{A_{V}}$} &
   \colhead{Q\tablenotemark{d}}
}
\startdata
e108 & 42.370 & 0.140 & 52979.48 & 0.59   &  -11.768  &    9.954   &  -0.338 & 0.111 &  0.097 & 0.102 & T    \\
k425 & 41.207 & 0.254 & 53335.19 & 0.43   &   -9.535  &   19.618   &  -0.087 & 0.177 &  0.310 & 0.236 & T    \\
q002 & 41.212 & 0.359 & 54002.83 & 0.75   &   -6.340  &   15.192   &   0.594 & 0.279 &  0.549 & 0.369 & T    \\
x080 & 42.006 & 0.440 & 54384.94 & 1.54   &   -4.178  &   17.699   &   0.132 & 0.343 &  0.315 & 0.276 & T    \\
\enddata
\tablenotetext{a}{MLCS2k2 reports distance moduli with H$_{0} = 65$~km s$^{-1}$ Mpc$^{-1}$.}
\tablenotetext{b}{$\Phi_{\text{First|Last}}$ is the rest-frame phase of the first and last observation (respectively) and is dependent on the $B$-band time of maximum brightness, $T^{B}_{max}$. }
\tablenotetext{c}{Mag units. We use the Galactic reddening law of \citet{odonnell94}, with $R_{V}$ fixed to 3.1, to model the extinction in the host galaxy of the SN.} 
\tablenotetext{d}{Flag describing if the object passes (T) or fails (F) the light-curve quality cuts described by WV07.} 
\end{deluxetable*}

\subsubsection{SALT2 Light-Curve Analysis}

Additionally, we employ version 2.2.0b of SALT2 released together with G10.
SALT2 is a flux-based fitter and employs measurements of the baseline
flux to restrict the search range for fitted parameters. However, only data
within the rest-frame phase range $-15 \lt \Phi \lt 45$~days are used in the
\chisq\ minimization. Measurements in observer-frame filters that map to the
rest-frame wavelength range $3000 \lt \lambda \lt 7000$~\AA\ are used in the
fit. Model and $K$-correction uncertainties are propagated into the error
matrix, and an additional $U$-band calibration uncertainty of 0.1~mag is added
in quadrature for low-$z$ near-UV data.

\begin{figure*}[htpb]
\includegraphics[width=\textwidth,keepaspectratio=true]{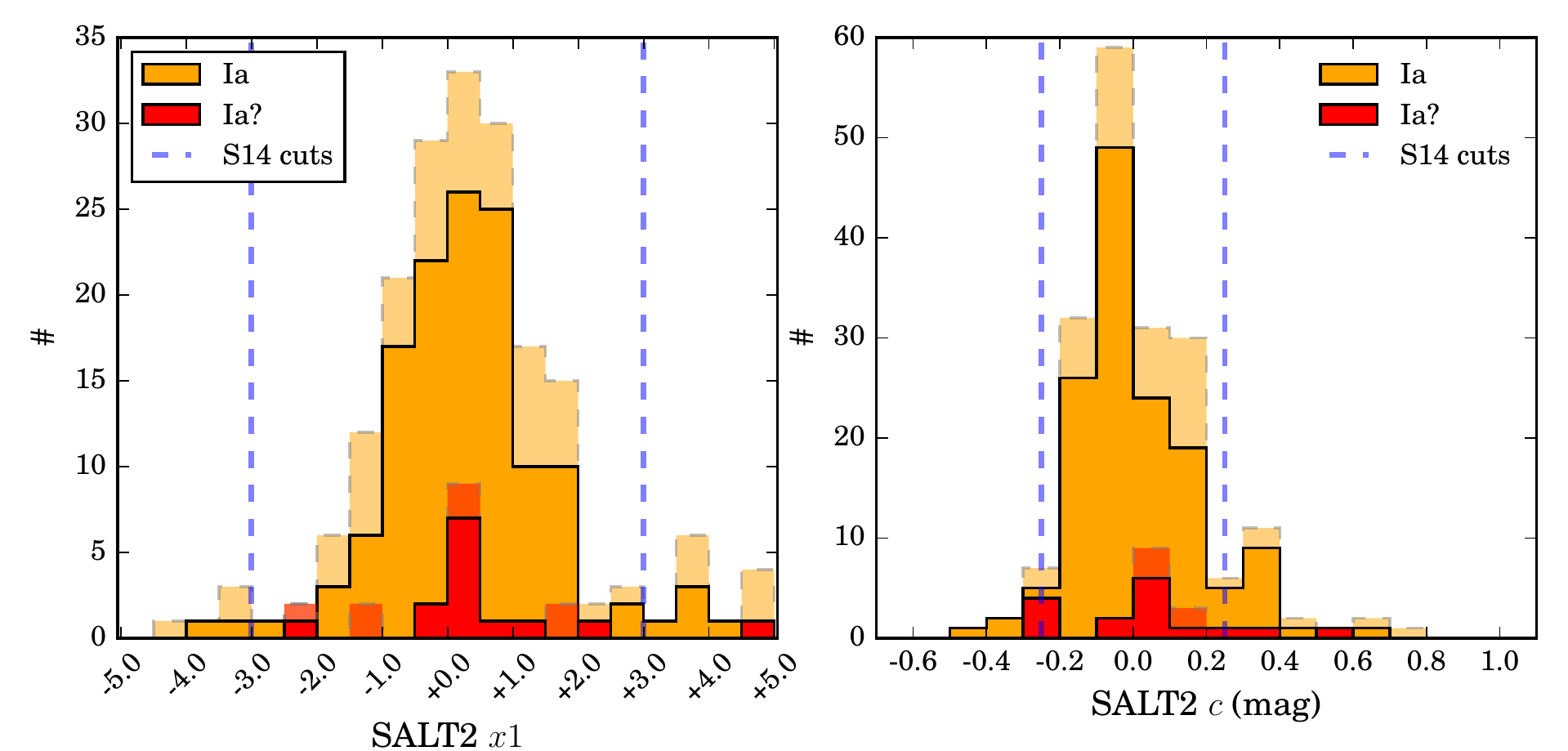} \caption[SALT2 $x_{1}$ and $c$ distributions for ESSENCE \snia]
{The light-curve shape
($x_{1}$, left panel) and color ($c$, right panel) distributions
estimated by SALT2 for ESSENCE \snia\ (orange) and ``Ia?'' (red)
objects. Objects that pass the original SALT selection cuts (except the cuts
on the parameter itself being plotted) imposed by WV07 are indicated in the
solid regions, while objects that fail are shown in the light regions bounded
by dashed lines. Only the WV07 cuts relating to the sampling and \chisqnu\ of
the fit are used here. Based on a visual inspection of all the light-curve
fits, we required that the fits used at least 8 epochs, rather than the weaker
cut of at least 5 epochs employed for SALT by WV07. We adopt the same
cuts as \citet{conley11} and \citet{Scolnic13} on $c$ and $x_{1}$,
respectively.}\label{fig:SALTfitp} \end{figure*}

The SALT2 light-curve shape ($x_{1}$) and color ($c$) distributions for
ESSENCE \snia\ and ``Ia?'' objects are shown in Figure~\ref{fig:SALTfitp}. The
SALT2 light-curve fit parameters for the ESSENCE sample are provided in
Table~\ref{tab:SALT2fitp}. Additionally, we have indicated whether the objects
pass a combination of the light-curve quality cuts used by WV07 for the 4-year
sample, as well as shape and color cuts employed by \citet{conley11} and
\citet{Scolnic13}.

Objects with $x_{1} < -3.0$ and $x_{1} > 3.0$ are poorly represented in the
SALT2 training sample. Fits with these values often have unconstrained
rises or peaks and provide unreliable distance estimates.

\citet{conley11} required that $-0.25 \lt c \lt 0.25$~mag to eliminate blue
objects that were not adequately represented in the training sample, as well as
objects with very red colors, which they believe are the result of a
combination of different effects. It is possible that extinction in the
host galaxy in the \snia\ is one of these effects. Three objects (m040,
m070, and m075) are catastrophic outliers and are not displayed
here. All three were discovered near the start of the 2005 observing season,
and none has any pre-maximum photometry.

\begin{deluxetable*}{lccccccrcrcrc}
\tabletypesize{\scriptsize}
\tablecolumns{13}
\tablewidth{0pt}
\tablecaption{SALT2 Light-Curve Fit Parameters for ESSENCE \snia\ and ``Ia?'' Objects\label{tab:SALT2fitp}}
\tablecomments{This table is published in its entirety in the electronic edition of the journal. A portion is shown here for guidance regarding its form and content.}
\tablehead{
    \colhead{ID} &
    \colhead{$m_{B}$\tablenotemark{a}}&
    \colhead{$\sigma_{m_{B}}$ }&
    \colhead{$m_{V}$}&
    \colhead{$\sigma_{m_{V}}$ }&
    \colhead{$T^{B}_{\text{max}}$\tablenotemark{b}}&
    \colhead{$\sigma_{T^{B}_{\text{max}}}$ }&
    \colhead{$x_{1}$}&
    \colhead{$\sigma_{x_{1}}$}&
    \colhead{$c$\tablenotemark{c}} &
    \colhead{$\sigma_{c}$ }&
    \colhead{${\rm Cov}(c,x_{1})$\tablenotemark{d}}&
    \colhead{Q\tablenotemark{e}}
}
\startdata
b010 & 23.4583 & 0.0666 & 23.5138 & 0.0937 & 52593.2578 & 0.8841 &  0.9105 & 0.7618 & -0.0781 & 0.1012 &  0.0248 & T     \\
g050 & 23.2239 & 0.0702 & 23.4278 & 0.1084 & 53302.2280 & 0.5811 & -0.2400 & 0.5682 & -0.2226 & 0.1069 &  0.0153 & T     \\
h323 & 23.4921 & 0.0594 & 23.3174 & 0.0873 & 53329.8365 & 0.6337 &  0.6074 & 0.5493 &  0.1485 & 0.0930 &  0.0094 & F    \\
n322 & 24.3561 & 0.0919 & 24.4587 & 0.2019 & 53707.8798 & 1.4726 &  0.1779 & 1.0967 & -0.1237 & 0.1342 &  0.0518 & T     \\
\enddata
\tablenotetext{a}{SALT2 does not directly report distance estimates. The distance modulus is determined using a global fit for all \snia\ together with other cosmological parameters.}
\tablenotetext{b}{While the SALT2 shape estimates are strongly affected by measurements pre-maximum brightness, it uses significantly more low-S/N measurements on the decline, as well as measurements for faint and/or high-$z$ objects. Consequently, the cut on the number of measurements post-maximum has very little impact. We instead require that a total of 8 observations be used in the fit, to ensure that the measured parameters are reliable.}
\tablenotetext{c}{We use the updated SALT2 color law described by \citet{guy10}. This differs significantly from the \citet{odonnell94} extinction law in the near-UV.} 
\tablenotetext{d}{Covariances between all the fit parameters --- Cov($m_{B}$,$x_{1}$), Cov($m_{B}$,$c$), and Cov($x_{1}$,$c$) --- are calculated, and these values will be included in the machine-readable tables provided with this work.} 
\tablenotetext{e}{Flag describing if the object passes (T) or fails (F) the light-curve quality cuts for SALT described by WV07 and the shape and color cuts described by \citet{conley11}.} 
\end{deluxetable*}                                           
\section{Systematics Affecting the ESSENCE Survey Photometry}\label{sec:syserr}

Here we identify and assess the size of each effect using empirical tests of
internal and absolute photometric calibration. Wherever possible, we quantify
systematics by directly introducing a bias at either the image or catalog level
and propagating the bias through our pipeline to measure the effect on output
photometry. We also compare our photometry to SDSS photometry converted to the
Landolt system, to set an upper limit on our systematic error budget and
evaluate our absolute photometric calibration in the different ESSENCE fields. 

The systematic effects can be divided into two categories:
\begin{enumerate}
\item Effects that cause errors in individual photometric measurements
\emph{and correlate with distance, leading to a bias} in cosmological
inference.
\item Effects that cause errors in individual photometric measurements, but are
not correlated with distance, and therefore do not bias the cosmological
result but nevertheless lead to \emph{increased dispersion} in Hubble
residuals.
\end{enumerate}

We list the various sources of photometric error in Table~\ref{tab:syserr}, and
we detail and estimate the effect of each in the subsections that follow.

\begin{center}
\scriptsize
\begin{ThreePartTable}
\begin{TableNotes}
\footnotesize
\item[a] Italicized entries are sources of increased dispersion on distance moduli but do not introduce systematic bias.
\end{TableNotes}

\begin{longtable}{lrr} 
\caption{Summary of the ESSENCE Systematic Uncertainties\tnote{a}\label{tab:syserr}}\\

\hline \hline 
Effect & 
$\Delta{R}$ &
$\Delta{I}$ \\
&
(mag)        &
(mag)        \\
\hline
\endfirsthead

\hline 
\hline
\insertTableNotes
\endlastfoot

\multicolumn{3}{c}{Errors in the Measurement of Flux}\\
\emph{Shutter precision}                      & $<0.001$   & $<0.001$   \\
Detector linearity                            & $ 0.005$   & $ 0.005$   \\
\emph{Image Detrending}                      & $\pm0.005$ & $\pm0.005$ \\
\emph{Astrometric Uncertainties}              & $ 0.005$   & $ 0.005$   \\
\multicolumn{3}{c}{Errors in the Photometric Calibration}\\
$\pm10$\% error in airmass relation           & $\mp0.001$ & $\mp0.002$ \\
Uncertainties in color term                   & $\pm0.005$ & $\pm0.005$ \\
Uncertainties in the zero point               & $ 0.003$   & $ 0.001$   \\
Uncertainties in extrapolating zero points    & $<0.001$   & $<0.001$   \\
Magnitudes of BD+17\arcdeg4708                & $\pm0.002$ & $\pm0.002$ \\
\hline 
Total                                         & $\pm0.011$ & $\pm0.010$ \\
\hline 
SED of BD+17\arcdeg4708                       & $\pm0.002$ & $\pm0.003$ \\
\hline 
Total                                         & $\pm0.012$ & $\pm0.011$ \\       
\end{longtable}
\end{ThreePartTable}
\end{center}

\subsection{Shutter Precision}
The MOSAIC~II shutter is described in \S\ref{sec:survey}. The shutter blades
take 23~ms to cover the entire field, leading to a $\pm0.5$\% nonuniformity
for a 1~s exposure. The correction is negligible for the exposure
times of all ESSENCE science (200~s in $R$ and 400~s in $I$) and calibration
frames ($>10$~s in both filters).

\subsection{Detector Linearity}
We imaged the Ru149 field in $R$ and $I$, varying the exposure times from
2~s to 400~s\footnote{Exposures under 10~s are not used outside this
analysis of detector linearity.}. Fluxes are measured for isolated stars using
a fixed 20~pixel aperture radius and corrected for extinction. These stars span
the dynamic range of the detector below saturation. We compute residuals to the
average magnitude for each star and the 3$\sigma$ clipped average
residuals for all stars. We find that these average residuals are $< 0.005$~mag
over the entire range of exposure times for both filters, so we infer that the
detector is linear to \about0.5\%. We also examined the difference between our
measured instrumental magnitudes and catalog magnitudes, at constant exposure
time, to check if there was any departure from linearity with flux. We do not
see any evidence of nonlinearity with flux below saturation.

\subsection{Systematic Uncertainties with Image Detrending}
We avoid most long-period systematic errors with image preprocessing by using
biases and flat fields obtained nightly, rather than a global bias or flat
field for a full observing season. Any systematic errors caused by a
misestimation of the bias or the flat field will only affect measurements made
on a single night.  While photometric measurements of objects observed on those
nights will be systematically biased, this error does not affect photometric
measurements of the same objects from other nights. Consequently, the effect is
very unlikely to correlate with distance modulus and will not lead to a bias
in cosmological measurements.

The systematic uncertainties associated with the illumination correction are
estimated in Appendix~\ref{sec:illum} and found to be $<0.3$\%. We typically
obtain \about10 bias and dome-flat images in each filter each observing night.
Comparing the combined bias and flat-field images of consecutive nights, we
find differences of \about0.1\%. As these errors can occur simultaneously, and
affect the image processing additively, we adopt a 0.5\% error as the
systematic error associated with our image detrending.

\subsection{Astrometric Uncertainties}
The astrometric uncertainty of a single detection is composed of a systematic
floor and a term that is inversely proportional to the S/N and directly
proportional to the FWHM of the detection:

\begin{equation}
\sigma^2_{a} = \sigma^{2}_{\rm sys} + \sigma^{2}_{d}\left(\frac{\rm FWHM}{\rm S/N}\right)^{2}.
\end{equation}

We use the procedure detailed by \citet[Appendix A]{Rest13} and find that the
single-epoch positions for supernovae are accurate to within 0.02\arcsec. M07
measured the impact of such an offset by identifying sources of known flux with
FWHM typical for the survey, and measuring their flux through a PSF offset by
1\arcsec. As this is much larger than the typical astrometric uncertainty, we
adopt the procedure used by \citet{Rest13} and find that an offset of
0.3~pixels produces a subpercent impact on photometry. Since the uncertainty is
related to the S/N, we expect increased dispersion at high $z$; however, our
cadence provides S/N $> 10$ for even our highest-redshift objects, and
we do not find any net bias in the recovered astrometry of known sources with
magnitude. We adopted a fixed valued of 0.005~mag in both filters to
account for the systematics arising from astrometric uncertainties.

\subsection{Uncertainties in Determining the Airmass Relation}
An error in the slope of the airmass relation would lead to an error in
extrapolating the zero points from the Landolt standard fields to the ESSENCE
fields, as well as between the ESSENCE fields. We mitigate this uncertainty by
requiring the images we use to extrapolate the zero point for a given image to
have a difference in airmass smaller than 0.5. We introduce a 10\% error in the
airmass relation and propagate the error to our photometric catalogs. Such a
large error is extremely unlikely, and would be visually apparent as we
obtained images of standard fields over an extended range in airmass, but
allows us to place an upper limit on the resulting systematic error in
magnitudes.

\begin{figure*}[htb] 
\centering
\includegraphics[width=0.45\textwidth]{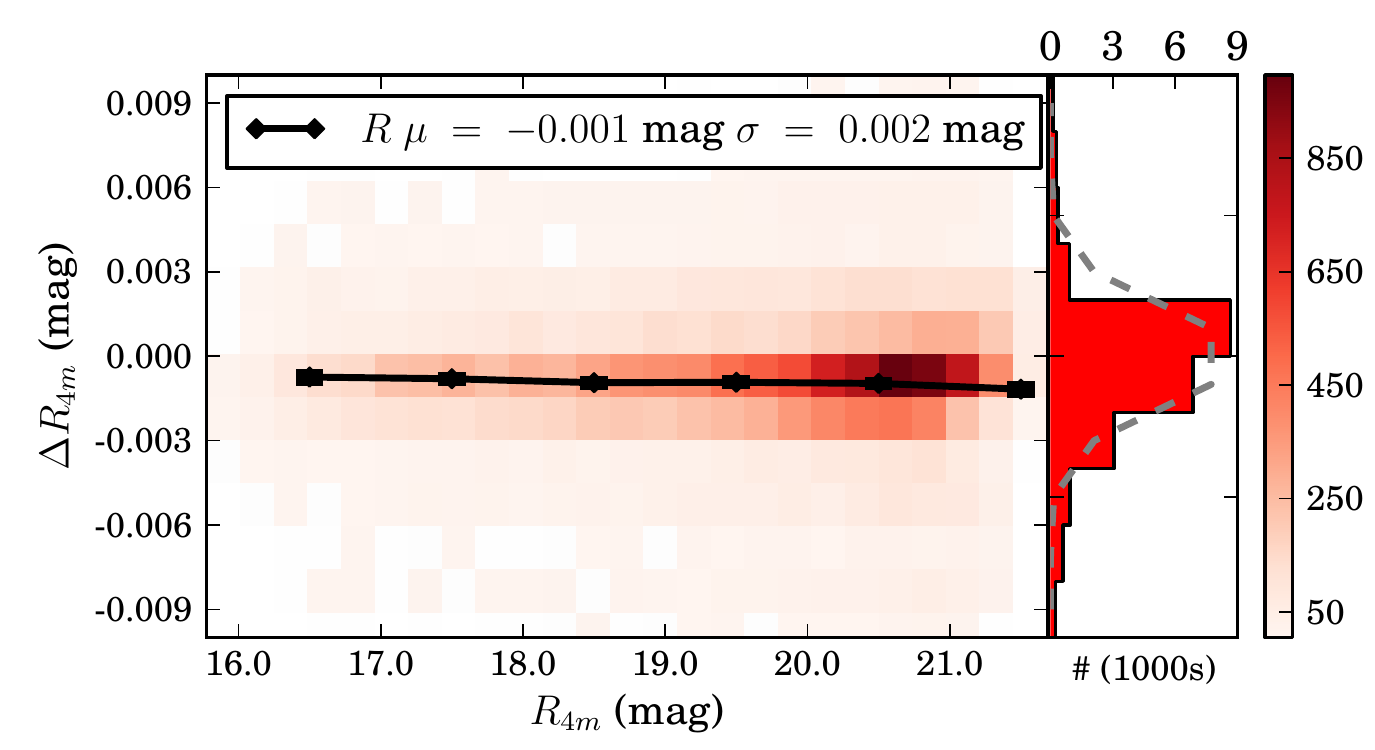}
\hfill
\includegraphics[width=0.45\textwidth]{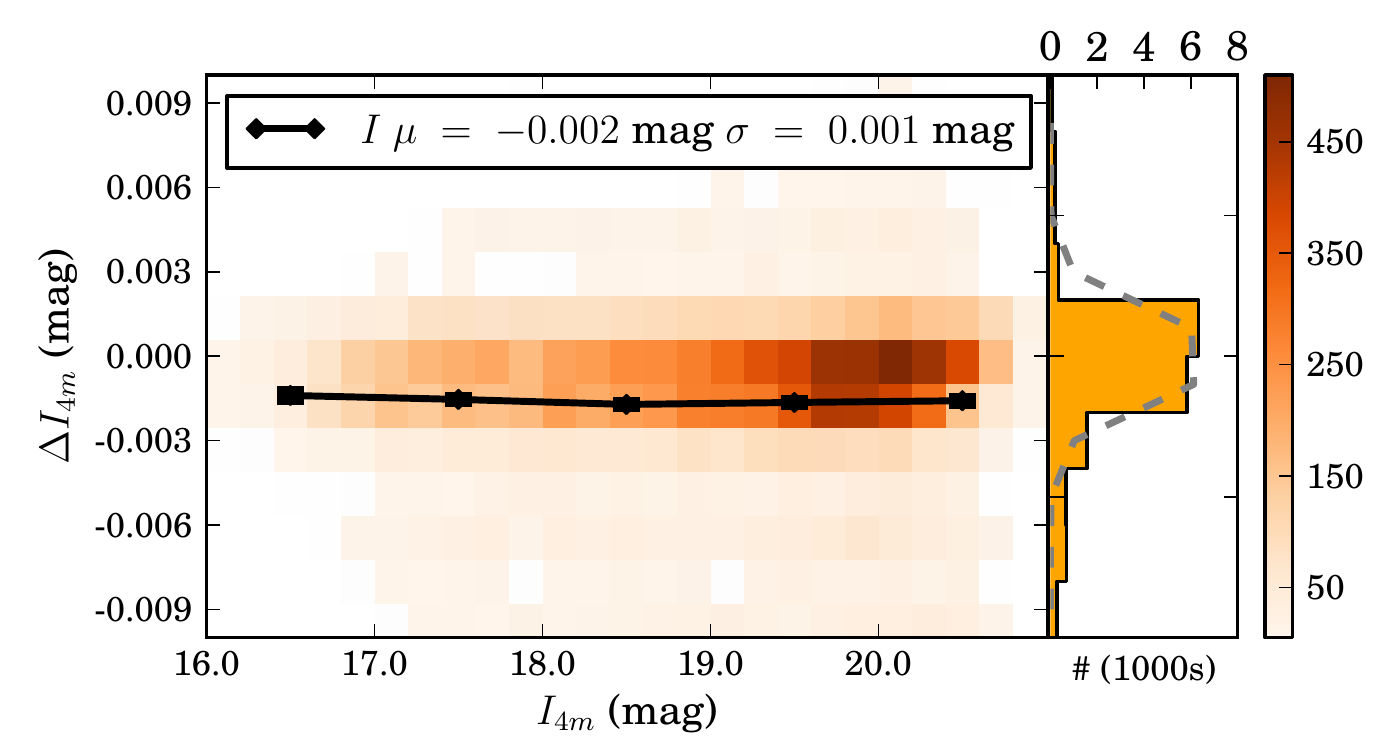} 
\caption[Effects of 10\% Extinction Errors on Tertiary Catalogs]{Differences in
$R$-band (left) and $I$-band (right) photometry with a 10\% error in the slope of
the airmass relation (in the sense of with offset {\it minus} no offset),
in each case binned as a function of magnitude (left large panel) and for all amplifier images
(right small panel).  The color bars indicate the number of amplifier images in each
bin.  The means in bins with 1~mag widths are overplotted (black diamonds). We
find $-0.001$~mag and $-0.002$~mag differences in the $R$ and $I$ bands, respectively. 
Dispersions are computed using a Gaussian fit to
the histograms shown in the right small panels.}\label{fig:airmasssyst} 
\end{figure*}

We find that a $\pm10$\% error in the airmass term causes a $\mp0.001$~mag
error in $R$ and a $\mp0.002$~mag error in $I$ (see
Fig.~\ref{fig:airmasssyst}). The slightly larger effect in $I$, despite the
weaker extinction coefficient, is a result of the  smaller number of $I$ images
overall, and the larger fraction of $I$ calibration images that were observed
at high airmass, relative to $R$. 

We also compute the mean difference as a function of magnitude to look for any
residual trends.  This is critical for \snia\ measurements which span a wide
range in magnitude as a function of redshift, light-curve shape, and
host-galaxy extinction. We find weak ($<0.1$\%) trends as a function of magnitude.

\subsection{Uncertainties in Determining the Photometric Transformation to the Landolt System}

Extinction caused by dust in the host galaxies of the supernovae makes them appear
fainter than predicted for their redshift, mimicking the effect of dark
energy. Accurate measurements of \snia\ color are critical in constraining the
reddening and allow us to disentangle the effect of dust from the dark-energy
signal. 

Because high-$z$ \snia\ surveys are deep but cover a small solid angle,
they are inefficient at finding large numbers of nearby supernovae.  Analysis
of the high-$z$ samples requires low-$z$ \snia\ from the literature as an anchor
for cosmological measurements
\citep{hamuy93b,Riess99,Jha06,Hicken09a,Contreras10,Ganeshalingam10,Stritzinger11,Hicken12}.
Thus, any \emph{absolute} zero-point offsets, even if arising from inaccuracies
in the nearby sample, are a common source of systematic error for high-$z$
surveys. Because most nearby surveys are tied to the Landolt network, we can
estimate a lower bound to this offset by examining how the ESSENCE data are
tied to the Landolt system.

In addition, any errors in our $R$ and $I$ flux scaling \emph{relative} to each
other would distort the observed color of the entire sample. At the typical
redshifts probed by ESSENCE, our $RI$ photometry covers the rest-frame $BV$,
and our inferred host-galaxy extinction is related to the measured rest-frame
color excess, $E(B-V)$. If we assume that the slope of the reddening law,
$R_{V}$,  in the host galaxies of our \snia\ is similar to that in our Milky 
Way Galaxy
($R_{V}\approx3.1$), then any error in our measured color would lead to an
error \about3 times larger in the extinction ($A_{V}$) and the distance
modulus ($\mu$). 

\snia\ surveys like ESSENCE are therefore particularly sensitive to systematics
affecting measured colors. An error in the photometric transformation can take
the form of an error in determining the slope of the color law, or a residual
difference in magnitudes around the intercept. The effect of an error in the
slope of the color relation is small, as the error is on the order of the
product of the error in the color term and the difference between the mean
color of our field stars and the color of BD+17\arcdeg4708. We measure the
effect of an error in the color term using synthetic photometry, as described
in Appendix~\ref{sec:synphot}, and find that a $\pm0.02$ error in the slope of
the color relation would lead to a \about$\pm0.003$~mag systematic error in the
magnitudes of field stars and derived zero points. We conservatively adopt an
error of 0.005~mag as the systematic error resulting from an error in the
estimate of the color term.  

\begin{figure*}[htb]
\centering
\plotone{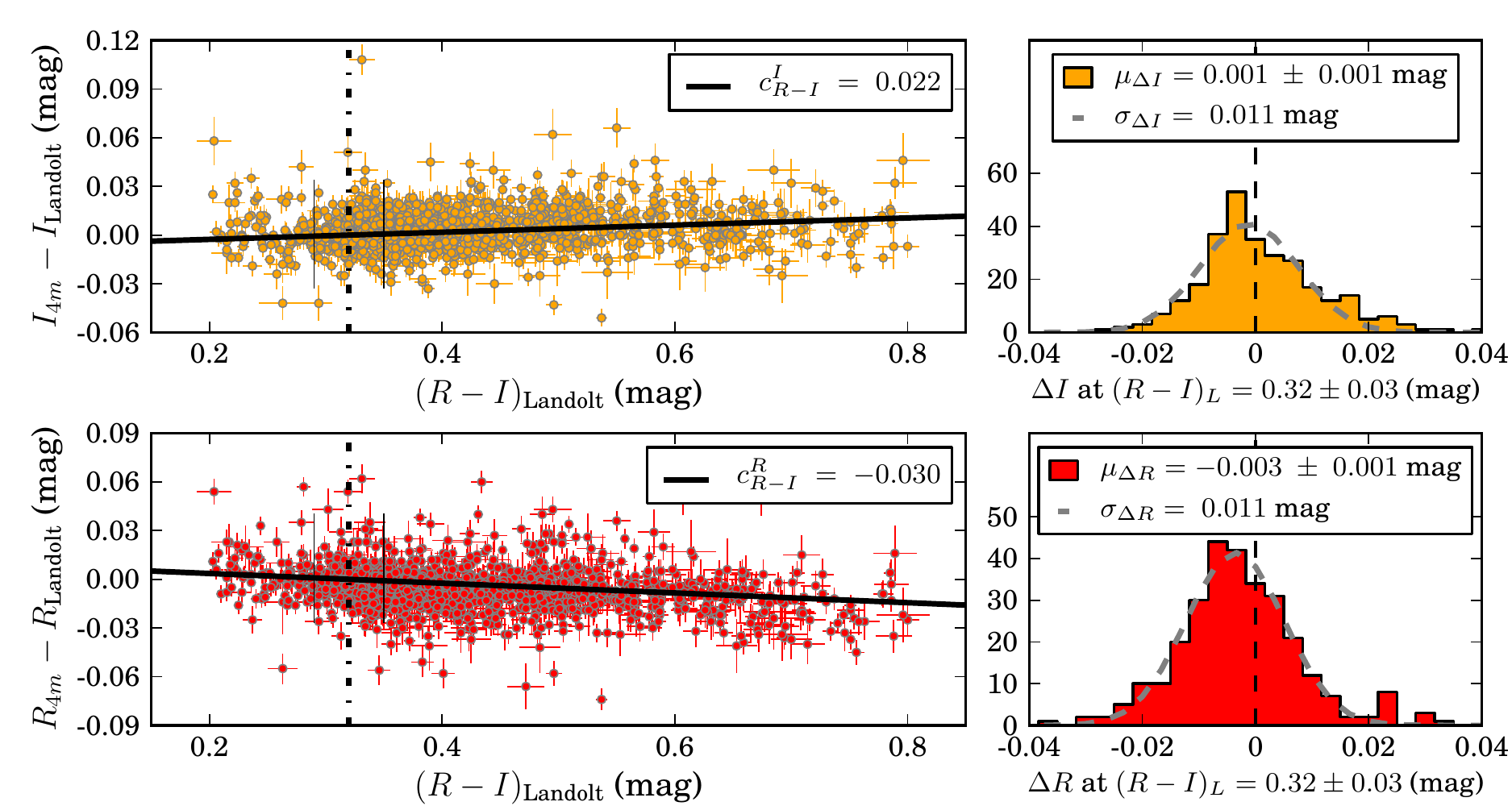}
\caption[Differences Between Landolt and CTIO 4m Natural System
Photometry]{Differences in the $I$-band (upper left) and $R$-band (lower left)
magnitude between the CTIO Blanco ($4m$) and the Landolt network in the standard fields.
The solid line indicates the slope of the color relation. The slopes are
determined at an intercept of $R-I = 0.32$~mag. We examine the residuals in a
range of $\pm0.03$~mag (this range is indicated by short vertical lines in the
left panels) around the color intercept. We find the differences in both $R$
(lower right) and $I$ (upper right) are consistent with zero, indicating that
there is no significant residual offset after the color transformation is
determined. }\label{fig:colormag} 
\end{figure*}

We measure the residual difference in magnitudes, $m_{4m} -
m_{\text{Landolt}}$, for our calibration fields around $R-I = 0.32$~mag in our
standard fields and find these to be $-0.003$~mag and $-0.001$~mag in $R$ and
$I$, respectively, with an uncertainty of \about0.001~mag in both bands (see
Fig.~\ref{fig:colormag}). The dispersions about the mean residual are \about1\%
in both $R$ and $I$. The residual is consistent with zero in $I$ and of low
significance in $R$. We adopt these values as systematic uncertainties in the
absolute zero point.

\subsubsection{Comparison to the Sloan Digital Sky Survey}

While we cannot directly compare our magnitudes to Landolt magnitudes in our
science fields, we use stars selected from SDSS DR7 and converted onto the
Landolt network using transformation equations. This procedure has some
limitations: the SDSS imaging is not as deep as MOSAIC~II images of the ESSENCE
fields, and SDSS magnitudes converted onto the Landolt network have large
statistical uncertainties associated with the transformation between two
dissimilar photometric systems. In addition, our ``wcc'' field is outside 
the SDSS footprint and is not included in the analysis.  However, as the SDSS
photometry was not used in determining our photometric calibration, it provides
a useful, independent test of our photometric accuracy. 

\begin{figure}[htb]
\centering
\plotone{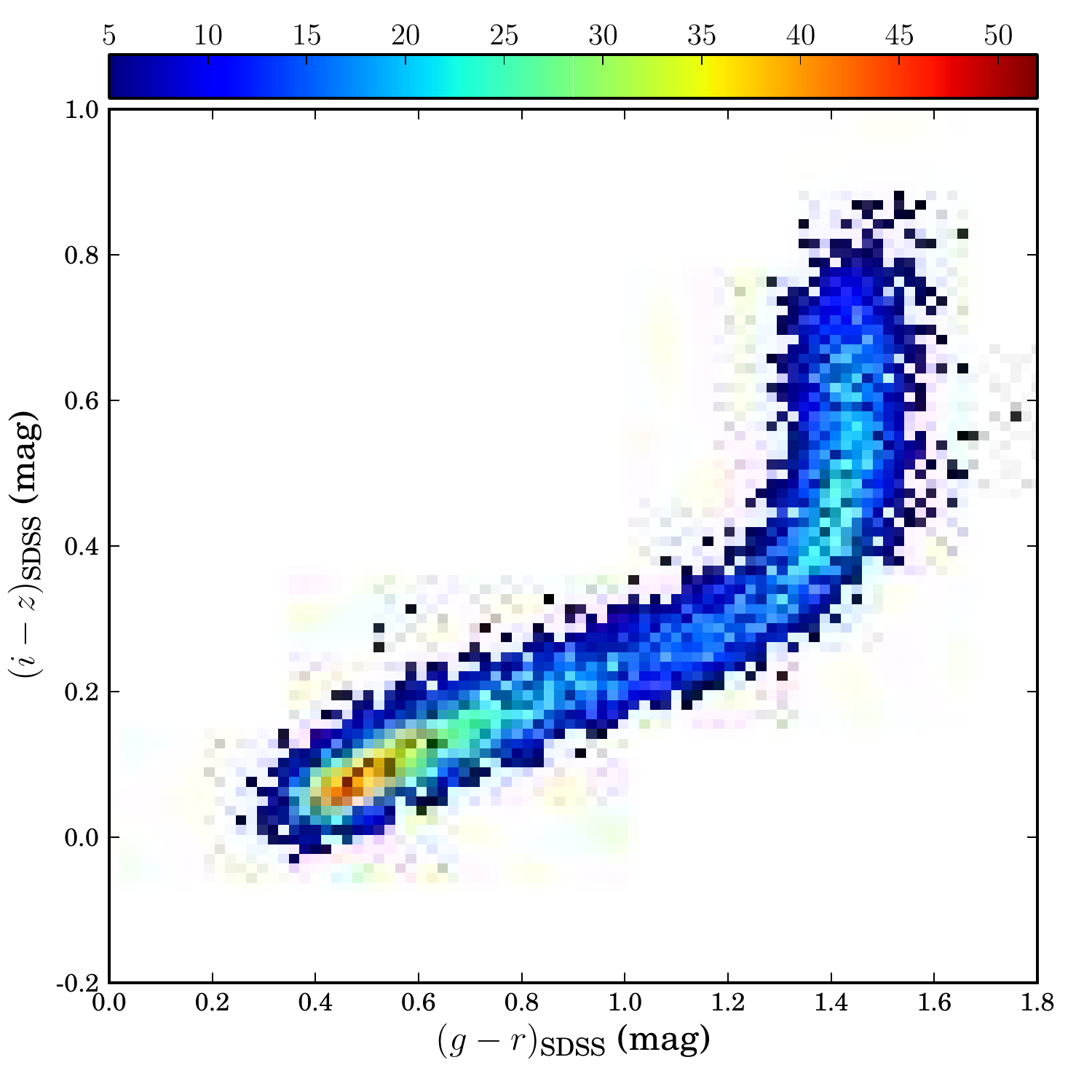} 
\caption[SDSS Stars selected in ESSENCE fields]{Color-color diagram of SDSS
stars in ESSENCE fields. Only stars with $r-i < 0.8$~mag are used to derive
transformations to Landolt and  assess our absolute photometric
consistency.}\label{fig:SDSSstars} \end{figure}

We cross-match stars from \citet{stetson2005} in SDSS, and extract \"{U}bercal
\citep{Padmanabhan08} corrected magnitudes. We select from SDSS only 
objects with clean photometry, point-source PSFs, \gri\ uncertainties 
$< 0.1$~mag, and $\sigma_{z} < 0.15$~mag without a corresponding entry 
in the DR7 QSOBest catalog, satisfying 

\begin{eqnarray*}
\begin{aligned}
\displaystyle  0.95 &< u-g < 2.75~{\rm mag},\\ 
\displaystyle -0.01 &< g-r < 1.78~{\rm mag},\\
\displaystyle -0.12 &< r-i < 2.74~{\rm mag,~and},\\
\displaystyle -0.13 &< i-z < 1.58~{\rm mag}, 
\end{aligned}
\end{eqnarray*}

\noindent in close proximity to the stellar locus 
(see Fig.~\ref{fig:SDSSstars}).

\begin{figure*}[htb]
\includegraphics[width=\textwidth]{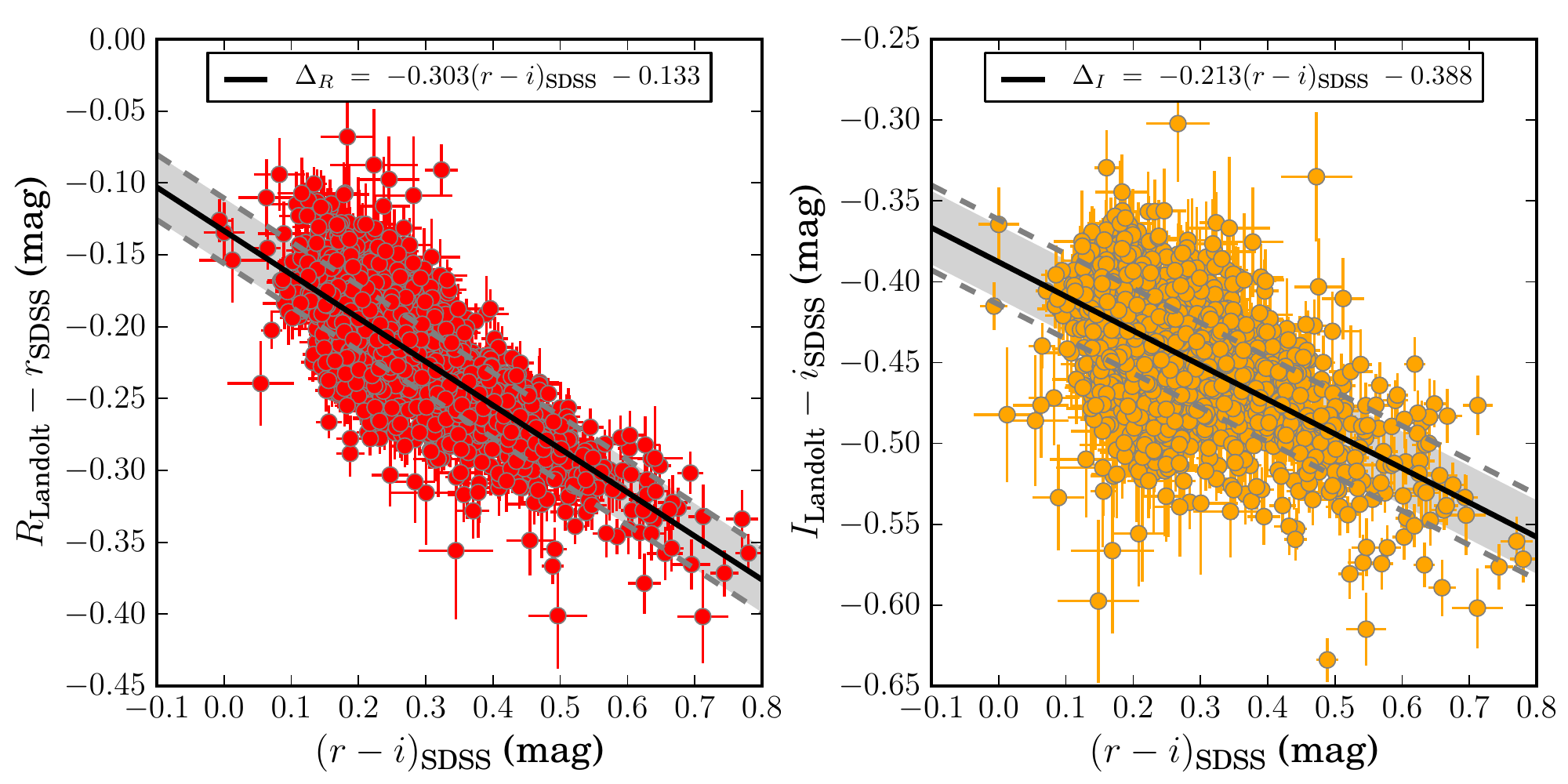} 
\caption[Transformations between SDSS DR7 and Landolt
Photometry]{Transformations between the Landolt and SDSS photometric systems
using stars observed by \citet{stetson2005} for $R$ (left) and $I$ (right). The shaded gray
region enclosed between dashed gray lines indicates the intrinsic dispersion in
the fit and is \about0.025~mag for both transformations.}\label{fig:SDSStrans}
\end{figure*}

We use simple linear transformations for stars with $r-i < 0.8$~mag, determined
using the ``\texttt{LINMIX\_ERR}'' routine \citep{Kelly07} available in the \texttt{IDL}
Astronomy Library\footnote{\url{http://idlastro.gsfc.nasa.gov/}}. We derive the
following transformations (see Fig.~\ref{fig:SDSStrans}) using $>1300$ measured
stars:

\begin{eqnarray}
\begin{aligned}
\displaystyle R_{L} &= r - (0.303 \pm 0.006)(r-i) - (0.133 \pm 0.002)~{\rm mag}, \\
\displaystyle I_{L} &= i - (0.213 \pm 0.007)(r-i) - (0.388 \pm 0.002)~{\rm mag}.
\end{aligned}
\end{eqnarray}

We find large intrinsic dispersions of \about0.025~mag in the relations between
the SDSS and Landolt photometry for both $R_{L}$ and $I_{L}$. This dispersion
is inherent in the transformation between two photometric systems having very
dissimilar transmissions and significantly different dynamic ranges, and
further justifies our choice to base the calibration of the Blanco natural system
on the Landolt standard network. We do not find any significant trend in the
residuals of transformed $R_{L}$ with $g-r$, or of $I_{L}$ with $i-z$. We apply
these transformations to SDSS stars in our science fields, selected using the
same criteria, to derive their Landolt magnitudes. 

We compare our tertiary photometric catalogs for the science fields to SDSS
stars, selected using the same criteria as above, converted to Landolt using
our derived transformations. No significant field-to-field differences
are found. We measure the offset between the CTIO Blanco natural system and
these transformed stars around $R-I = 0.32$~mag as in the standard fields, and
find offsets (in the sense of Blanco magnitude \emph{minus} transformed SDSS
magnitude) of $0.009 \pm 0.03$~mag in $R$ and $0.013 \pm 0.03$~mag in $I$,
consistent with zero. The large uncertainties arise from the intrinsic
dispersion in the transformation to $R$ and $I$, as well as the $r$ and $i$
uncertainties that are propagated into the uncertainty in $R-I$.

\subsection{Uncertainties in Extrapolating Photometric Zero Points}

\begin{figure*}[htb]
\centering
\includegraphics[width=\textwidth]{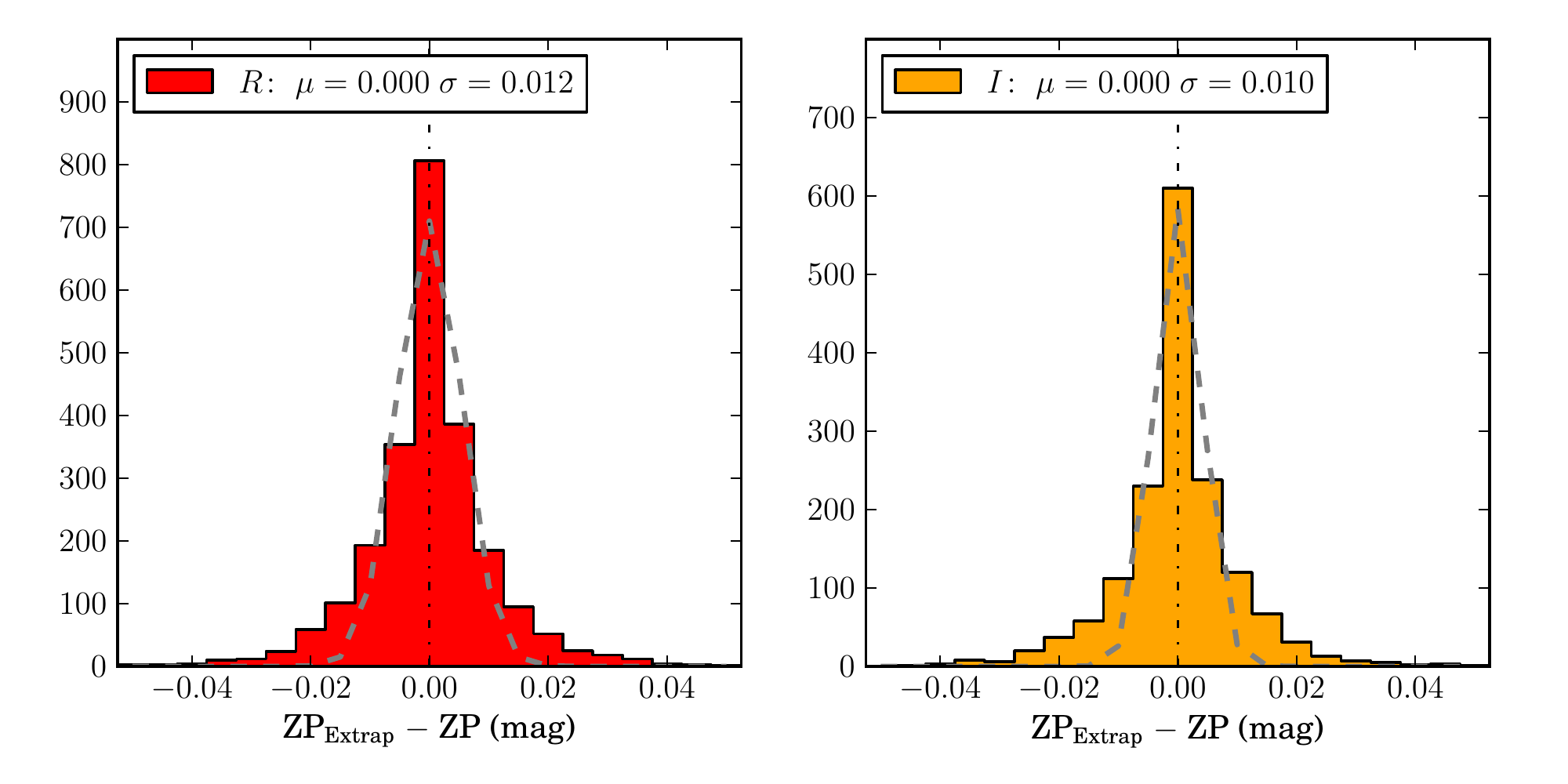}
\caption[Histogram of Extrapolated vs Measured Zero Points]{Differences between
the fitted zero point of an amplifier image and the zero point extrapolated
from the average of the other amplifiers of the same image, in $R$ (left) and
$I$ (right). We find no net offset between the directly fitted and extrapolated
zero points. In addition, the standard deviation of the residuals normalized by
the uncertainties is close to 1, indicating that the uncertainties are well
modeled. }\label{fig:zptextp}
\end{figure*}

We evaluate the uncertainty in determining the photometric zero point for a single
amplifier by extrapolating the zero point of the image using the average of all
other amplifiers of the same image, and the average of all other images that
are within $\pm0.5$ in airmass and $\pm100$~s in exposure time, adjusted
for both the difference in airmass and the difference in exposure time. We find
the difference between the zero point and the extrapolated zero point to be
$<0.001$ mag. We also construct this statistic field by field and amplifier by
amplifier, finding no significant difference in these subsamples.  The
histograms of differences between the extrapolated and directly fitted zero
points are shown in Fig.~\ref{fig:zptextp}.

Additionally, we find that the images with the largest differences between
fitted and extrapolated zero points are typically taken in nonphotometric
conditions and fail quality tests for difference imaging. We find no
significant trends in the difference between direct and extrapolated zero
points with airmass, aperture correction error, exposure time, FWHM, or sky
background.  The standard deviations of the unclipped data are \about0.01~mag
in both $R$ and $I$. Either $3\sigma$ clipping extreme outliers or using a
Gaussian to model the data reduces the estimate of the standard deviation to
$<0.01$ mag.  This is a strong indication that our internal photometric
calibration is no worse than $1$\%.

\subsection{Uncertainties in Determining the Natural-System Magnitudes of BD+17\arcdeg4708}

The CTIO Blanco natural-magnitude system adopted in this work utilizes
BD+17\arcdeg4708 as the fundamental spectrophotometric standard;
consequently, the magnitudes of BD+17\arcdeg4708 in the natural system are
close to its Landolt magnitudes by construction. However, there are several
astrophysical differences between BD+17\arcdeg4708 and the ``typical'' Landolt
standard star.  We determine the corrections to the first-order magnitudes of
BD+17\arcdeg4708 in Appendix~\ref{sec:bd17mags}.  Systematic errors in the
magnitudes of BD+17\arcdeg4708 would lead to an error in the synthetic zero
points and K-corrections, and the uncertainty budget is dominated by the
impact of a potential unresolved binary companion.  

\subsection{Uncertainties in the SED of BD+17\arcdeg4708}

While the derivation of the magnitudes of BD+17\arcdeg4708 in the Appendix
relies on the PHOENIX synthetic spectral library \citep[][and references
therein]{PhoenixOld,PhoenixLessOld}, the derivation of synthetic zero points
requires its true SED. We use the CALSPEC determination of the SED of the
BD+17\arcdeg4708 and adopt a 0.5\% uncertainty over 3000--10,000~\AA. These
translate into 0.002~mag and 0.003~mag differences in the synthetic $R$ and $I$
Blanco ($4m$) magnitudes.


\section{Conclusions}\label{sec:conclusions}

We have recalibrated the CTIO Blanco and MOSAIC~II system, with a focus on
minimizing the systematic errors that originate from photometry and affect the
high-redshift \snia\ measurements from ESSENCE. This calibration supersedes
that presented by \citet{miknaitis07}, and improves on it by deriving
photometric transformations between the Landolt network and the Blanco natural
system without employing any observations from the CTIO 0.9~m telescopes, 
thereby avoiding cross-telescope systematics.

Additionally, in this work we selected BD+17\arcdeg4708 as the fundamental
spectrophotometric standard star. The $R-I$ color of this standard is
considerably closer to the average color of Landolt network stars as well as of
\snia\ around the median redshift of ESSENCE. This choice  minimizes systematic
errors arising from errors in determining the photometric transformation
between the Landolt and Blanco natural systems. 

We employed these transformations to derive secondary photometric catalogs 
for our Landolt calibration fields that span the MOSAIC~II field of view. We
demonstrated that we could accurately extrapolate zero points between the
different amplifiers of the imager. Tertiary catalogs were derived for ESSENCE
fields and zero points were established for our imaging. The zero points in 
both passbands are stable relative to each other over the
entire duration of the survey. We provided a model of the system response
of the $R$ and $I$ ESSENCE passbands, and we made a comprehensive estimate
of the effect of various systematics on magnitudes in both passbands.

The primary application of this work is the calibration of light curves of
\snia\ discovered by ESSENCE, to derive the equation-of-state parameter ($w$) 
of the dark energy. We outlined our spectroscopic follow-up and
classification program to identify \snia\ within survey data, presenting
calibrated light curves in the CTIO Blanco natural system of \nsn\ \snia\
discovered by ESSENCE. 

There remain several potential areas for improvement where the calibration
given in this paper may be further refined. Our imaging of Landolt standard
fields was obtained during our 2006--2007 observing seasons. Consequently, we 
use science images obtained only on the same nights to derive our tertiary 
catalogs, rather than all survey images.  While there are clear changes in the
zero points over the course of the survey, the lack of standard-field imaging
covering the same range of time prevents us from deriving the absolute
CTIO-to-Landolt photometric transformation as a function of time. As our
fundamental spectrophotometric standard, BD+17\arcdeg4708, was not directly
observed using the CTIO Blanco, we derived estimates of its natural-system
magnitudes  using the CTIO-to-Landolt transformations, together with Landolt
photometry and the PHOENIX synthetic spectral library to characterize the
effect of metallicity, surface gravity, and extinction. However, the principal
shortcoming of the ESSENCE \snia\ photometry remains its lack of multicolor
information. This increases our sensitivity to priors on the colors or
extinction of \snia. Nevertheless, this work demonstrates that the systematic
errors from photometry are \about1\% in both $R$ and $I$. This represents a
better understanding of the systematic errors arising from photometric
calibration and an overall reduction of its impact on the ESSENCE systematic
error budget.

In future work, we will combine our calibrated light curves with our
spectroscopic observations, as well as \snia\ host-galaxy information
\citep[][submitted]{Tucker16}, to derive accurate distance moduli from ESSENCE.
We will combine our measurements with those from other low-$z$ and high-$z$
\snia\ surveys to place constraints on cosmological parameters.

\begin{acknowledgments} 

Based in part on observations obtained at the Cerro Tololo
Inter-American Observatory (CTIO), part of the National Optical
Astronomy Observatory (NOAO), which is operated by the Association of
Universities for Research in Astronomy (AURA), Inc. under a
cooperative agreement with the US National Science Foundation (NSF);
the European Southern Observatory, Chile (ESO Programmes 170.A-0519
and 176.A-0319); the Gemini Observatory, which is operated by AURA,
Inc.  under a cooperative agreement with the NSF on behalf of the
Gemini partnership [the NSF (United States), the Particle Physics and
  Astronomy Research Council (United Kingdom), the National Research
  Council (Canada), CONICYT (Chile), the Australian Research Council
  (Australia), CNPq (Brazil) and CONICET (Argentina) (Programs
  GN-2002B-Q-14, GS-2003B-Q-11, GN-2003B-Q-14, GS-2004B-Q-4,
  GN-2004B-Q-6, GS-2005B-Q-31, GN-2005B-Q-35)]; the Magellan
Telescopes at Las Campanas Observatory; the MMT Observatory, a joint
facility of the Smithsonian Institution and the University of Arizona;
and the F.  L. Whipple Observatory, which is operated by the
Smithsonian Astrophysical Observatory. Some of the data presented
herein were obtained at the W. M. Keck Observatory, which is operated
as a scientific partnership among the California Institute of
Technology, the University of California, and the National Aeronautics
and Space Administration. The Observatory was made possible by the
generous financial support of the W. M.  Keck Foundation.  The ESSENCE
survey team is very grateful to the scientific and technical staff at
the observatories we have been privileged to use.

The ESSENCE survey is supported by the NSF through grants AST-0443378 and
AST-0507475.  G.N. is supported by NSF award AST--0507475 and the Department of
Energy.  A.V.F.'s group at UC Berkeley received additional assistance from NSF
grants AST-0908886 and AST-1211916, the TABASGO Foundation, and the Christopher
R. Redlich Fund. The Dark Cosmology Centre is funded by the Danish National
Research Foundation.  J.M.S. is supported by an NSF Astronomy and Astrophysics
Postdoctoral Fellowship under award AST-1302771. A.C. acknowledges support from
grant IC120009 awarded to the Millennium Institute of Astrophysics, MAS, by the
Ministry of Economy, Development and Tourism, and grant Basal CATA PFB 06/09
from CONICYT.

Our project was made possible by the survey program administered by
NOAO, and builds upon the data-reduction pipeline developed by the
SuperMACHO collaboration.  

We made extensive use of the Odyssey Cluster administered by the FAS-IT
Research Computing Group at Harvard and are very grateful to the staff there.

\end{acknowledgments}

{\it Facilities:}
\facility{Blanco (MOSAIC II)}, 
\facility{CTIO:0.9m (CFCCD)}, 
\facility{Gemini:South (GMOS)}, 
\facility{Gemini:North (GMOS)}, 
\facility{Keck:I (LRIS)},
\facility{Keck:II (DEIMOS, ESI)},
\facility{VLT (FORS1)},
\facility{Magellan:Baade (IMACS)}, 
\facility{Magellan:Clay (LDSS2)}.


\clearpage
\newpage
\LongTables
\begin{deluxetable*}{llllrccccccc}
\tabletypesize{\scriptsize}
\tablewidth{0pt}
\tablecolumns{14}
\tablecaption{Transient Objects Considered for Follow-up Observations by the ESSENCE Survey\label{tab:object}}
\tablehead{
    \colhead{ESSENCE ID} &
    \colhead{IAU ID} &
    \colhead{Type} &
    \colhead{Subfield} &
    \colhead{Amp} &
    \colhead{$\alpha$ (h:m:s)} &
    \colhead{$\delta$(J2000) ($^\circ:':''$)} &
    \colhead{$z_{\rm SNID}$ } &
    \colhead{$\sigma_{z}$} &
    \colhead{$z_{\rm Gal}$} &
    \colhead{$\sigma_{z_{\rm Gal}}$} &
    \colhead{$E(B-V)_{\rm MW}$} \\
}
\startdata
a002 &  ---    & Gal    & wxc1   &  4 & 23:34:48.580 & $-$10:11:02.260 & ---    & ---   & 0.3155  & 0.0002 & 0.02469   \\
b001 &  ---    & Unk    & wxc1   & 14 & 23:36:07.860 & $-$10:02:26.660 & ---    & ---   & ---     & ---    & 0.02580   \\
b002 &  ---    & Star   & wxh1   &  1 & 00:12:51.140 & $-$10:38:27.050 & ---    & ---   & $-$0.0003 & 0.0002 & 0.03826   \\
b003 &  2002iu & Ia     & wx06   & 14 & 00:13:33.100 & $-$10:13:09.920 & 0.115  & 0.005 & ---     & ---    & 0.03980   \\
b004 &  2002iv & IaT    & wx17   &  6 & 02:19:16.110 & $-$07:44:06.720 & 0.226  & 0.006 & 0.2283  & 0.0006 & 0.02329   \\
b005 &  2002iw & Gal    & wxd1   & 11 & 23:43:07.250 & $-$09:48:05.200 & ---    & ---   & 0.2048  & 0.0003 & 0.02989   \\
b006 &  2002ix & II?    & waa7   & 16 & 23:31:00.863 & $-$09:29:07.205 & ---    & ---   & ---     & ---    & 0.02854   \\
b008 &  2002jq & Ia     & wx01   &  5 & 23:35:57.960 & $-$10:05:56.880 & 0.477  & 0.005 & ---     & ---    & 0.02647   \\
\enddata
%
%
\end{deluxetable*}
\begin{minipage}{\textwidth}
\scriptsize
\noindent{
\begin{itemize}\itemsep1pt                                           
\item Ia  = Type Ia supernova, no subtype reported
\item IaT = similar to the overluminous Type Ia SN~1991T or SN~1999aa
\item IaP = similar to peculiar Type Ia supernovae SN~2000cx or SN~2002cx
\item Ib = Type Ib supernova, no subtype reported
\item Ib-pec = Type Ib supernova with peculiar spectral features
\item Ic = Type Ic supernova, no subtype reported
\item II  = Type II supernova, no subtype reported
\item IIn = Type II supernova with relatively narrow emission lines
\item IIP = Type II supernova with a ``plateau'' in the light curve 
\item II-pec = Type II supernova with peculiar spectral features
\item Classifications followed by a ``?'' are not definitive
\item PISN? = Possible pair-instability supernova (P. Garnavich, priv. communication)
\item Gal = Galaxy, subtypes are reported by \citet{Tucker16}
\item AGN = active galactic nucleus
\item Unk = not observed or could not be classified based on spectra.
\end{itemize}
}
\noindent{Notes: $z_{\text{SNID}}$ and $z_{\text{Gal}}$ are reported in the heliocentric frame, and must be converted into the cosmic microwave background (CMB) frame, while accounting for local peculiar velocities at low $z$. For this work, we have employed the Milky Way reddening values from \citet{schlegel98}, rather than the updated values provided by \citet{schlafly12}, to facilitate the combination of our objects with literature samples.}
\end{minipage}

\clearpage
\newpage

\begin{deluxetable*}{cclc}
\tabletypesize{\scriptsize}
\tablecolumns{4}
\tablewidth{0pt}
\tablecaption{Photometry of ESSENCE Objects\label{tab:lightcurve}}
\tablecomments{This table is published in its entirety in the electronic edition of the journal. A portion is shown here for guidance regarding its form and content.}
\tablehead{
    \colhead{MJD} & 
    \colhead{Passband} & 
    \colhead{Flux$_{25}$} & 
    \colhead{$\sigma_{\text{Flux}}$}}
\startdata
\multicolumn{4}{c}{k425}\\
52990.0582 & R & -0.331500 & 0.561100 \\
52990.0745 & I & -0.013500 & 1.108300 \\
52994.0601 & R & 0.468100 & 0.468700 \\
52994.0772 & I & 1.395600 & 0.782300 \\
53268.1072 & R & 0.029500 & 0.836900 \\
53268.1484 & I & -0.254100 & 1.248500 \\
53283.1311 & R & -0.815500 & 0.746300 \\
53283.1629 & I & 1.064200 & 1.631700 \\
53289.0561 & R & 0.149500 & 0.764900 \\
53289.0726 & I & -1.360000 & 1.253400 \\
53293.0558 & I & 1.089600 & 1.544700 \\
53297.0640 & R & -0.521300 & 0.616900 \\
53297.0809 & I & -1.181900 & 0.945800 \\
53301.0728 & R & 0.085300 & 1.094000 \\
53301.0973 & I & -1.519200 & 1.085500 \\
53315.0736 & R & 0.777700 & 0.493500 \\
53315.0906 & I & 0.207800 & 0.852200 \\
53323.0829 & R & 10.472000 & 0.477600 \\
53323.1006 & I & 11.788600 & 0.649700 \\
53329.0363 & R & 17.830900 & 0.655000 \\
53329.0533 & I & 24.461000 & 0.892200 \\
53342.0814 & R & 24.361700 & 0.582500 \\
53342.0984 & I & 30.781000 & 0.963700 \\
53346.0734 & R & 20.966300 & 0.543200 \\
53346.0919 & I & 25.465800 & 1.022600 \\
53350.0550 & R & 17.523900 & 1.083800 \\
53350.0724 & I & 21.250500 & 1.189000 \\
53358.0431 & R & 10.427300 & 0.720000 \\
53358.0720 & I & 16.345000 & 0.915900 \\
53360.0755 & R & 10.464500 & 1.257200 \\
53360.1073 & I & 14.251400 & 2.462600 \\
53385.0554 & R & 1.924400 & 0.784900 \\
53385.0589 & I & 4.661300 & 1.151400 \\
53639.0843 & R & -0.178200 & 0.629700 \\
53639.1007 & I & 0.131800 & 0.934900 \\
\enddata
\end{deluxetable*}
\clearpage
\bibliographystyle{apj}
\bibliography{W6yr}
\newpage
\begin{appendix}
\section{Estimation and Properties of the Illumination Correction}\label{sec:illum}

Flat-field images obtained with the CTIO Blanco are corrected using an
illumination correction derived from science images as described in
\S\ref{sec:detrend}. Here we detail the estimation of the illumination
corrections and their time dependence, and quantify the associated 
systematic errors.

\subsection{Deriving the Illumination Correction}

We create our illumination corrections using the following prescription:

\begin{enumerate}
\item Create a master dome flat from the set of dome flats, $F_D$,
\item Calibrate science frames, $F_S$, with the master dome flats.
\item Mask out all stars in the resulting science frames.
\item Normalize the masked science frames to the same average sky value.
\item Average the resulting frames to produce one combined image.
\item Normalize the combined image to a mean of unity, and take the multiplicative inverse.
\item Smooth the normalized combined image with a large kernel to generate the final illumination correction.
\item Multiply the dome flat with the illumination correction.
\end{enumerate}

Mathematically, we can describe our final illumination correction, $I(t)$, as

\begin{equation}\label{eqn:illum}
I = {\mathrm S}_K\left( \left< \frac{F_S}{\langle F_D \rangle} \right>^{-1} \right),
\end{equation}

\noindent where $S_K$ represents the smoothing kernel used in the stage, and
angled braces denote the average. We bin each $1024\times4096$~pixel
amplifier image by a factor of 4, and we smooth the binned image with a
$30\times30$~pixel area before re-expanding the binned image to the original
dimensions. This effective 120~pixel scale is larger than the small-scale
structures of the flat field, such as out-of-focus dust ``donuts,'' while
retaining the large-scale gradients that we seek to remove. Finally, we
construct a master illumination-corrected flat-field image, $F_I$, via

\begin{equation}\label{eqn:illcor}
F_I(t) = F_D(t) \times I(t),
\end{equation}

\noindent where $I$ and $F_D$ are normalized to an average value of $1.0$ and 
$t$ denotes the night of observation. The illumination-corrected
flat-field image is used to flatten the science images from the night.

\subsection{Temporal Stability of the Illumination Correction}

We distinguish two types of changes affecting the optical system:

\begin{enumerate}
\item Global changes that affect all images, including new dust grains on the
optics, changes in instrument mounting, and mechanical changes in the mirror
support. 
\item Flat-field changes that only affect our dome-flat images, including
ghosting, nonuniformity of the flat-field screen, and instances where a
flat-field lamp burned out. 
\end{enumerate}

\begin{figure*}[htb]
\centering
\includegraphics[width=\textwidth]{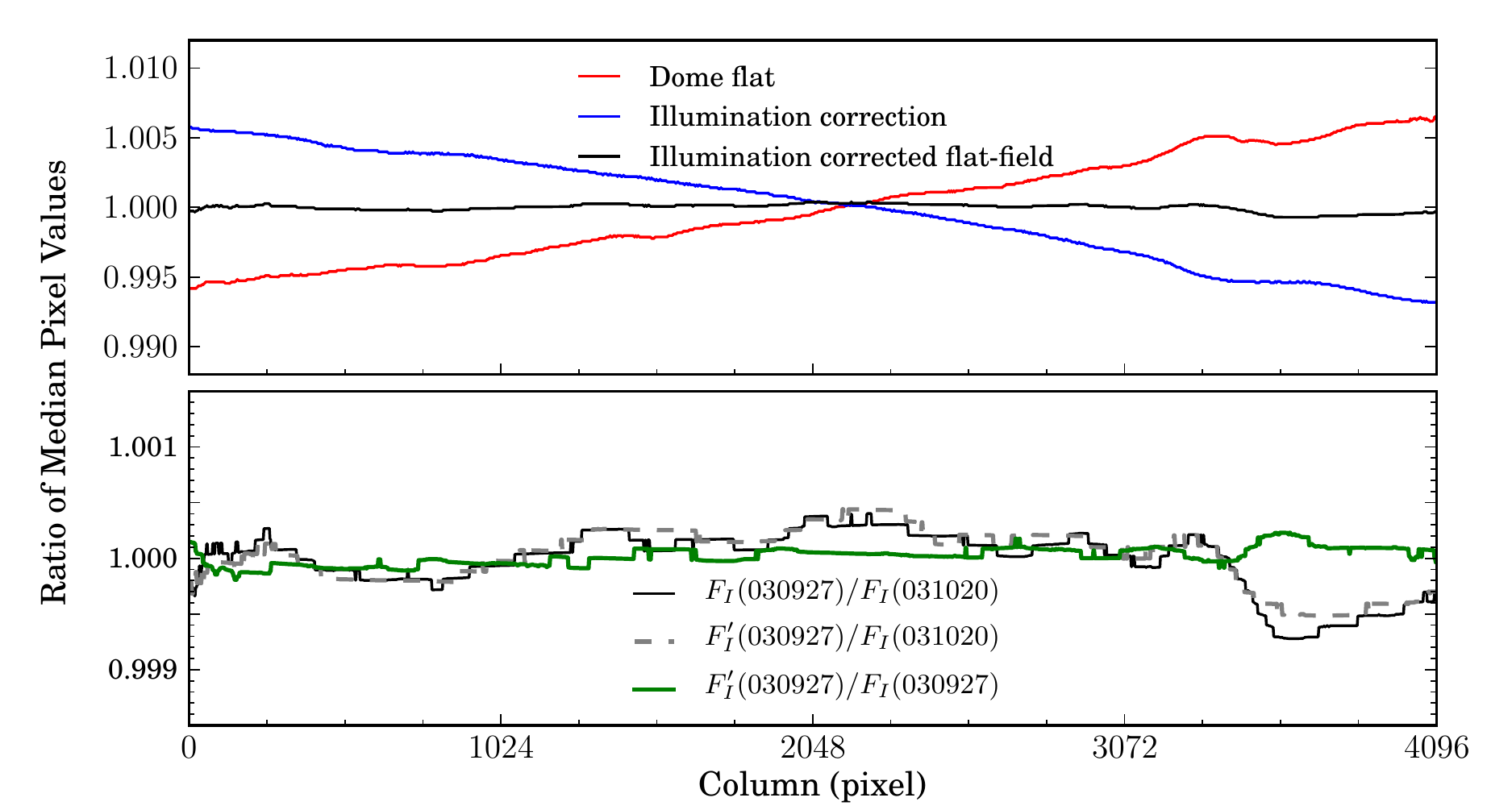}
\caption[Stability of Illumination Corrected Flat Fields]{An illustration of
the stability of the illumination corrections from 20030927 (randomly selected)
and 20031020 (\about1~month later). (Top) The ratio of the dome-flat frames is
shown in red, while the ratio of the illumination-correction frames is shown in
blue. Both ratios indicate that there are differences at the \about0.5\% level
between these two nights. The ratio of the illumination-corrected flat field
between the two nights, shown in black, is within 0.1\% of unity, indicating
that the illumination correction is accurately accounting for the variations in
the dome-flat images, despite them being separated by almost a month. (Bottom)
The ratio of the illumination-corrected flat fields between 20030927 and
20031020 is again shown in black on a finer scale to illustrate the structure.
We construct an estimated illumination correction for 20030927,
$F^{\prime}_{I}$, using the flat and bias images from 20031020 and the science
frames from 20030927. The ratio of the flat-field image processed with the
estimated illumination correction and the illumination-corrected flat field for
20031020 is shown as a dashed grey line. The ratio of the derived and the
estimated illumination-corrected flat fields on 20030927 is shown in green; it
illustrates that the illumination-corrected flat fields are stable to better
than 0.1\% between the two dates.
}\label{fig:ilstab}
\end{figure*}

We examine the temporal stability of the illumination-corrected flat fields
when subjected to both types of changes. From Equations~\ref{eqn:illum}
and~\ref{eqn:illcor}, provided the global changes are small, the product
$F_I(t)$ should not be sensitive to changes in the dome flats. We determine the
ratios of dome flat, illumination correction, and illumination-corrected
flat fields for all nights within an observing run. This is illustrated in
Fig.~\ref{fig:ilstab}, where we compare frames between 20030927 and 20031020.
The ratio of the illumination-corrected flat field images is within 0.1\% of
unity, despite differences at the 0.5\% level between the flat field and
illumination-correction frames. We calculate the standard deviation of the
ratio images, as well as the maximum difference between the ratio image and the
average of the ratio. Nights for which the standard deviation of the ratio is
consistently $>0.1$\%, or the maximum error of the ratio is consistently
$>0.3$\%, are flagged. Comparing our flagged nights to subjective observing
logs for the nights, we find that flagged nights have excessive moonlight.
This difference presumably arises from nonfocusing light paths producing stray
light illumination of the focal plane, with an intensity pattern different from
that of the light path for focused celestial sources. We find that the illumination
corrections degrade more rapidly toward full moon in $R$ than $I$, and we
attribute this to the steeper gradients in the sky brightness in $R$. This 
temporal stability is exploited to estimate an illumination correction for
flagged nights using other nights within the observing run. A 0.3\%
error is adopted as the systematic on the illumination-correction frames.
\section{Properties of the CTIO Blanco Natural System}\label{sec:natsys}

We describe the properties of the CTIO Blanco natural system in the following
subsections.  We derive the system transmission and compute synthetic color terms
to the Landolt system.  Our determined transmission and synthetic 
photometry of model SEDs are used to study the differences between our
fundamental spectrophotometric standard, BD+17\arcdeg4708, and ``typical''
Landolt stars at similar colors. Finally, we establish synthetic zero points to
derive natural-system magnitudes from flux-calibrated SEDs.

\subsection{Transmission}\label{sec:systrans}

We model the transmission ($T$) of the CTIO Blanco system by the product of four
components: the atmosphere (Atm), optics (Opt), filter (PB), and
quantum efficiency of the MOSAIC~II CCDs (QE):

\begin{equation}
T(\lambda) = T_{\text{Atm}}(\lambda) \times T_{\text{Opt}}(\lambda) \times T_{\text{PB}}(\lambda) \times \text{QE}(\lambda). 
\end{equation}

\subsubsection{Detector Quantum Efficiency}
The eight Tek CCDs that comprise the MOSAIC~II have slightly different QEs
(listed in Table\ref{tab:mosaicQE}). However, we find the
differences in synthetic photometry from using different QE
curves is $< 0.001$~mag for both $R$ and $I$ over a wide range of color.
Consequently, we elect to use a single average value of the QE
for all the CCDs.

\begin{deluxetable*}{crrrrrrrrr}
\tabletypesize{\scriptsize}
\tablewidth{0pt}
\tablecolumns{10}
\tablecaption{Quantum Efficiency of the MOSAIC~II Imager\label{tab:mosaicQE}}
\tablehead{
    \colhead{Wavelength} &
    \multicolumn{9}{c}{Transmission (\%)} \\
    \colhead{(\AA)} &
    \colhead{CCD 1} &
    \colhead{CCD 2} &
    \colhead{CCD 3} &
    \colhead{CCD 4} &
    \colhead{CCD 5} &
    \colhead{CCD 6} &
    \colhead{CCD 7} &
    \colhead{CCD 8} &
    \colhead{Average} \\
}
\startdata
3000   &  8.90 &  9.70 &  7.60 &  7.80 &  9.50 &  9.40 &  8.40 &  9.60 &   8.86 \\
3200   & 18.00 & 18.90 & 15.80 & 16.10 & 18.50 & 18.50 & 18.50 & 19.30 &  17.95 \\
3340   & 22.70 & 27.90 & 22.40 & 23.00 & 26.40 & 27.10 & 25.40 & 27.60 &  25.31 \\
3650   & 48.40 & 52.60 & 42.10 & 43.50 & 53.00 & 52.70 & 49.20 & 54.90 &  49.55 \\
3800   & 62.80 & 56.10 & 56.20 & 57.10 & 62.10 & 61.90 & 58.20 & 65.80 &  60.02 \\
4050   & 67.50 & 68.80 & 57.90 & 60.90 & 63.90 & 66.90 & 63.40 & 72.10 &  65.17 \\
4500   & 74.00 & 74.40 & 63.60 & 65.60 & 70.70 & 72.50 & 70.20 & 78.30 &  71.16 \\
5000   & 77.90 & 79.40 & 69.90 & 73.60 & 76.10 & 77.80 & 75.00 & 81.40 &  76.39 \\
5500   & 83.20 & 83.90 & 75.30 & 77.70 & 81.60 & 81.40 & 80.30 & 86.50 &  81.24 \\
6000   & 86.80 & 87.00 & 80.30 & 84.40 & 88.30 & 87.70 & 85.10 & 89.70 &  86.16 \\
6500   & 87.80 & 88.70 & 82.70 & 86.60 & 89.70 & 89.20 & 87.10 & 90.60 &  87.80 \\
7000   & 84.70 & 86.20 & 82.80 & 84.50 & 88.40 & 86.30 & 85.70 & 88.10 &  85.84 \\
7500   & 78.20 & 78.30 & 76.40 & 77.00 & 81.50 & 80.50 & 79.00 & 80.30 &  78.90 \\
8000   & 68.40 & 68.80 & 67.00 & 68.00 & 71.90 & 68.10 & 69.50 & 70.10 &  68.97 \\
8500   & 54.00 & 54.50 & 54.70 & 55.30 & 57.60 & 54.00 & 56.50 & 56.10 &  55.34 \\
9000   & 39.30 & 40.30 & 40.20 & 44.10 & 41.90 & 39.20 & 41.30 & 40.90 &  40.90 \\
9500   & 24.50 & 25.40 & 25.40 & 25.50 & 26.20 & 24.50 & 26.20 & 26.00 &  25.46 \\
10000  & 10.90 & 11.60 & 11.80 & 12.20 & 12.00 & 10.60 & 11.80 & 11.70 &  11.57 
\enddata
\end{deluxetable*}

\subsubsection{$R$ and $I$ Optical Filters}
The MOSAIC~II uses filters that are $146 \times 146$~mm and \about12~mm thick.
The transmissions of the $R$ (NOAO code c6004) and $I$ (c6028) filters were
measured by CTIO
staff\footnote{\url{http://www.ctio.noao.edu/noao/content/mosaic-filters}} using an
OceanOptics S2000 spectrometer. The S2000 is a crossed Czerny-Turner
spectrometer, configured with a 600~line mm$^{-1}$ grating blazed at 7500~\AA\ for
measurements over 6000--12,000~\AA. Measurements were obtained through a 10~$\mu$m
wide slit coupled to a fiber optic with 400~$\mu$m core diameter. The resulting
optical resolution is \about100~\AA\ FWHM. The filters are illuminated with a
General Electric 787 halogen lamp with quartz bulb, identical to those used to
illuminate the Blanco flat-field screen, through a ground-glass diffuser. The
spectrum is projected onto a $1\times2048$~pixel CCD array and digitized.  An
OceanOptics HG-1 He-Ar lamp produces reference spectral features to determine
the pixel-to-wavelength transformation. The transformation is modeled as a
simple cubic polynomial. The central wavelength of the filters is shifted
\about15~\AA\ to the blue when mounted in the prime focus of the $f/2.87$ beam
with ADC, relative to measurements at normal incidence. The shift is included
in the provided transmission curve. 

\subsubsection{Telescope Optics}
As the MOSAIC~II is mounted at prime focus, the transmission of the optics is
dominated by the wavelength-dependent reflectivity of the primary mirror, and
is well modeled by the reflectivity of aluminum. The transmission of the
ADC\footnote{\url{http://www.ctio.noao.edu/mosaic/manual/pfadc\_paper.ps}} was
measured to above 85\% in the range 3500--8500~\AA.  The transmission of the ADC
does fall significantly in the UV, but this has no effect on our $RI$
photometry. The dropoff at the red end is very gradual and the transmission at
10,000\AA\ is \about75\%. 

\subsubsection{Atmospheric Transmission}
M07 used a model of the atmospheric transmission derived from observations of
spectrophotometric standards, with removal of telluric features. The resulting
atmospheric model, while reasonably precise, depends on the standard used and
the details of the reduction, particularly on the fit of a smooth
psuedocontinuum. We generate an atmospheric model using the \texttt{MODTRAN4} code.  The
generated atmosphere is appropriate for an airmass of 1, and consists of 2~mm
PMW of water vapor at an altitude of 2~km, convolved with the atmospheric
scattering function and the transmission from aerosols.  The differences
between our atmospheric model and that employed by M07 are primarily in the
strength of the absorption features, with the largest differences on the
red wing of the $I$ band ($>9500$~\AA). The differences result in a
$<0.001$~mag change in synthetic colors over a wide range (note that the M07
transmission is provided in erg \AA$^{-1}$ and must be divided by 
$\lambda$ for comparison with this work).

The total system throughput at an airmass of unity is listed in
Table~\ref{tab:systemthroughput}.  Measurements of the system throughput using
a tunable laser, calibrated to a NIST photodiode, were consistent with the
product of each component \citep{stubbs07}. We could not measure the system
throughput of the $I$ filter (c6005) used very early in the survey and replaced
after significant damage in November 2002. 

\subsection{Synthetic Color Relations}\label{sec:synphot}

\begin{figure*}
\centering
\includegraphics[width=\textwidth]{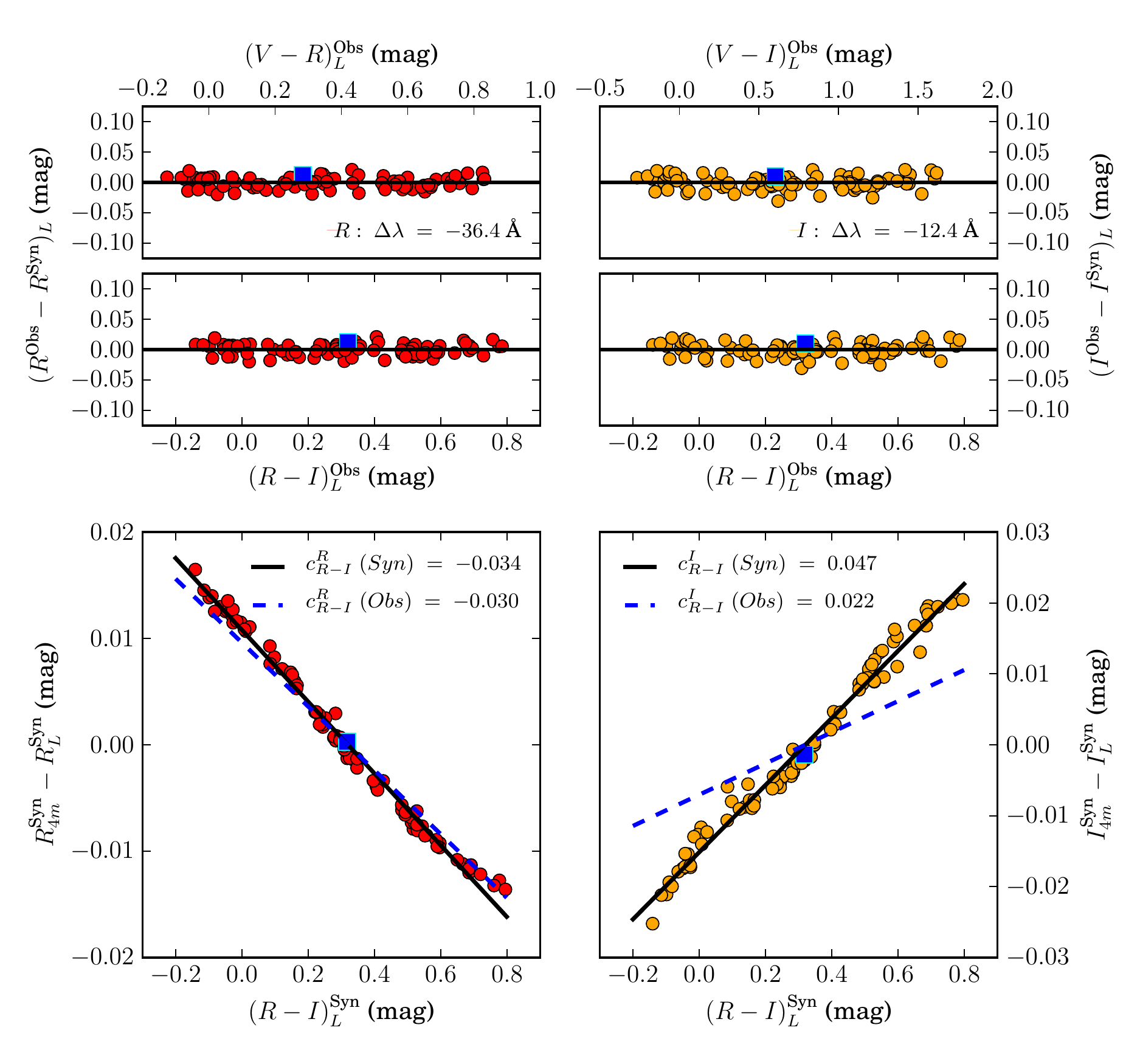}
\caption[Synthetic Color Relations between Landolt and the CTIO 4m Natural
System]{(Top and Middle) Residuals between observed (Obs) Landolt magnitudes
and synthetic (Syn) magnitudes of 99 nonvariable stars in the spectral library
of \citet{stritzinger05}, as a function of the observed Landolt color indicated
for $R$ (left panels) and $I$ (right panels). The Landolt passbands are
modeled by shifting the \citet{Bessell90} determinations in wavelength by
$\Delta \lambda = -36$~\AA\ and $-12$~\AA\ in $R$ and $I$, respectively. A
solid black line at $\Delta M = 0$ is included as a visual guide. (Bottom)
Synthetic color transformations between our determination of the CTIO system
throughput and the model Landolt system throughput for $R$ (left) and $I$
(right). The observed color relations from photometric
measurements is indicated by dashed blue lines, while the best-fit relation to
the synthetic photometry is indicated by a solid black line. There is excellent
agreement in $R$. We believe that the disagreement in $I$ is a result of not
modeling the rolloff in the detector QE for the model Landolt
throughput. A blueward shift of $-40$~\AA\ is sufficient to recover the observed
Landolt-to-CTIO color term in $I$, but it introduces a small color term between
the observed and synthetic Landolt measurements in $R-I$. The observed and
synthetic photometry of BD+17\arcdeg4708, using the CALSPEC SED, is indicated
by a blue square in all the plots. There is a \about1\% offset between the flux
calibration of the CALSPEC BD+17\arcdeg4708 SED and the mean flux calibration
of the \citet{stritzinger05} spectral library.}\label{fig:syncolor}
\end{figure*}

We derive synthetic color terms between the CTIO natural system and the Landolt
network, using a procedure similar to that of \citet{stritzinger05}. We
approximate the Landolt passbands using the Cousins $R_{C}$ and $I_{C}$
transmissions published by \citet{Bessell90}, convolved with a model
atmosphere, and shifted in wavelength by a small amount ($\Delta \lambda$). The
shifts are determined by comparing the observed Landolt photometry of the
nonvariable standards in the spectral library of \citet{stritzinger05} to
their synthetic photometry, and shifting the passbands without shifting the
atmospheric features, until the $R$ and $I$ synthetic and observed photometry
agreed, with a color term consistent with zero in $V-R$, $V-I$, and $R-I$. We
find that the $R$ and $I$ Bessell filters have to be \emph{blueshifted} by
36~\AA\ and 12~\AA\ in $R$ and $I$, respectively. Using our determination of the
CTIO system throughput in Table~\ref{tab:systemthroughput}, we compare
synthetic photometry of the spectral library to synthetic photometry through
the shifted Bessell passbands. We derive synthetic Landolt-to-CTIO color
transformations, finding $c^{R}_{R-I} (\text{Syn}) = -0.033$ and $c^{I}_{R-I}
(\text{Syn}) = 0.047$. The results of this analysis are presented in
Figure~\ref{fig:syncolor}.

The synthetic color term in $R$ is in excellent agreement with the color term
determined from photometric observations, but there is a significant
discrepancy in $I$. A blueshift of \about40~\AA\ to the $I$ Bessell
transmission is required to reproduce the observed Landolt-to-CTIO color term
in $I$, but a shift of this size introduces a nonzero $R-I$ color term between
the observed and synthetic Landolt magnitudes. There is no wavelength shift for
the Bessell determination of $I$ such that the synthetic and observed Landolt
magnitudes and the synthetic CTIO and synthetic Landolt magnitudes
simultaneously agree with nonzero color terms. Fundamentally, approximating
the Landolt $I$ passband by a shifted Bessell $I$ filter is not accurate, as
the shapes of these filters differ. Specifically, the transmission in the $I$
band is significantly affected by the rolloff in the detector QE,
which is not included in the Bessell determination. The detector
QE is effectively constant over $R$, and therefore has an
insignificant effect on the shape of the transmission. Current and future
surveys observing in \griz\ will be able to calibrate to photometric systems
such as SDSS, Pan-STARRS, and SkyMapper, which have well-measured system
responses.

\subsection{The Magnitudes of BD+17\arcdeg4708 in the CTIO Blanco Natural System}\label{sec:bd17mags}

\begin{figure*}
\centering
\includegraphics[width=0.95\textwidth]{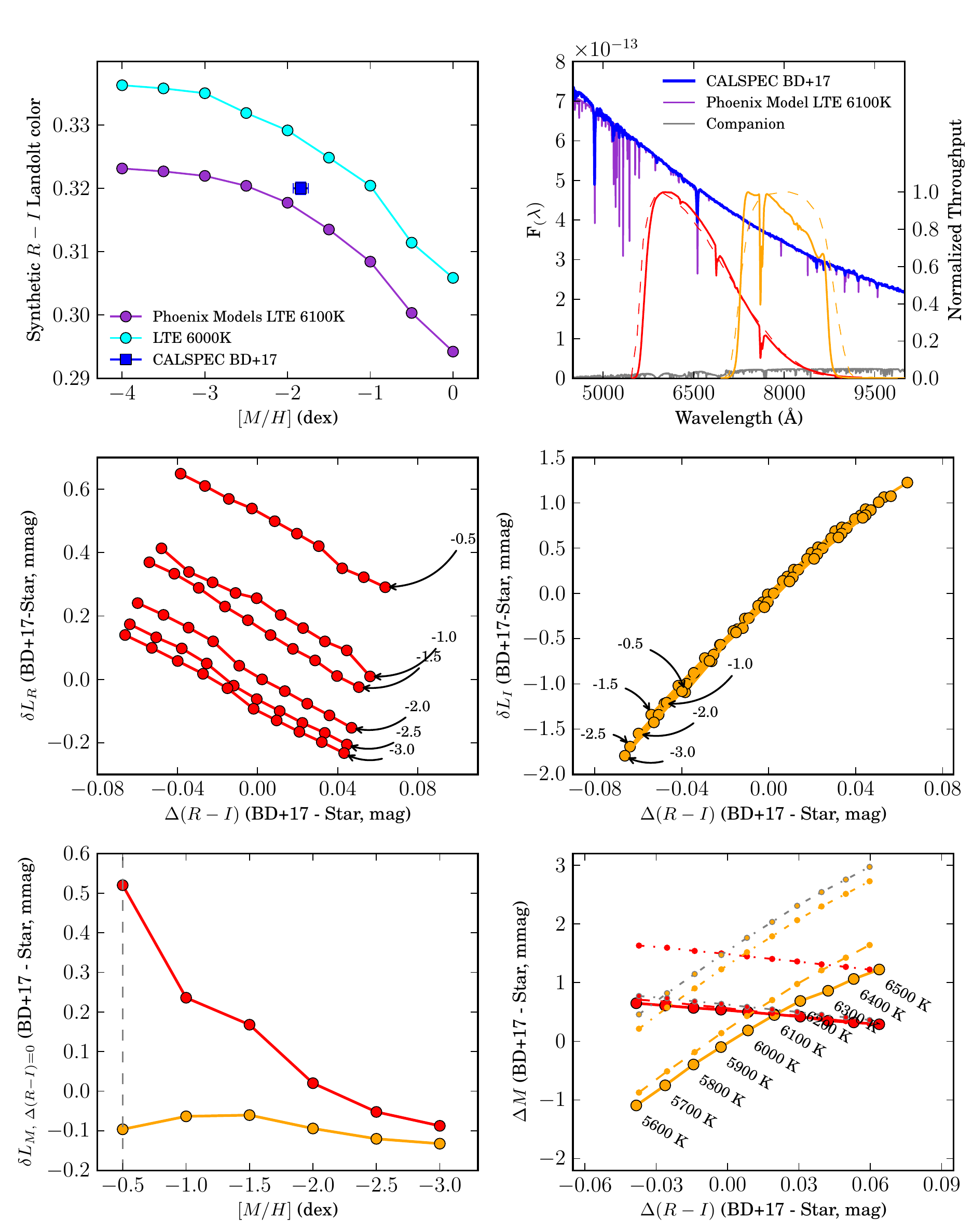}
\caption[Synthetic Magnitudes of BD+17\arcdeg4708 in the CTIO 4m Natural
System]{(Top left) Synthetic colors of Phoenix SEDs as a function of
metallicity, compared to the synthetic color of the CALSPEC SED of
BD+17\arcdeg4708. The best model has $T_{\text{eff}} = 6100$~K and [M/H] $=
-2.0$~dex. (Top right) Comparison of the CALSPEC SED and the adopted Phoenix
model. The adopted model for the companion of BD+17\arcdeg4708 has
$T_{\text{eff}} = 3000$~K and [M/H] $= -2.0$~dex. Normalized CTIO Blanco (solid)
and Bessell (dashed) transmissions in $R$ (red) and $I$ (orange) are shown for
comparison. (Middle) Difference in photometric residual, $\delta L$ [in the
sense of BD+17\arcdeg4708 mag \emph{minus} Landolt mag ($T_{\text{eff}}$,[M/H])] over a
range of temperature and metallicity for $R$ (Mid left) and $I$ (Mid right) vs. the
difference in $R-I$ color (in the sense of BD+17\arcdeg4708 color \emph{minus} Landolt
color). The effect of changing metallicity is negligible in $I$. (Bottom left)
The values of $\delta L$ at $\Delta(R-I) = 0$ for the different metallicities.
The typical metallicity of Landolt stars ([M/H] $= -0.5$~dex) is indicated by
the vertical line. (Bottom right) Deviations from the relation of $\delta L$ at
[M/H] $= -0.5$~dex are shown for changes in surface gravity (dashed),
extinction (dot-dashed), and the addition of a companion (dot-dashed grey) for
$R$ (red) and $I$ (orange).
}\label{fig:synbd17}
\end{figure*}

The transformations defined by Equation~\ref{eqn:phot} are constructed such
that, to first order, the natural-system magnitudes of BD+17\arcdeg4708 are
equal to its Landolt magnitudes in $R$ and $I$. However, since we could not
observe BD+17\arcdeg4708 directly, we determine the coefficients of the
transformation equations using the Landolt network of stars. In this subsection,
we quantify the difference in natural-system magnitudes between
BD+17\arcdeg4708 and Landolt stars having $R-I$ color similar to it by 
examining the photometric residual:

\begin{equation}
\begin{split}
\delta L_{T} = M_{4m} - M_{L} - c^{M_{4m}}_{(R-I)_{L}}((R-I)_{L}-0.32).
\end{split}
\end{equation}

\noindent By construction, the average residual $\langle \delta L \rangle \ = \
0$ for average Landolt stars. Following \citet{regnault09}, we consider the
photometric residual arising from metallicity and surface gravity, extinction 
differences between BD+17\arcdeg4708 and the average Landolt star, and 
the systematic effect of a possible faint, unresolved companion. The various
effects considered are illustrated in Figure~\ref{fig:synbd17}.

\subsubsection{Metallicity and Surface Gravity}

We determine the difference in synthetic $R$ and $I$ Blanco ($4m$) magnitude residuals
between BD+17\arcdeg4708 and ``typical'' Landolt stars with metallicity [M/H]
$= -0.5$ and log$(g) = 4.0$ as a function of the difference in synthetic $R-I$
color, over 5600~K$ < T_{\text{eff}} < $6500~K. We find the relationship
between the mean magnitude residual and difference in $R-I$ color to be linear
for both $R$ and $I$. We determine the intercept at $R-I = 0.32$ mag, and find (in
the sense of BD+17\arcdeg4708 mag \emph{minus} Landolt mag) that $\delta L <
0.001$~mag for both $R$ and $I$.

To measure the effect of surface gravity alone, we select synthetic SEDs with
the same parameters as above, except at log$(g) = 4.5$. We measure the
difference in the residual to normal Landolt stars, $\delta L_{M}$, caused by
perturbing the synthetic SEDs from log$(g) = 4.0$ to log$(g) = 4.5$. We find
the effect of changing surface gravity on the difference in residual (in the
sense of residual at log$(g)=4.5$ \emph{minus} residual at log$(g)=4.0$) is
$\delta L_{R} < 0.001$~mag, while $\delta L_{I} \approx +0.002$~mag. The
combined effect of metallicity and surface gravity leads to a negligible
difference in $R$ and a net $\delta L_{I}$ of \about0.001~mag.

\subsubsection{Extinction}

\citet{regnault09} express the distance of BD+17\arcdeg4708 from the stellar
locus in $V-R$ vs. $R-I$ in terms of the effect of the difference in extinction and
the difference in metallicity (the effect of surface gravity being negligible
over the color range in question). Having determined the effect of a difference
in metallicity using a procedure similar to that above, they found the
difference in the reddening between BD+17\arcdeg4708 and Landolt stars of
similar color to be $\Delta E(B-V) \approx 0.045$~mag. 

We redden the synthetic SED of BD+17\arcdeg4708 by this amount and examine the
difference in the residual to normal Landolt stars (in the sense of residual
with reddened SED \emph{minus} residual with unreddened SED) to be less than
0.001~mag in $R$ and $\sim0.001$~mag in $I$.

The combined effect of the difference in metallicity, surface gravity, and
extinction is found to be $\delta L_{R} = 0.001$~mag and $\delta L_{I} =
0.002$~mag. These offsets are added to the first-order estimates of the
magnitudes of BD+17\arcdeg4708.

\subsubsection{Binarity}

Using the estimates from \citet{ramirez06} for the possible companion of
BD+17\arcdeg4708 ($T_{\text{eff}} = 3000$~K, log$(g) = 4.5$, and $\left[{\rm M/H}\right] =
-2$), we compute the difference in photometric residuals and find (in the sense
of with companion \emph{minus} without companion) that $\delta L_{R,I} \
\approx \ 0.001$~mag. As we do not know the fraction of Landolt stars that are
also in binaries, we treat these offsets as systematic errors.

\subsection{Photometric Zero Points}

With the Landolt magnitudes of BD+17\arcdeg4708 ($R=9.166$~mag and
$I=8.846$~mag) and the photometric residuals caused by the differences in
metallicity, surface gravity, and extinction to typical Landolt stars computed
in the previous subsection, we invert Equation~\ref{eqn:synphot} to derive
synthetic passband zero points for $R$ and $I$ and find

\begin{eqnarray}
\begin{aligned}
\displaystyle \text{ZP}_{R_{4m}} &= -21.649 \pm 0.001~\text{mag}, \\
\displaystyle \text{ZP}_{I_{4m}} &= -22.305 \pm 0.002~\text{mag}.
\end{aligned}
\end{eqnarray}

These values differ from those determined using the \citet{stritzinger05}
SED library by 0.012~mag. This discrepancy is likely the result of a difference
in flux calibration between CALSPEC and \citet{stritzinger05} as illustrated in
Figure~\ref{fig:syncolor}. We use the CALSPEC SED of BD+17\arcdeg4708 for the
determination of passband zero points, as its flux calibration is not affected
by atmospheric transmission and has been carefully studied by several groups.

\subsection{Differences in Natural-System Definition from the 4~year Data Release}\label{sec:photdiff}

M07 tied the natural system of the Blanco to Landolt using $\alpha$~Lyrae as their
fundamental standard, with $(R-I)_{\rm Landolt} = 0$~mag. In addition, a slightly
steeper $c^{I}_{R-I}$ color term was employed in that work, and we expect a
difference on the order of the product of color-term differences and
the average color of Landolt stars, $\langle(R-I)_{\rm Landolt}\rangle$. 

To first order, the differences between the photometry of stars in this work
and M07 are the result of the differences between the \emph{definition} of the
photometric system in Equation~\ref{eqn:phot} and the M07 definition:

\begin{eqnarray}
\begin{aligned}
\displaystyle \Delta R & \approx c^{R}_{R-I}\times (R-I)_{\rm BD+17} \\
\displaystyle & \approx -0.030 \times 0.32 \\ 
\displaystyle & \approx -0.01~\text{mag}, \\
\displaystyle \Delta I & \approx c^{I}_{R-I}\times (R-I)_{\rm BD+17} \\
\displaystyle & + \Delta c^{I}_{R-I} \times (\langle (R-I)_{\rm Landolt}\rangle - (R-I)_{\rm BD+17}) \\
\displaystyle & \approx (0.030 \times 0.32) - 0.008 \times (0.47 - 0.32) \\
\displaystyle & \approx 0.009~\text{mag}.
\end{aligned}
\end{eqnarray}
\noindent
However, we have taken various measures to improve the calibration of the
natural system, with a view to minimizing our overall photometric error budget,
as discussed in \S\ref{sec:datared}. Consequently, the methodology used in this
paper differs substantially from that used by M07. In particular, this work
uses observations of the ESSENCE fields tied directly to Landolt fields,
whereas M07 tied the Blanco photometry to 0.9~m observations of field stars that
were in turn tied to Landolt. This is a potential source of additional
differences above the expected 1\% level. 

\begin{deluxetable*}{cccccc}
\tabletypesize{\scriptsize}
\tablewidth{0pt}
\tablecolumns{6}
\tablecaption{Photon Transmission Function of the ESSENCE Survey\label{tab:systemthroughput}}
\tablecomments{This table is published in its entirety in the electronic edition of the journal. A portion is shown here for guidance regarding its form and content.}
\tablehead{
    \colhead{Wavelength} &
    \multicolumn{5}{c}{Transmission (\%)} \\
    \colhead{(\AA)} &
    \colhead{QE} &
    \colhead{Filter} &
    \colhead{Optics} &
    \colhead{Atmosphere} &
    \colhead{Total} \\
}
\startdata
\multicolumn{6}{c}{$R$ (c6004)}\\
\hline \\
5470 & 0.8018 & 0.0000 & 0.9087  &    0.8400 & 0.0000 \\
...  & ...    & ...    & ...     &    ...    & ...    \\
6240 & 0.8760 & 0.7764 & 0.9021  &    0.8700 & 0.5338 \\
...  & ...    & ...    & ...     &    ...    & ...    \\
7005 & 0.8546 & 0.4846 & 0.8904  &    0.9160 & 0.3378 \\
...  & ...    & ...    & ...     &    ...    & ...    \\
7775 & 0.7454 & 0.1784 & 0.8725  &    0.9410 & 0.1092 \\
...  & ...    & ...    & ...     &    ...    & ...    \\
8540 & 0.5431 & 0.0409 & 0.8719  &    0.9530 & 0.0185 \\
...  & ...    & ...    & ...     &    ...    & ...    \\
9310 & 0.3111 & 0.0000 & 0.9133  &    0.8530 & 0.0000 \\
\hline \\
\multicolumn{6}{c}{$I$ (c6028)}\\
\hline \\
 6940 & 0.8592 & 0.0000 & 0.8916   &   0.9060 & 0.0000 \\
...  & ...    & ...    & ...     &    ...    & ...    \\
 7635 & 0.7721 & 0.9348 & 0.8763   &   0.5660 & 0.3580 \\
...  & ...    & ...    & ...     &    ...    & ...    \\
 8330 & 0.6052 & 0.9563 & 0.8674   &   0.9420 & 0.4729 \\
...  & ...    & ...    & ...     &    ...    & ...    \\
 9030 & 0.3959 & 0.0070 & 0.9007   &   0.9330 & 0.0023 \\
...  & ...    & ...    & ...     &    ...    & ...    \\
 9725 & 0.1912 & 0.0047 & 0.9275   &   0.9490 & 0.0008 \\
...  & ...    & ...    & ...     &    ...    & ...    \\
10425 & 0.0000 & 0.0047 & 0.9365   &   0.9690 & 0.0000 
\enddata
\end{deluxetable*}
\end{appendix}
\end{document}